\documentclass[twocolumn]{aastex631}

\usepackage{graphicx}
\usepackage{amsmath}

\usepackage{natbib}

\newcommand{\av}{$A_V$}

\newcommand{\izw}{I\,Zw\,18}
\newcommand{\msunyr}{M$_\odot$\,yr$^{-1}$}
\newcommand{\zsun}{$Z_\odot$}
\newcommand{\zzsun}{$Z/Z_\odot$}
\newcommand{\hi}{H{\sc i}}
\newcommand{\hii}{H{\sc ii}}
\newcommand{\htwo}{H$_2$}

\newcommand{\mhi}{M$_\mathrm{HI}$}

\newcommand{\msun}{M$_\odot$}

\newcommand{\logoh}{12$+$log(O/H)}
\newcommand{\cmtwo}{cm$^{-2}$}
\newcommand{\cmthree}{cm$^{-3}$}
\newcommand{\hst}{{HST}}
\newcommand{\jwst}{JWST}

\newcommand{\cii}{C\,{\sc ii}}
\newcommand{\oi}{O\,{\sc i}}

\newcommand{\oiv}{[O\,{\sc iv}]}

\newcommand{\nev}{[Ne\,{\sc v}]}

\newcommand{\spit}{\textit{Spitzer}}

\newcommand{\kms}{km\,s$^{-1}$}
\newcommand{\Ntot}{$N_\mathrm{tot}$}
\newcommand{\Nhi}{$N_\mathrm{HI}$}
\newcommand{\tu}{$T_u$}
\newcommand{\tl}{$T_l$}

\newcommand{\wms}{W\,m$^{-2}$\,sr$^{-1}$}

\newcommand{\sopr}{$S_\mathrm{OPR}$}
\newcommand{\nh}{n$_\mathrm{H}$}

\newcommand{\gnot}{$G_0$}

\newcommand{\al}{Al$_2$O$_3$}

\begin{document}

\title{The Interstellar Medium in \izw\ seen with JWST/MIRI: II. Warm Molecular Hydrogen and Warm Dust}

\correspondingauthor{Leslie Hunt}
\email{leslie.hunt@inaf.it}

\author[0000-0001-9162-2371]{L.~K. Hunt}
\affiliation{INAF -- Osservatorio Astrofisico di Arcetri, Largo E. Fermi 5, 50125 Firenze, Italy}

\author[0000-0002-0846-936X]{B.~T. Draine}
\affiliation{Dept. of Astrophysical Sciences, Princeton University, Princeton, NJ 08544, USA}

\author[0000-0002-1860-2304]{M.~G. Navarro}
\affiliation{INAF -- Osservatorio Astronomico di Roma, Via di Frascati 33, 00040 Monteporzio Catone, Italy}

\author[0000-0003-4137-882X]{A. Aloisi}
\affiliation{Astrophysics Division, Science Mission Directorate, NASA Headquarters, 300 E Street SW, Washington, DC 20546, USA}
\affiliation{Space Telescope Science Institute, 3700 San Martin Drive, Baltimore, MD 21218, USA}

\author[0000-0001-9719-4080]{R.~J. Rickards Vaught}
\affiliation{Space Telescope Science Institute, 3700 San Martin Drive, Baltimore, MD 21218, USA}

\author[0000-0002-8192-8091]{A. Adamo}
\affiliation{Department of Astronomy, The Oskar Klein Centre, Stockholm University, AlbaNova, SE-10691 Stockholm, Sweden}

\author[0000-0003-3758-4516]{F. Annibali}
\affiliation{INAF -- Osservatorio di Astrofisica e Scienza dello Spazio, Via Gobetti 93/3, 40129 Bologna, Italy}

\author[0000-0002-5189-8004]{D. Calzetti}
\affiliation{Department of Astronomy, University of Massachusetts Amherst, 710 North Pleasant Street, Amherst, MA 01003, USA}

\author[0000-0003-4857-8699]{S. Hernandez}
\affiliation{AURA for ESA, Space Telescope Science Institute, 3700 San Martin Drive, Baltimore, MD 21218, USA}

\author[0000-0003-4372-2006]{B.~L. James}
\affiliation{AURA for ESA, Space Telescope Science Institute, 3700 San Martin Drive, Baltimore, MD 21218, USA}

\author[0000-0003-2589-762X]{M. Mingozzi}
\affiliation{AURA for ESA, Space Telescope Science Institute, 3700 San Martin Drive, Baltimore, MD 21218, USA}

\author[0000-0001-9317-2888]{R. Schneider}
\affiliation{Dipartimento di Fisica, 'Sapienza' Universit{\`a} di Roma, Piazzale Aldo Moro 2, I-00185 Roma, Italy}
\affiliation{INAF -- Osservatorio Astronomico di Roma, Via di Frascati 33, I-00040 Monte Porzio Catone, Italy}
\affiliation{INFN, Sezione Roma1, Dipartimento di Fisica, 'Sapienza' Universit{\`a} di Roma, Piazzale Aldo Moro 2, I-00185 Roma, Italy}
\affiliation{Sapienza School for Advanced Studies, Viale Regina Elena 291, I-00161 Roma, Italy}

\author[0000-0002-0986-4759]{M. Tosi}
\affiliation{INAF -- Osservatorio di Astrofisica e Scienza dello Spazio, Via Gobetti 93/3, 40129 Bologna, Italy}

\author[0000-0001-9737-169X]{B. Brandl}
\affiliation{Leiden Observatory, Leiden University, PO Box 9513, 2300 RA Leiden, The Netherlands}
\affiliation{Faculty of Aerospace Engineering, Delft University of Technology, Kluyverweg 1, 2629 HS Delft, The Netherlands}

\author[0000-0002-0191-4897]{M.~G. del Valle-Espinosa}
\affiliation{Space Telescope Science Institute, 3700 San Martin Drive, Baltimore, MD 21218, USA}

\author[0000-0002-6460-3682]{F. Donnan}
\affiliation{Department of Physics, University of Oxford, Keble Road, Oxford, OX1 3RH, UK}

\author[0000-0002-2954-8622]{A.~S. Hirschauer}
\affiliation{Department of Physics \& Engineering Physics, Morgan State University, 1700 East Cold Spring Lane, Baltimore, MD 21251, USA}

\author[0000-0002-0522-3743]{M. Meixner}
\affiliation{Jet Propulsion Laboratory, California Institute of Technology, 4800 Oak Grove Dr., Pasadena, CA 91109, USA}

\author[0000-0001-6854-7545]{D. Rigopoulou}
\affiliation{Department of Physics, University of Oxford, Keble Road, Oxford OX1 3RH, UK}
\affiliation{School of Sciences, European University Cyprus, Diogenes street, Engomi, 1516 Nicosia, Cyprus}



\begin{abstract}
We present JWST/MIRI spectra from the Medium-Resolution Spectrometer of \izw,
a nearby dwarf galaxy with a metallicity of $\sim$3\% Solar.
Here, we investigate warm molecular hydrogen, \htwo, 
observed in spectra extracted in $\sim 120$\,pc apertures centered on eleven regions of interest.
We detect 7 \htwo\ rotational lines, some 
of which are among the weakest ever measured. 
The \htwo\ population diagrams are fit with 
local-thermodynamic-equilibrium models and models of photodissociation regions.
We also fit the ortho-/para-\htwo\ ratios (OPRs); 
in three of the six regions for which it was possible to fit the OPR,
we find values significantly greater than 3, the maximum value for local thermodynamic equilibrium.
To our knowledge, although predicted theoretically, this is the first time 
that OPR significantly $> 3$ has been measured in interstellar gas.
We find that OPR tends to increase with decreasing \htwo\ column density,
consistent with the expected effects of self-shielding in advancing photodissociation fronts.
The population diagrams are consistent with H nucleon densities of $\sim 10^5$\,\cmthree,
and an interstellar radiation field scaling factor, \gnot, of $\sim 10^3$. 
This warm, dense \htwo\ gas co-exists with the 
same highly ionized gas that emits \oiv\ and \nev.
Emission from $T\,\ga\,50$\,K dust is detected, 
including an as-yet unidentified dust emission feature near 14\,\micron;
possible identification as \al\ is discussed.
The continuum emission from several regions requires that a considerable
fraction of the refractory elements be incorporated in dust.
Despite stacking spectra in the SE where \htwo\ is found,
no significant emission from polycyclic aromatic hydrocarbons is detected.
\end{abstract}



\section{Introduction} \label{sec:intro}

Metal-poor dwarf galaxies at redshifts near or beyond the Epoch of Reionization (EoR)
are now being found routinely by \jwst\
\citep[e.,g.][]{trump23,furtak23,heintz23,rhoads23,vanzella23,morishita24,vanzella24}.
These galaxies are characterized by an intense star-formation rate (SFR) well above
galaxies with similar mass locally, consistently with the evolution of
the ``main sequence'' of star formation 
\citep{furtak23,calabro24}.
Given their high SFR, we would expect copious amounts of molecular gas to fuel their
star-formation activity, but this is difficult to test, 
since no observations of \htwo\ in similar environments are yet available.

Although molecular hydrogen, \htwo, is the most abundant molecule in the Universe,
it is not directly observable in emission from cold gas.
The rotational levels of \htwo\ are widely separated, 
with little or no thermal excitation in cold clouds. 
Fortunately, the heavier trace elements such as carbon and oxygen can combine 
in cold, dense clouds to form carbon monoxide, CO,
with rotational lines that are easily excited.
Thus, despite its low absolute abundance, $\sim 10^{-4}$ per H nucleon, CO has become
the commonly used local proxy for \htwo.

However, it is well known that in low-metallicity environments, CO is notoriously difficult to 
detect \citep{gondhalekar98,barone00,leroy07,cormier14,hunt14,hunt15}.
This is mainly due to photodissociation of CO, 
because in metal-poor and dust-poor environments \citep[e.g.,][]{vandishoeck88,bolatto99,wolfire10}, 
it is relatively unshielded from the strong radiation field (RF) produced by
massive stars. 
Consequently, the regions that contain CO shrink to the densest portions of the cloud, 
while other gas components such as singly ionized carbon C$^+$ and atomic carbon
become more dominant \citep[e.g.,][]{wolfire10}. 

It has been proposed that, in extremely metal-deficient environments,
stars can form directly from cold atomic hydrogen, \hi, rather than from the
conversion of \hi\ to dense molecular clouds \citep[e.g.,][]{glover12,krumholz12}.
Such a process would be extremely important, given that \hi\ tends to dominate
the gas budget in the interstellar medium (ISM) at low
metallicity locally \citep[e.g.,][]{bothwell13,hunt20},
and at high redshift \citep[e.g.,][]{popping14,walter20}. 

\jwst\ now offers the possibility to test the notion that star formation
can occur in extremely low-metallicity environments without molecular gas.
Here we present \jwst\ MIRI Medium Resolution Integral Field Unit Spectrometer (MRS)
observations of \izw, a relatively close by (18.2\,Mpc), 
star-forming dwarf galaxy at 3\%\,\zsun\ \citep[\logoh\,=\,7.18,][]{izotov99},
among the lowest metallicities in the nearby universe.
Like the high-$z$ dwarf galaxies, \izw\ is a starburst,
with a stellar mass of $\sim 10^6 - 10^7$\,\msun\ \citep{fumagalli10,madden14,jano17,nanni20},
an \hi\ mass of $\sim 10^8$\,\msun \citep{vanzee98,lelli12}, 
and an SFR estimated from the radio free-free continuum of $\sim 0.2$\,\msunyr\ \citep{hunt05}.
The SFR is estimated to have been even larger over the last 10\,Myr,
$\ga\,1$\,\msunyr\ \citep{annibali13,bortolini24}.
Consequently, we would expect significant \htwo\ emission if \htwo\ is, indeed,
necessary for star formation.

Our \jwst\ MIRI/MRS maps
of \izw\ cover the two star-forming complexes in its main body \citep[e.g.,][]{dufour90}. 
With \jwst, we can characterize the \htwo\ content of \izw\ with high spatial definition,
in an almost pristine environment similar 
to those of the dwarf galaxies newly discovered by \jwst\ at 
or beyond the EoR.
Our MIRI observations reveal high-ionization gas and \hi\ recombination lines
\citep[][hereafter Paper\,I]{hunt25}, 
as well as
a full complement of warm \htwo\ transitions and dust continuum, but no detectable emission from polycyclic aromatic
hydrocarbons (PAHs). 
In this paper, we focus on the \htwo\ and the dust emission.
The observations, their reduction, and analysis are summarized in Sect. \ref{sec:data}.
We analyze the warm \htwo\ in Sect. \ref{sec:pop} by fitting 
the population diagrams with empirical models, 
and also compare the population diagrams with state-of-the-art 
photodissociation region (PDR) models. 
Sect. \ref{sec:dust} discusses the dust continuum and the detection of an unidentified dust emission 
feature near 14\,\micron; a search for PAH emission is described in Sect. \ref{sec:pahs}.
We discuss our results in Sect. \ref{sec:discussion},
and summarize them in Sect. \ref{sec:summary}.
Forthcoming papers will describe other aspects of the data including
metallicity maps with \oiv\ 
\citep[][hereafter Paper\,III]{rickards25},
emission-line maps (Hunt et al., in prep.),
and inferred RF hardness from line-ratio maps (Rickards Vaught et al., in prep.).


\section{MIRI/MRS data and analysis\label{sec:data}}

We observed \izw\ on 8 March 2024 with the
\jwst\ MIRI/MRS \citep{argyriou23}
through the GO program \#3353 (PI Aloisi/coPI Hunt). 
Two overlapping pointings were executed, each of $\sim$8\,hrs duration, 
so as to encompass both the NW and SE OB complexes in the main body;
in an uninterruptible sequence, 
a separate background of the same duration was also observed.
An orientation range for position angle of the dedicated background observations 
enabled us to acquire on-source MIRI imaging in
F560W, F1130W, and F2550W to complement the filters acquired by GTO program \#1233 
\citep[PI Meixner, see][]{hirschauer24,bortolini24}. 
All three MIRI gratings, SHORT, MEDIUM, and LONG, were observed, in order to provide
spectral coverage from
$\sim 5-28$\,\micron\ with a spectral resolving power ranging from $\sim\,1550–3250$.
A 4-point dither pattern optimized for extended sources was employed, together with the SLOWR1 readout pattern,
having 40 groups per integration and 2 integrations per exposure.

The MRS data were reduced using custom pipelines, adapted from the MIRI \texttt{jupyter} notebook,
using CRDS (Calibration Reference Data System) version 11.17.25, and context \texttt{jwst\_1281.pmap}. 
These correspond to the latest calibrations given by \citet{law25}.
Because of the high signal-to-noise of the background observations, 
the background was subtracted in pixel-based mode (Stage 2) in order to better correct for localized 
spectral trends.
The fringing correction was performed both according to the ``standard'' pipeline, but also in a 
subsequent customized script created for this purpose.
For more details, see Paper\,I.

\begin{figure*}[t!]
\includegraphics[width=0.48\textwidth]{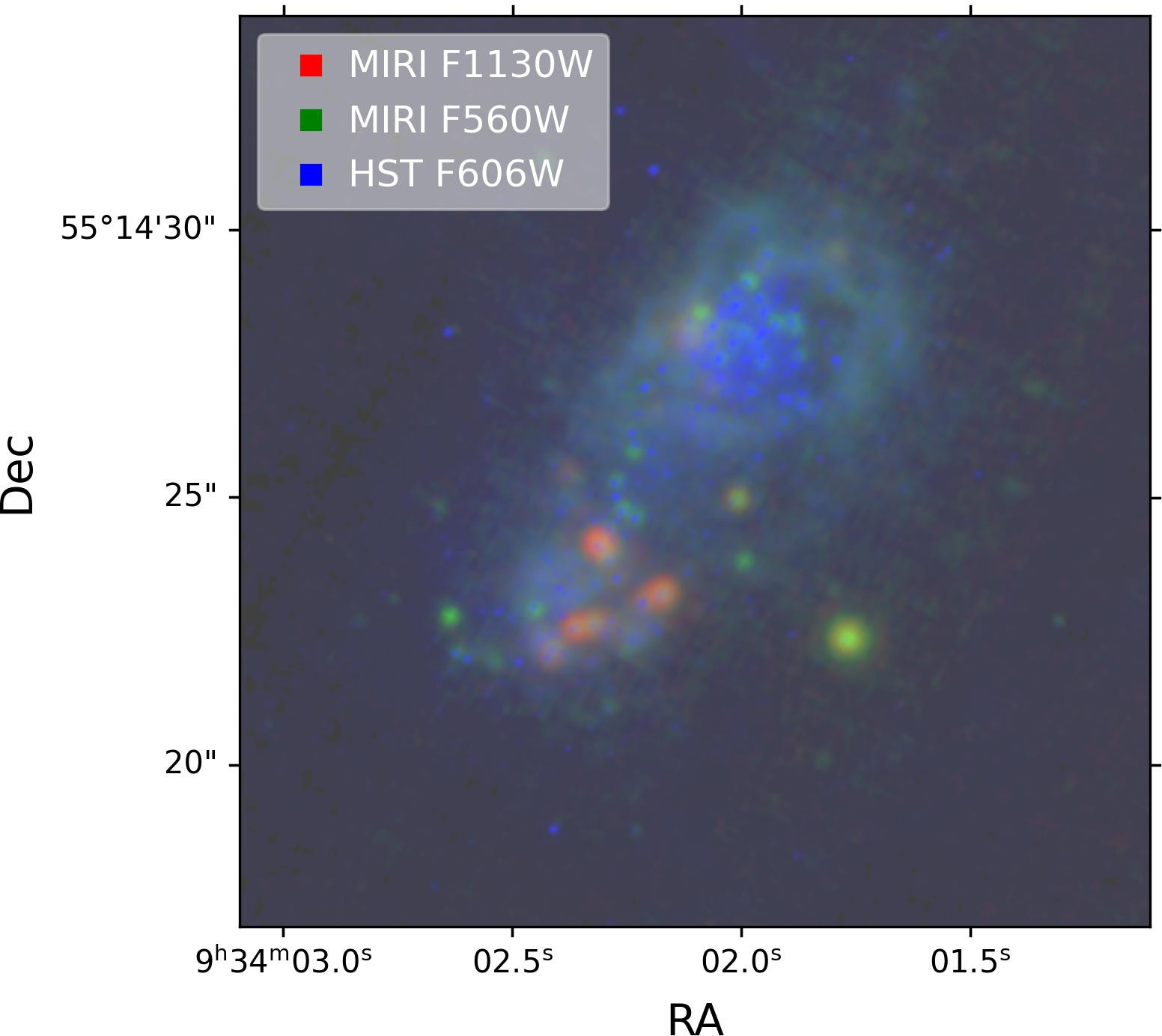}
\includegraphics[width=0.48\textwidth]{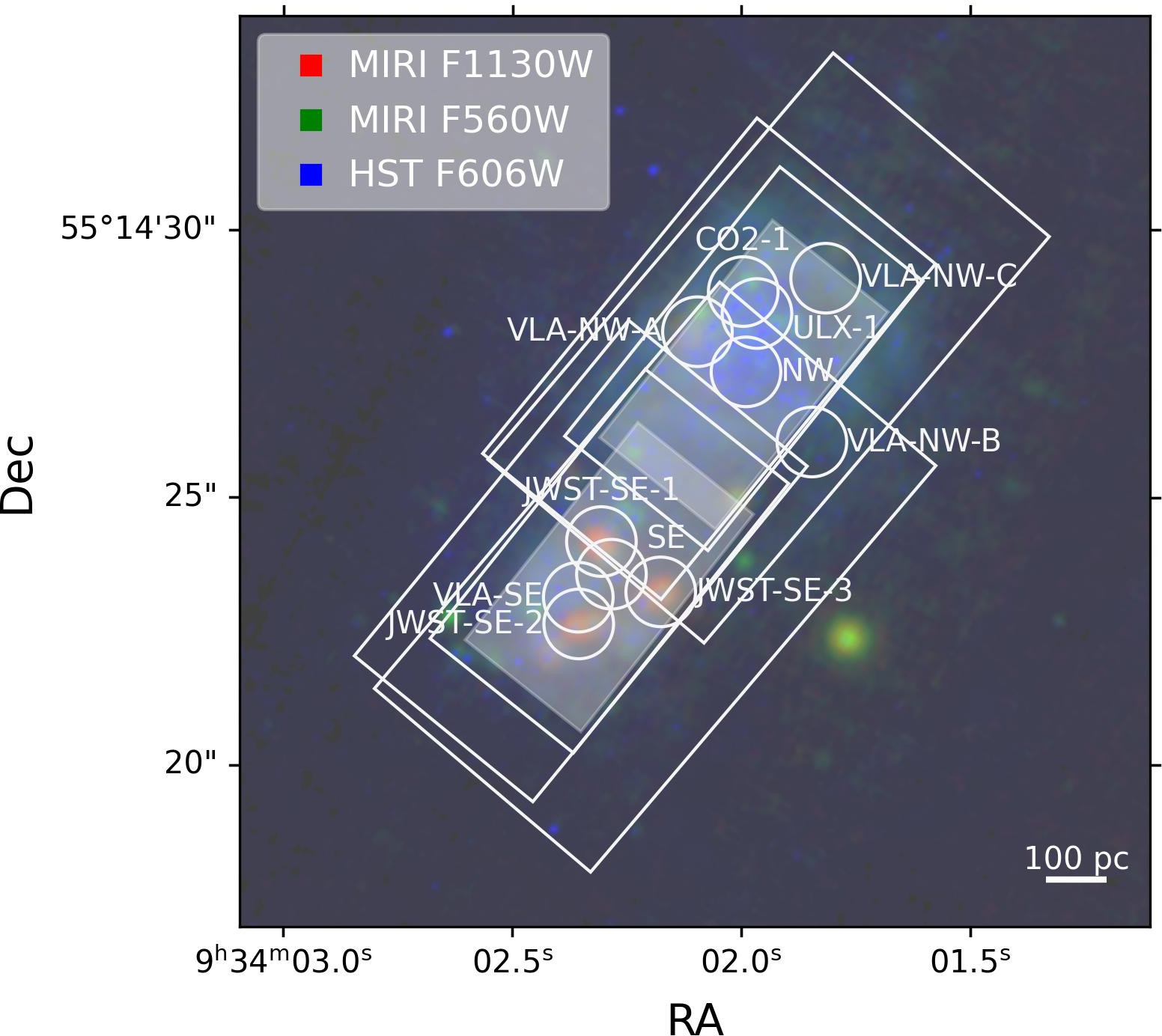}
\caption{17\arcsec$\times$17\arcsec\ RGB image of \izw\ showing the apertures of interest overlaid
in the right panel.
As in the legend, red is the MIRI F1130W background-subtracted image, green
the F560W background-subtracted image, and blue is the
\hst\ F606W image astrometrically corrected to Gaia.
In the right panel, the circles illustrate the apertures 
(0\farcs65 radius, $\sim 120$\,pc diameter) for spectral extraction;
the rectangles correspond to the MRS/IFU FoVs of the two separate pointings in the four channels.
The smallest FoV, Channel 1, is illustrated by a transparent shading; the two pointings overlap
allowing the construction of a complete spectral map even at the shortest wavelengths.
The MIRI 14\,\micron-continuum sources are apparent as red knots in the SE, 
and the faint knot in the NW associated with VLA-NW-A.
The horizontal line in the lower right corner corresponds to 100\,pc.
}
\label{fig:apertures}
\end{figure*}

\subsection{Convolved cubes \label{sec:convolved}}

The two pointings were combined with the \texttt{drizzle} method
\citep{law23} into 
a single set of cubes that consists of four channels, each conserving the 
native wavelength coverage and pixel sizes
(e.g., 0\farcs13, 0\farcs17, 0\farcs20, 0\farcs35, for Channels 1, 2, 3, and 4, respectively).
In order to 
determine the \htwo\ population diagrams,
it was necessary to convolve the original data cubes to the 
point-spread function (PSF) at a longer wavelength. 
We chose to convolve to 18\,\micron, slightly redward of the longest
wavelength of the \htwo\ transitions at $\lambda\,=\,17.03$\,\micron\ 
(see Table \ref{tab:htwo}).

To achieve matched spatial resolution at 18\,\micron,
we first assumed that the MIRI PSF is described by a 
Gaussian.
The MIRI PSF full-width half maximum (FWHM) at each wavelength was then estimated
according to \citet[][Eqn. (1)]{law23}, and the final Gaussian kernel was calculated,
assuming $\mathrm{FWHM}_\lambda = \sqrt{8\ln2}\sigma_\lambda$.
The kernel width was defined as the quadrature difference for each $\lambda$ plane in the cube,
such that it has a width $\sigma\,=\,\sqrt{\sigma_{18}^2  - \sigma_\lambda^2}$,
where $\sigma_\lambda$ is the intrinsic PSF $\sigma$ at the wavelength of interest.
These kernels were then convolved with the original cube to achieve 18\,\micron\ spatial resolution
at all planes with $\lambda\leq 18$\,\micron. 
Wavelength slices beyond 18\,\micron\ 
remain at their native resolution.

For the continuum fitting (Sect. \ref{sec:dust}),
we instead constructed a new set of cubes, convolved to the 27\,\micron\ PSF.
These are made in the same way as the 18\,\micron\ convolution, but 
for the $\lambda\,=27\,$\micron\ PSF.

\subsection{Aperture spectra \label{sec:spectra}}

Spectra were extracted from the convolved cubes
with apertures having radii of 0\farcs65 ($\sim 120$\,pc diameter),
centered on 11 regions of interest.
The spectral stitching was performed in a customized script,
followed by a one-dimensional (1D) residual fringe correction during the
spectral extraction phase to remove
high-frequency noise in Channels 3 and 4 \citep{kavanagh24}.
Additional details are given in Paper\,I.

The regions of interest comprise the four VLA \hii\ regions, denoted as VLA-NW-A, B, C, and VLA-SE, 
identified by the radio continuum maps 
with $\sim 2$\arcsec\ beam presented by \citet{cannon05}.
They also cover the peak of the $^{12}$CO(2--1) detection near the NW OB association 
by \citet{zhou21},
and the ultraluminous X-ray source ULX-1 identified by \citet{thuan04}.
In addition, apertures were placed on the approximate centers of the NW and SE 
OB complexes.
Visual inspection of the MIRI cubes showed four continuum sources.
Three are near the VLA-SE \hii\ region, although not exactly coincident with it.  
Thus, apertures for spectral extraction were also placed on the 14\,\micron\ peaks of this emission,
denoted as JWST-SE-1, JWST-SE-2, and JWST-SE-3. 
The fourth 14\,\micron\ source is coincident with VLA-NW-A. 
These continuum sources are also visible in the MIRI images presented by \citet{hirschauer24},
and discussed in Paper\,I.
Table \ref{tab:aper} reports the aperture centers used here (and in Paper\,I).

Figure \ref{fig:apertures} shows
an RGB image of \izw\ with blue taken as
the \hst/ACS F606W image of \izw\ from \citet{aloisi07}, but here
realigned to the Gaia DR3 \citep{gaiadr3} astrometric system.
The green color corresponds to the MIRI F560W, and the red to F1130W, both with a median
background subtracted.
The apertures shown in the right panel centered on the regions of interest are not independent,
but this does not interfere with our goal of sampling a variety
of individual regions with high signal-to-noise (S/N) to assess
differences in the properties of the \htwo.
The left panel of Fig. \ref{fig:apertures}
highlights clearly the presence of the reddish \jwst-identified 14\,\micron\ continuum sources
(Paper\,I).
The gas filaments toward the NW are also evident both in the \hst\ image \citep{aloisi07}, 
and in the MIRI F560W image,
illustrated by the greenish-blue color of the emission.

\begin{table}
\caption{Aperture centers for spectral extraction}
\begin{centering}
\begin{tabular}{lcc}
\hline
\hline
\multicolumn{1}{c}{\rule{0pt}{3ex} Region} &
\multicolumn{1}{c}{RA } &
\multicolumn{1}{c}{Dec.} \\
&
\multicolumn{2}{c}{(J2000)} \\
\hline
\\ 
NW        & 9:34:01.99 & +55:14:27.3 \\
VLA-NW-A  & 9:34:02.10 & +55:14:28.1$^a$\\
VLA-NW-B  & 9:34:01.85 & +55:14:26.0$^a$\\
VLA-NW-C  & 9:34:01.82 & +55:14:29.1$^a$\\
ULX-1     & 9:34:01.97 & +55:14:28.4$^b$\\
CO2-1     & 9:34:02.00 & +55:14:28.8$^c$\\
SE        & 9:34:02.29 & +55:14:23.5\\
VLA-SE    & 9:34:02.36 & +55:14:23.1$^a$\\
JWST-SE-1 & 9:34:02.31 & +55:14:24.1\\
JWST-SE-2 & 9:34:02.36 & +55:14:22.6\\
JWST-SE-3 & 9:34:02.18 & +55:14:23.2\\
\\
\hline
\hline
\end{tabular}
\end{centering}
\\
\newline
$^{a}$~\citet{cannon05}; \\
$^{b}$~\citet{thuan04,ott05,rickards21}; \\
$^{c}$~\citet{zhou21}.
\\
\label{tab:aper}
\end{table}

As in Paper\,I, we opted for an aperture radius of 0\farcs65 as a compromise between encompassing more light, 
and avoiding the danger of averaging over too many sub-structures within the aperture. 
\citet{rigby23} find that an aperture of 0\farcs65 radius encloses roughly 72\% of the energy in the 
F1800W filter,\footnote{See also \url{https://jwst-docs.stsci.edu/jwst-mid-infrared-instrument/miri-performance/miri-point-spread-functions\#gsc.tab=0}.} 
roughly the wavelength corresponding to the size of our convolution kernel.
Although a larger aperture might be desirable, 
the smaller field of view (FoV) at shorter wavelengths results in incomplete
spectra in the regions of interest close to the FoV border (see Fig. \ref{fig:apertures}).
Our adopted aperture diameter of 1\farcs30 is twice the PSF FWHM of MIRI/MRS at the 
longest \htwo\ wavelength (17\,\micron). 
Because we use spectra extracted after convolution to a common PSF, 
our \htwo\ population diagrams should not be affected by resolution, 
unless there are dramatic variations in the \htwo\ excitation within the aperture.

\begin{figure*}[t!]
\includegraphics[width=\textwidth]{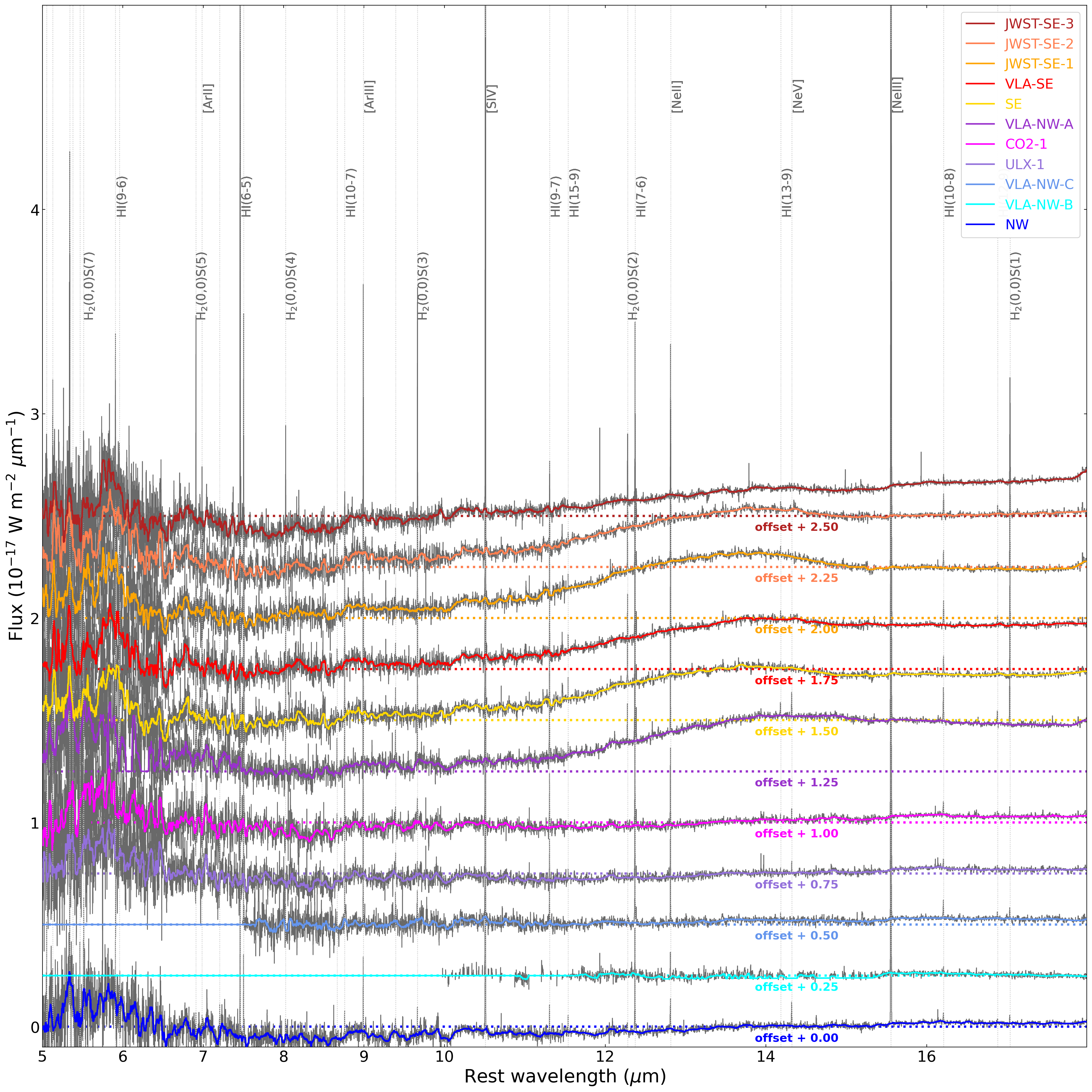}
\vspace{-\baselineskip}
\caption{One-dimensional spectra extracted from the convolved cubes within the
0\farcs65-radius apertures shown in Fig. \ref{fig:apertures}.
The vertical axis for flux density is in units of $10^{-17}$ W\,m$^{-2}$\,\micron$^{-1}$,
and the horizontal wavelength axis in \micron;
the horizontal axis has been restricted to show only the relevant \htwo\ lines.
The gray curves show the spectra, while the heavy colored lines show the smoothed continua;
the zero level for each spectrum is given as a horizontal dotted colored line.
For better visibility of the spectra, the spectra are offset by small increments (in 
units of $10^{-17}$\,W\,m$^{-2}$\,\micron$^{-1}$) as denoted in the figure.
The vertical dotted lines correspond to the detected transitions given in Tables \ref{tab:nwflux}
and \ref{tab:seflux}; to avoid overcrowding, not all lines are labeled. 
The spectra for VLA-NW-B and VLA-NW-C are missing the short-wavelength MIRI channels because 
part of the aperture falls outside the MIRI FoV at those wavelengths.
}
\label{fig:spectra}
\end{figure*}

Figure \ref{fig:spectra} shows the extracted spectra, over the spectral range of the \htwo\ lines, 
offset by arbitrary increments 
to enhance visibility. 
The original spectra are given by the light-gray curves, and
the smoothed continua by the heavy colored ones, as reported in the legend;
the zero level for each spectrum is illustrated as a horizontal dotted line.
As in Paper\,I,
we identified emission lines in the spectra automatically, using a 
custom script that relies on Python's \texttt{specutils/find\_lines\_threshold}.
First, the piece-wise fitted continuum was subtracted from each spectrum;
noise within these adjacent continuum regions was calculated for input to the line-finding algorithm. 
A S/N threshold of 3 was applied to identify potential line detections,
and each of these line candidates was fit with the sum of a linear continuum and
a Gaussian in a spectral window around the central wavelength.
The rest-frame central wavelength of the best-fit Gaussian was then compared with known line
lists, and a series of potential line detections was extracted for each spectrum.
We inferred the total line flux by 
integrating the Gaussian, and estimated flux uncertainties by propagating
the errors on the fitted parameters given by the Hessian matrix associated with the fit.
Ultimately, a list of detected lines was computed by requiring a S/N\,$\geq$\,3 on the fitted flux,
and by constraining the absolute value of the velocity offset to be 
$\leq\,$150\,\kms.
Vertical dotted lines in Fig. \ref{fig:spectra} show some of these identifications. 

As stated in Paper\,I,
the MIRI spectra of \izw\ show (S/N$\geq$3) detections for a total of
10 fine-structure lines, 15 \hi\ recombination lines, 
and 7 \htwo\ transitions.
In Paper\,I, we examined the ionized gas. 
Here we analyze the rotational \htwo\ transitions
in 
spectra extracted from the convolved cubes. 
The detected \htwo\ line fluxes from the convolved cubes
and their uncertainties are reported in Appendix \ref{sec:flux} in
Tables \ref{tab:nwflux} and \ref{tab:seflux};
Appendix \ref{sec:flux} Figure \ref{fig:linefits_h2} illustrates the fits of the \htwo\ lines and 
adjacent continua using the procedures described above.

\section{Warm molecular hydrogen\label{sec:pop}}

One of the main motivations for our \jwst/MIRI observations 
of \izw, with a metallicity of $\sim 3$\%\,\zsun, was to search 
for warm \htwo\ emission.
As mentioned in Sect. \ref{sec:spectra}, MIRI has detected (with S/N$\,\geq\,3$ 
and velocity offsets $\leq\,$150\,\kms) 7 
\htwo\ rotational transitions; here we model the resulting population diagrams 
as given in Table \ref{tab:htwo}.
These enable an estimation of the \htwo\ column density, and
the characterization of the properties of the emitting gas.

\subsection{The formalism \label{sec:formalism}}

First, assuming optical thinness, the observed 
S($J$) line flux is related to the \htwo\ column density $N$ by:

\begin{equation}
F_J\,=\,\left( \frac{\Omega}{4\pi} \right) h\nu_J\, A_J \, N_{J+2}\ ,
\label{eqn:flux}
\end{equation}
\noindent
where $\nu_J$ is the frequency of the transition,
$\Omega$ is the 
aperture solid angle, $A_J$ is the probability per unit time of emission,
and $N_{J+2}$ the associated column density.
If we assume that the \htwo\ gas is in local thermodynamic equilibrium (LTE),
namely with its excitation driven by collisions
(this may not be true for the high-$J$ transitions as we discuss in Sect. \ref{sec:excitation}),
then Boltzmann's equation relates the column density of a given \htwo\ level
$N_j$ to the total \htwo\ number column density \Ntot\ through the gas temperature:
\begin{equation}
N_j\,=\,\frac{g_j}{Z(T)} N_\mathrm{tot}\, \exp(-E_j/kT)
\label{eqn:ni_singleT}
\end{equation}
\noindent
where $g_j$ is the degeneracy of upper state $j$, 
$Z(T)$ is the temperature-dependent
partition function, $E_j$ is the energy level of the upper state, $k$ is Boltzmann's constant,
and $T$ is the kinetic temperature.
The degeneracy $g_j$ for even and odd values of $j$ is given by:

\begin{equation}
  g_j\,=\, 
  \begin{cases}
    2j + 1 & \text{for even $j$ (para-); }\\
    3\ (2j + 1) & \text{for odd $j$ (ortho-)\ . }
  \end{cases}
\end{equation}

\noindent
since \htwo\ has two orientations of its coupled proton spins:
one in which they are parallel (ortho-\htwo), and one in which they
are anti-parallel (para-\htwo). 
In ortho-\htwo, the total nuclear spin $F\,=\,1$, so
that the nuclear spin degeneracy $g_n\,=\,(2F\,+\,1)\,=\,3$.

Because ortho-para conversion is slow,
para-\htwo\ and ortho-\htwo\ behave like distinct species. 
It is therefore useful to define separate rotational partition functions, $Z$,
for para- and ortho-\htwo:

\begin{equation}
Z(T_\mathrm{rot})\,=\,
  \begin{cases}
    \sum\limits_{\mathrm{even}\ j} (2j + 1)\,e^{-E_j/kT_\mathrm{rot}} & \text{for even $j$ (para-); }\\
    \sum\limits_{\mathrm{odd}\ j} 3\,(2j + 1)\,e^{-E_j/kT_\mathrm{rot}} & \text{for odd $j$ (ortho-)\ . }
  \end{cases}
\label{eqn:Z}
\end{equation}
\noindent
These summations can be approximated by simple analytic functions: 
\begin{equation}
Z(T_\mathrm{rot})\,\approx\,
  \begin{cases}
  \left[1 + \left(\frac{T}{170\,\mathrm{K}}\right)^4\right]^{1/4} & \text{for $Z_\mathrm{para}$; }\\
  9\,e^{-170\,K/T} + \frac{T}{57\,\mathrm{K}}\,e^{-510\,\mathrm{K}/T} & \text{for $Z_\mathrm{ortho}$\ . }
  \end{cases}
\label{eqn:Zanalytic}
\end{equation}

\bigskip
\begin{table}
\caption{MIRI \htwo\ lines used for population diagrams$^a$}
\begin{centering}
\resizebox{\linewidth}{!}{
\begin{tabular}{lcrrc}
\hline
\hline
\multicolumn{1}{c}{\rule{0pt}{3ex} Transition} &
\multicolumn{1}{c}{\rule{0pt}{3ex} Notation} &
\multicolumn{1}{c}{Rest} &
\multicolumn{1}{c}{$E_u/k$} &
\multicolumn{1}{c}{$A$} \\
&&
\multicolumn{1}{c}{wavelength} \\
\multicolumn{1}{c}{\rule{0pt}{3ex} $v\,=\,0$} &
&
\multicolumn{1}{c}{(\micron)} &
\multicolumn{1}{c}{(K)$^{b}$} &
\multicolumn{1}{c}{($10^{-10}$\,s$^{-1}$)$^{b}$} 
\\
\hline
\\ 
$J\,=\,3\,\rightarrow\,1$ & S(1) & 17.03485 & 1015 & 4.76 \\
$J\,=\,4\,\rightarrow\,2$ & S(2) & 12.27861 & 1681 & 27.55 \\
$J\,=\,5\,\rightarrow\,3$ & S(3) & 9.66491 & 2503 & 98.36 \\
$J\,=\,6\,\rightarrow\,4$ & S(4) & 8.02504 & 3473 & 264.3 \\
$J\,=\,7\,\rightarrow\,5$ & S(5) & 6.90951 & 4585 & 587.9 \\
$J\,=\,8\,\rightarrow\,6$ & S(6) & 6.10856 & 5828 & 1142 \\
$J\,=\,9\,\rightarrow\,7$ & S(7) & 5.51118 & 7196 & 2001 \\
$J\,=\,10\,\rightarrow\,8$ & S(8) & 5.05312 & 8677 & 3236 \\
\\
\hline
\hline
\end{tabular}
}
\end{centering}
\newline
\noindent
$^{a}$~For completeness, we include the S(6) transition, even though it is not detected in \izw\ (see Tables \ref{tab:nwflux}, \ref{tab:seflux}). \\
$^{b}$~Taken from \citet{roueff19}. \\
\label{tab:htwo}
\end{table}

Following an empirical approach \citep{zakamska10,pereira14,togi16}, 
we model the \htwo\ column densities
using a continuous single power-law temperature $T$ distribution such that

\begin{equation}
dN \propto T^{-n}\, dT\ ,
\label{eqn:temp}
\end{equation}

\noindent
for \tl$\,\leq T \leq\,$\tu, with \tl\ and \tu\ 
the lower and upper limits of the distribution, respectively;
$N$ is the column density, and $n$ is the power-law index.
Considering the continuous temperature distribution, we can derive a new expression
for the relation of the column density of the $i^\mathrm{th}$ transition 
with the total \htwo\ column \Ntot:
\begin{equation}
\frac{N_j}{g_j}\,=\,\left[\ \frac{(n - 1) N_\mathrm{tot}}{T_{l}^{1-n} - T_{u}^{1-n}}\ \right]
\int_{T_{l}}^{T_{u}}\ \frac{S_\mathrm{OPR}}{Z(T)} e^{\frac{-E_j}{kT}}\ T^{-n} dT
\label{eqn:niT}
\end{equation}

\noindent
where 
$Z(T)$ is given by Eq. \eqref{eqn:Zanalytic}. 
With OPR defined as the ortho/para abundance ratio, 
\sopr\ is the scaling factor for the column densities that takes into account the OPR, namely
\sopr\,=\,1/(1\,+\,OPR) for para-\htwo, and \sopr\,=\,OPR/(1\,+\,OPR) for ortho-\htwo. 
\htwo\ in LTE has OPR\,=\,OPR$_\mathrm{LTE}$\,=\,$Z_\mathrm{ortho}(T)/Z_\mathrm{para}(T)$.
In LTE at temperatures $T\,\la\,40$\,K, \htwo\ is primarily in the ground state $J\,=\,0$, so that
(OPR)$_\mathrm{LTE} \rightarrow 0$.
In Eq. \eqref{eqn:niT}, the implicit assumption is that the gas at different temperatures
has the same OPR.

Operationally, by fixing OPR\,=\,3,
Eq. \eqref{eqn:niT} has four free variables, \Ntot, $n$, \tl, and \tu.
To simplify the fit, we therefore assume a given \tu: 
in one set of fits \tu\,=\,2000\,K as used by \citet{togi16} and others, 
and in another set \tu\,=\,3500\,K, in order to accommodate a potential excess of warm-hot \htwo\ in \izw.
We also normalize the fit to the brightest \htwo\ line, S(1);
this leaves two free parameters in the fit, namely \tl\ and $n$.
After fitting, it is straightforward to estimate the total \htwo\ column density \Ntot:
\begin{equation}
N_\mathrm{tot}\,=\, \frac{N_3}{(n-1) g_3}\ \frac{(1 + \mathrm{OPR})}{\mathrm{OPR}}\ \frac{ \left( T_l^{1-n} - T_u^{1-n} \right)} { \int_{T_{l}}^{T_{u}} \frac{e^{\frac{-E_3}{kT}}\ T^{-n} dT }{Z_\mathrm{ortho}(T)} }
\label{eqn:ntot}
\end{equation}

\noindent
where $N_3$ is related to the integrated flux of S(1) through Eq. \eqref{eqn:flux}.
Finally, the total \htwo\ mass in the aperture can be calculated:

\begin{equation}
\mathrm{M_{H2}}\,=\, N_\mathrm{tot}\ \Omega\ D^2 \mathrm{m_{H2}}\ ,
\label{eqn:h2mass}
\end{equation}

\noindent
where $D$ is the distance to \izw, and $\mathrm{m_{H2}}$ is the mass of the \htwo\ molecule.

\begin{figure*}[t!]
\includegraphics[width=0.48\linewidth]{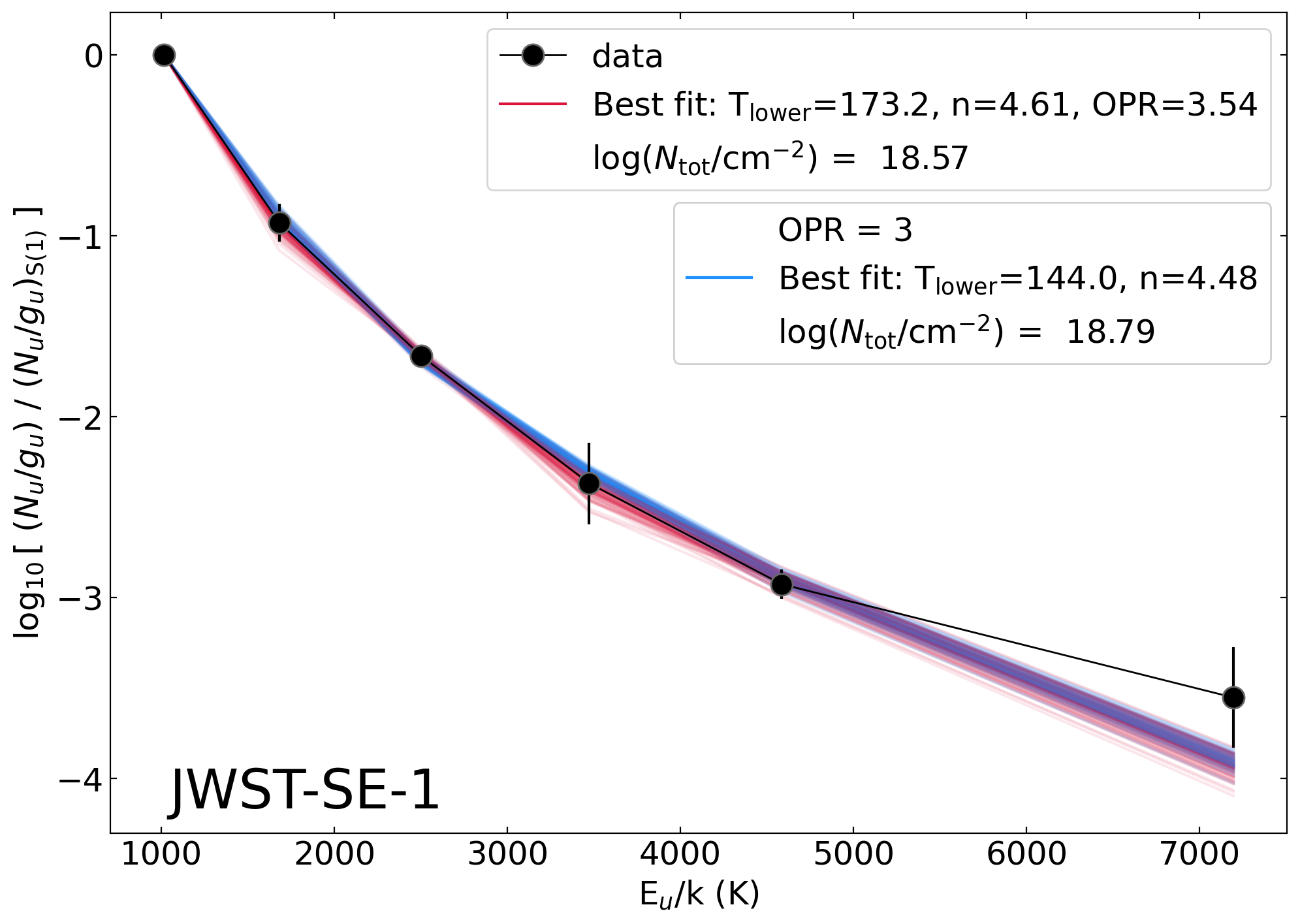} 
\includegraphics[width=0.48\linewidth]{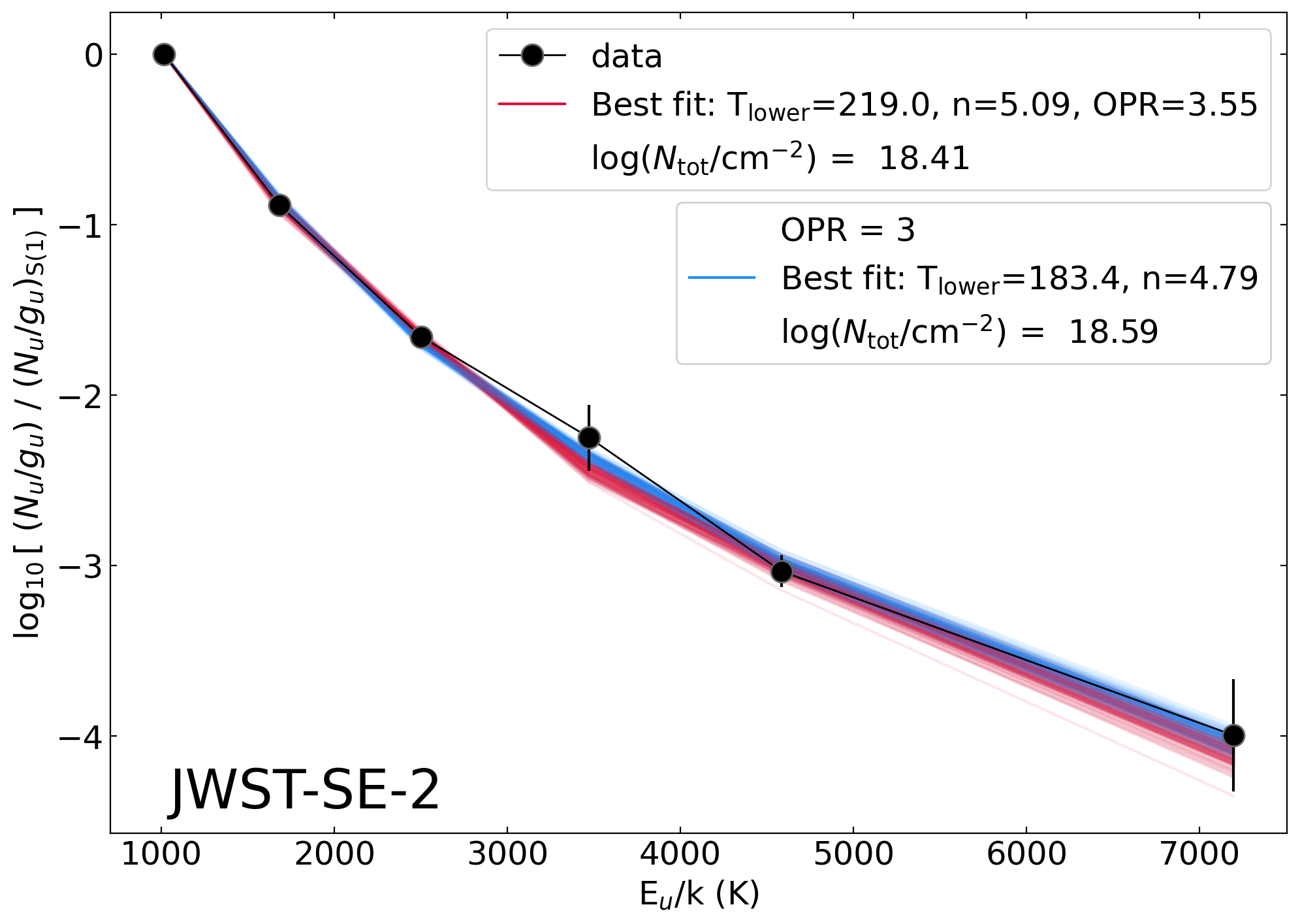} \\
\includegraphics[width=0.48\linewidth]{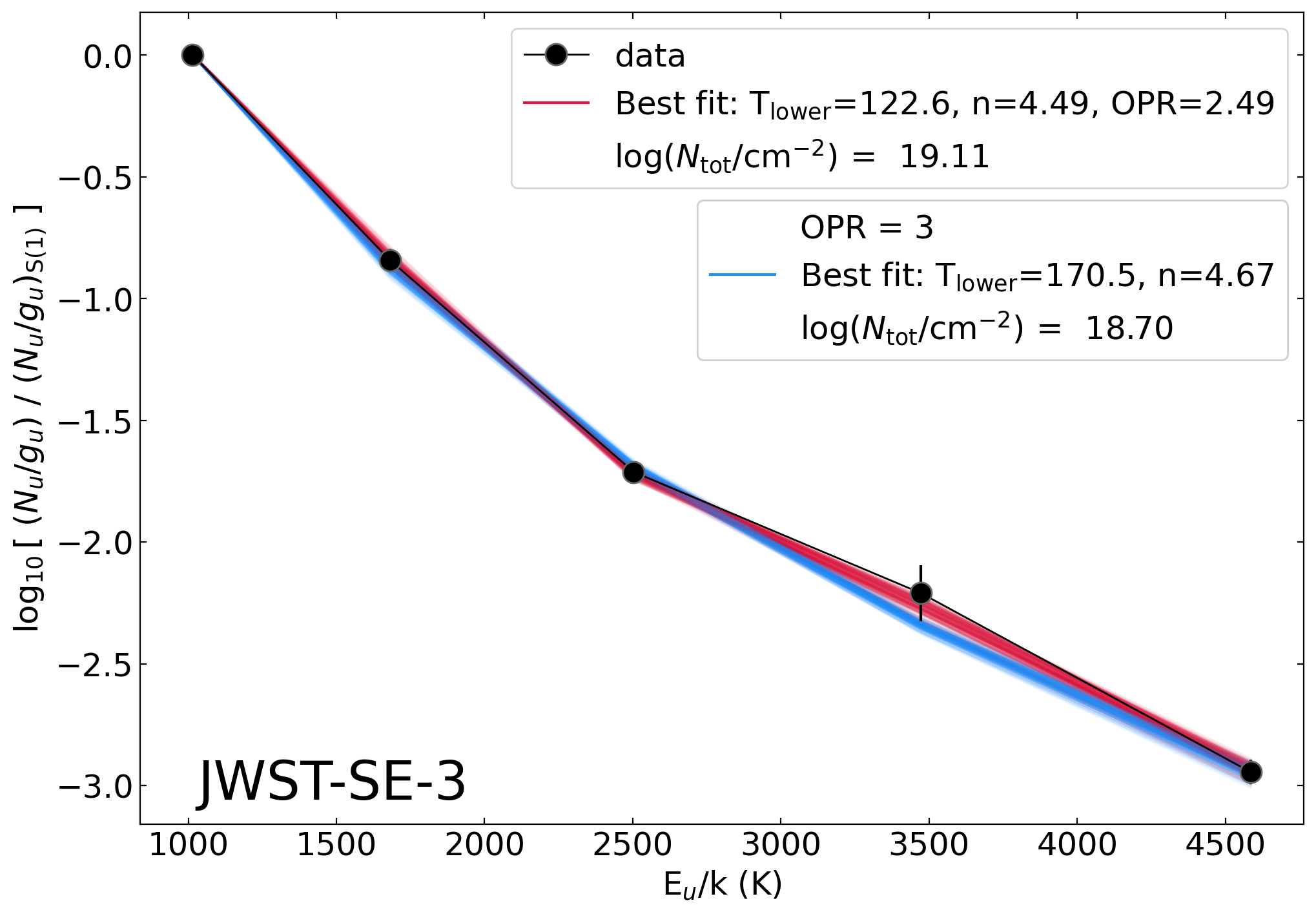} 
\includegraphics[width=0.48\linewidth]{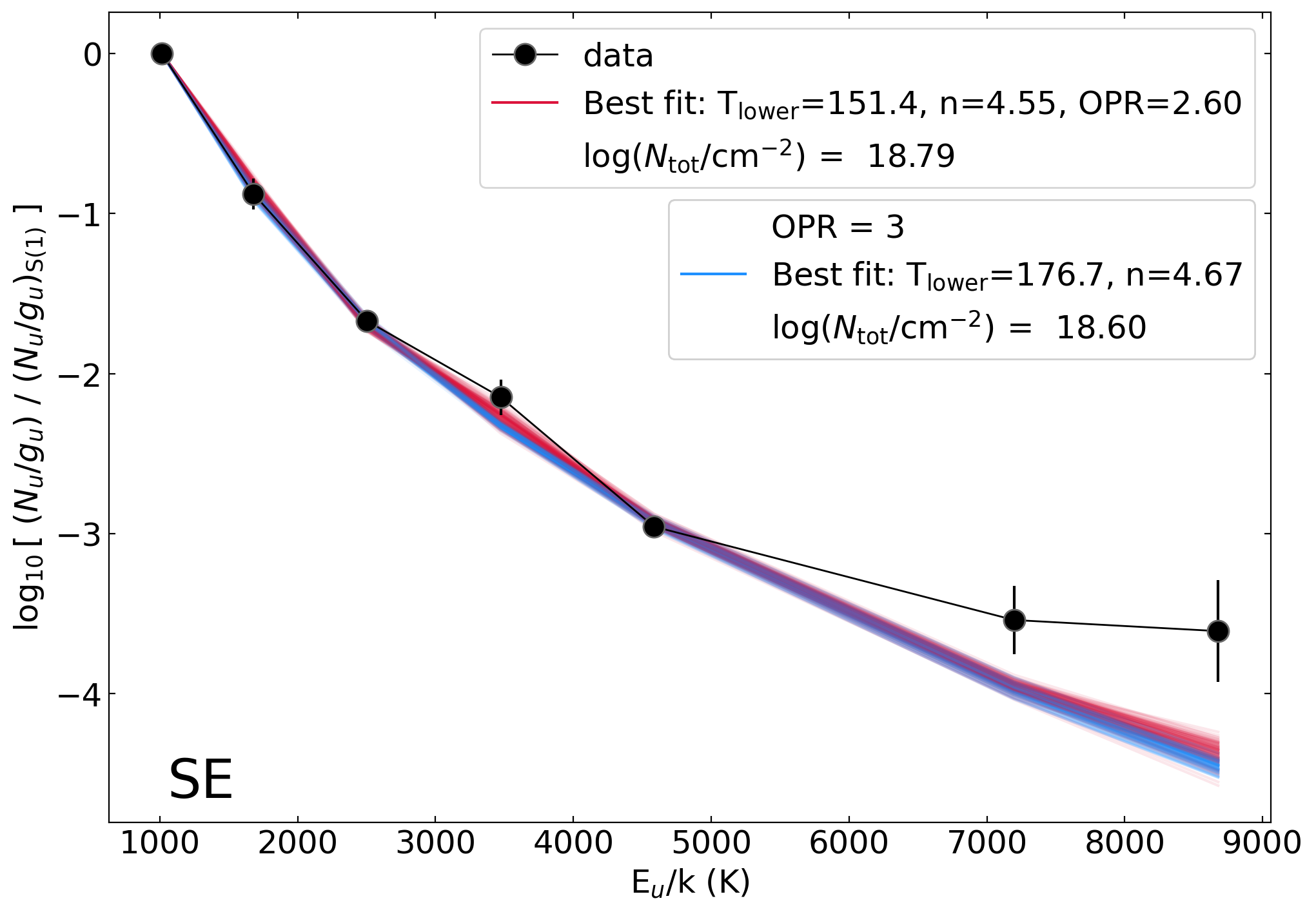} \\
\includegraphics[width=0.48\linewidth]{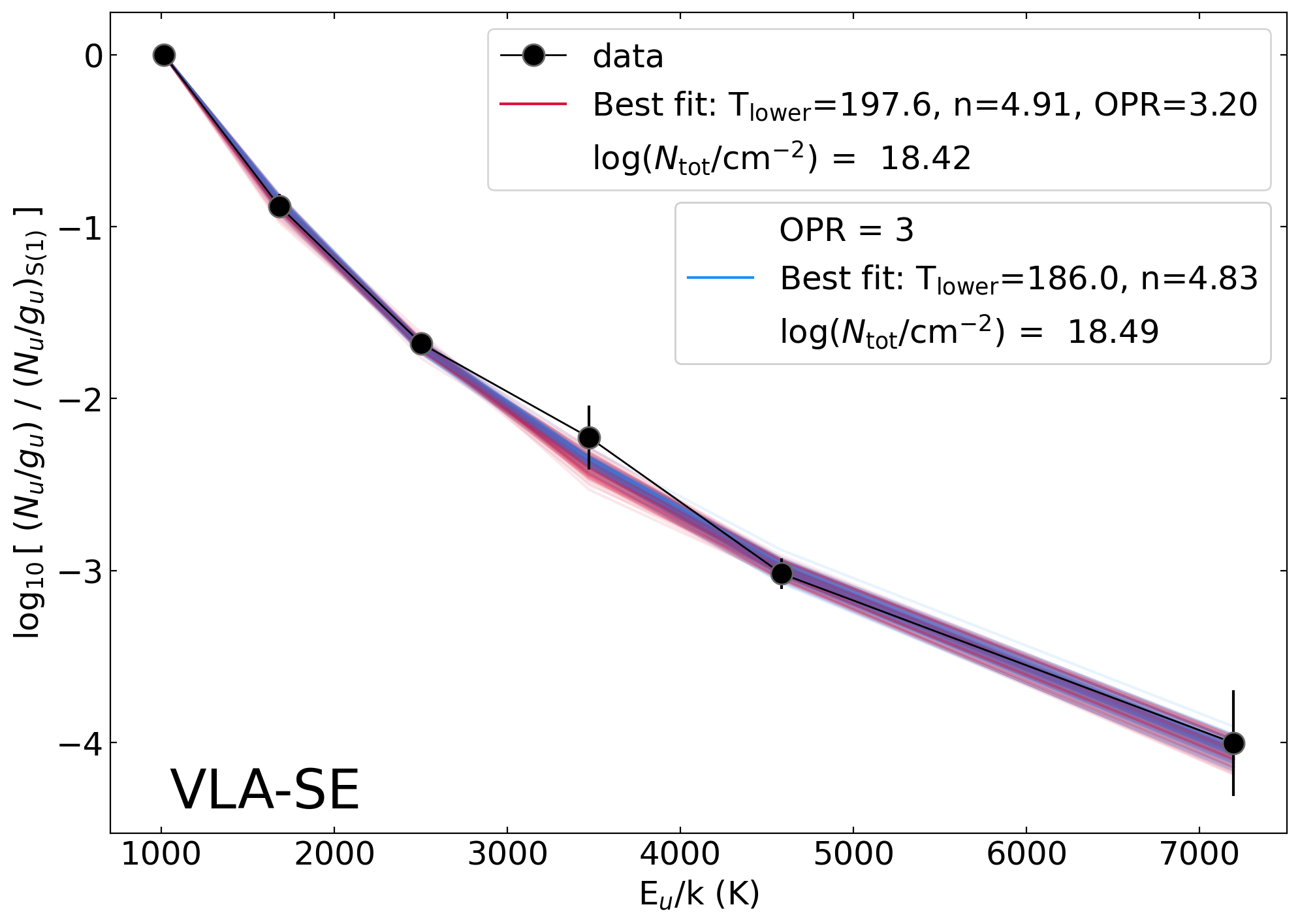} 
\includegraphics[width=0.48\linewidth]{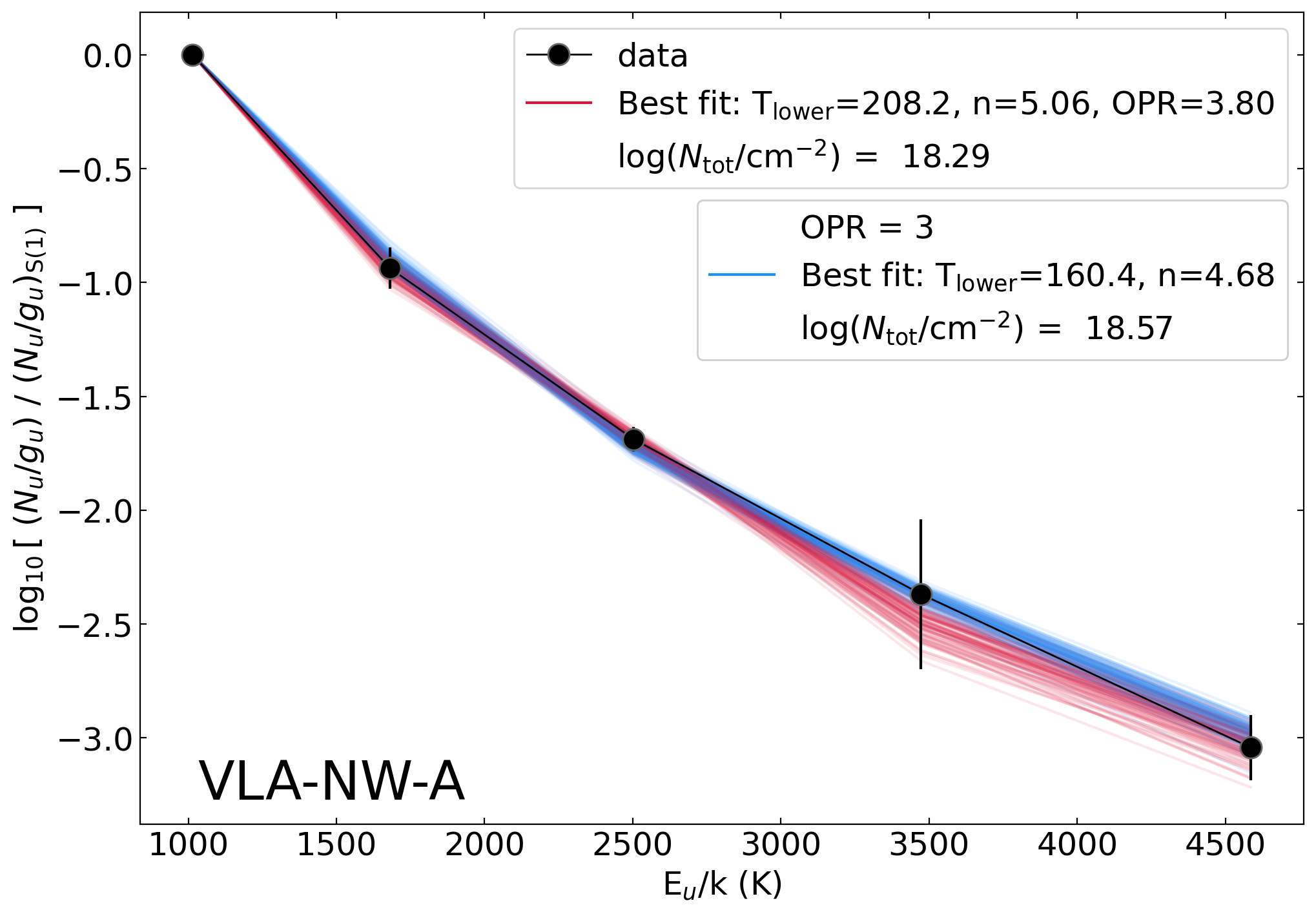} \\
\caption{Molecular hydrogen population diagrams for the apertures with 5 or more \htwo\ detections
at a $3\sigma$ level or greater, together with the best-fit models with \tu\,=\,2000\,K.
Fits with OPR\,=\,3  letting \tl\ and $n$ vary are shown as blue curves, according to the MCMC sampler spread;
the analogous fits letting OPR, \tl\ and $n$ vary are shown as red curves (see Sect. \ref{sec:opr}).
The total \htwo\ column density shown in the legends is given by Eq. \eqref{eqn:ntot}.
}
\label{fig:pop}
\end{figure*}

\begin{figure*}[t!]
\vbox{
\textbf{JWST-SE-1}\hspace{3.9cm}\textbf{JWST-SE-2}\hspace{4.0cm}\textbf{JWST-SE-3}\\ 
\centering
\includegraphics[width=0.32\linewidth]{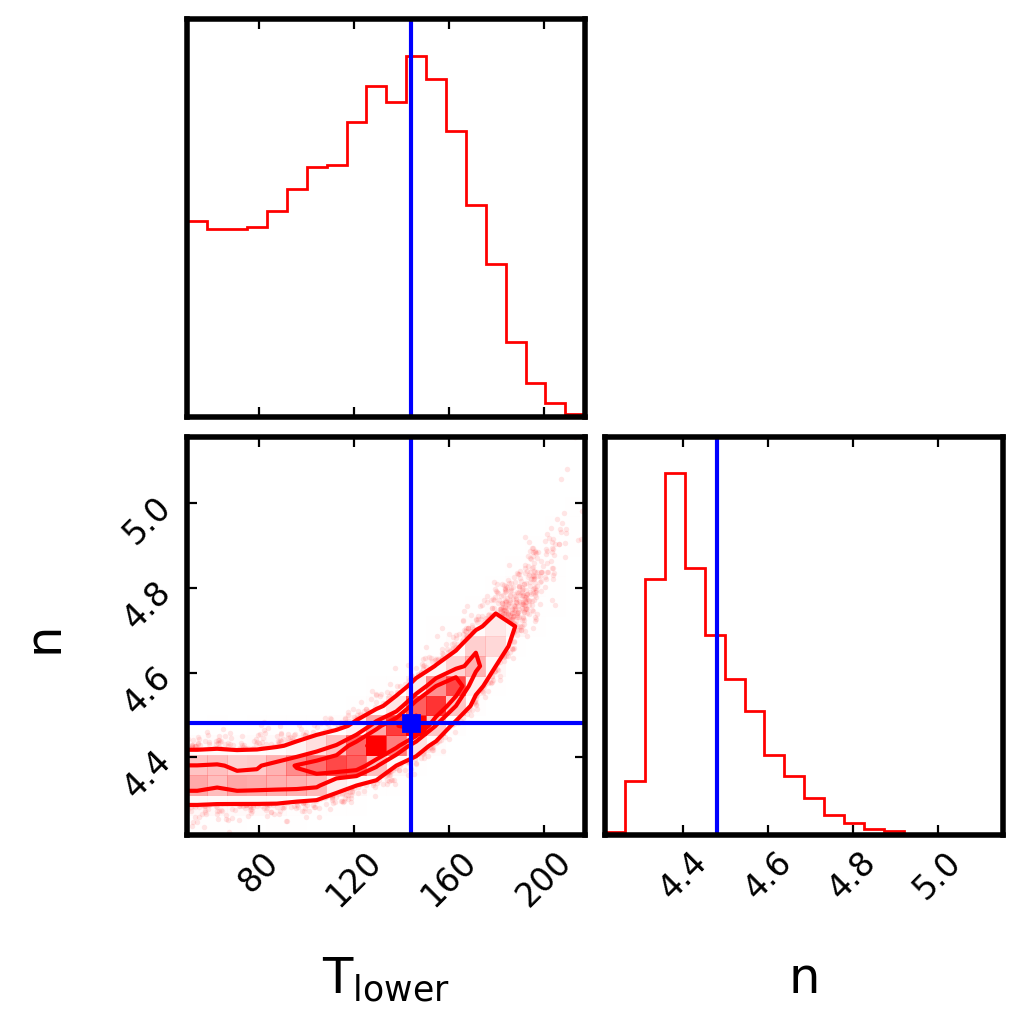}
\includegraphics[width=0.32\linewidth]{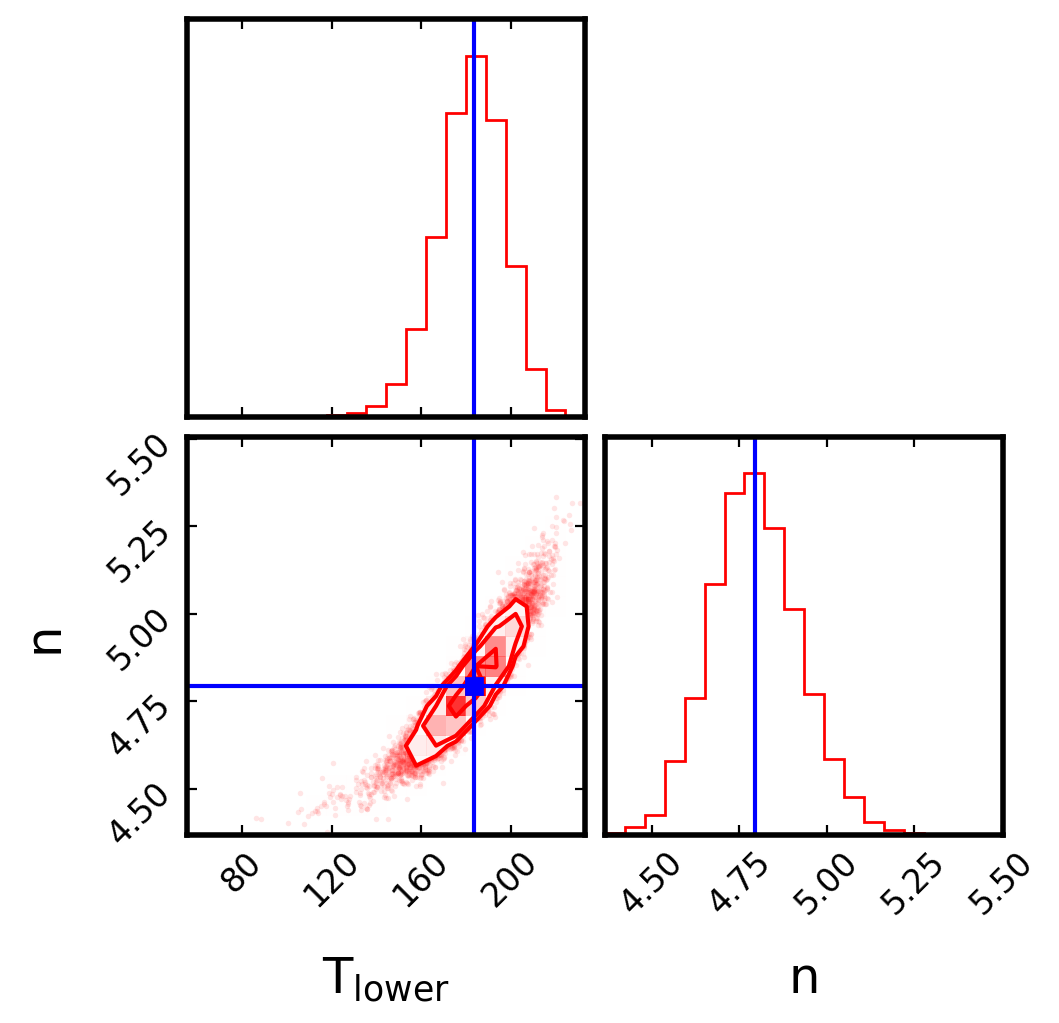} 
\includegraphics[width=0.32\linewidth]{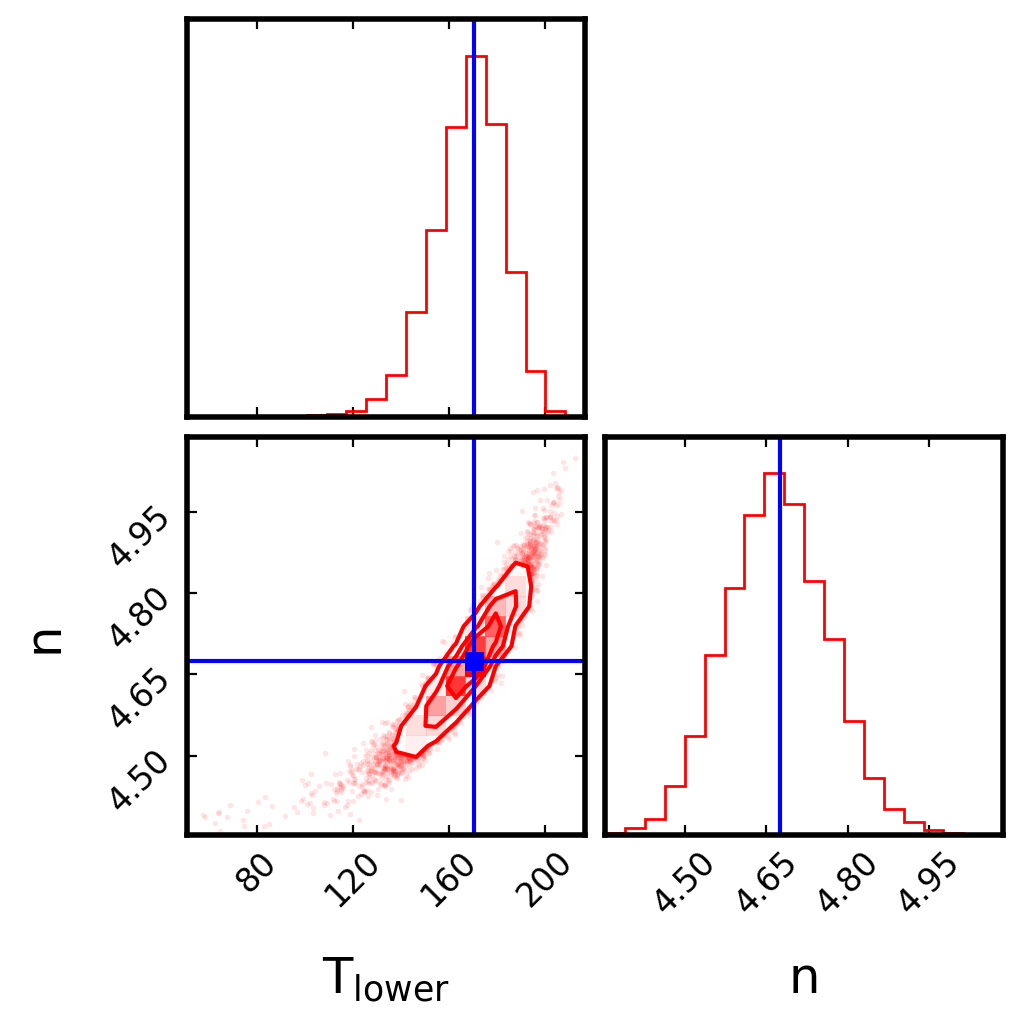}\\
}
\vbox{
\hspace{0,7cm}\textbf{SE}\hspace{4.7cm}\textbf{VLA-SE}\hspace{4.3cm}\textbf{VLA-NW-A}\\ 
\centering
\includegraphics[width=0.32\linewidth]{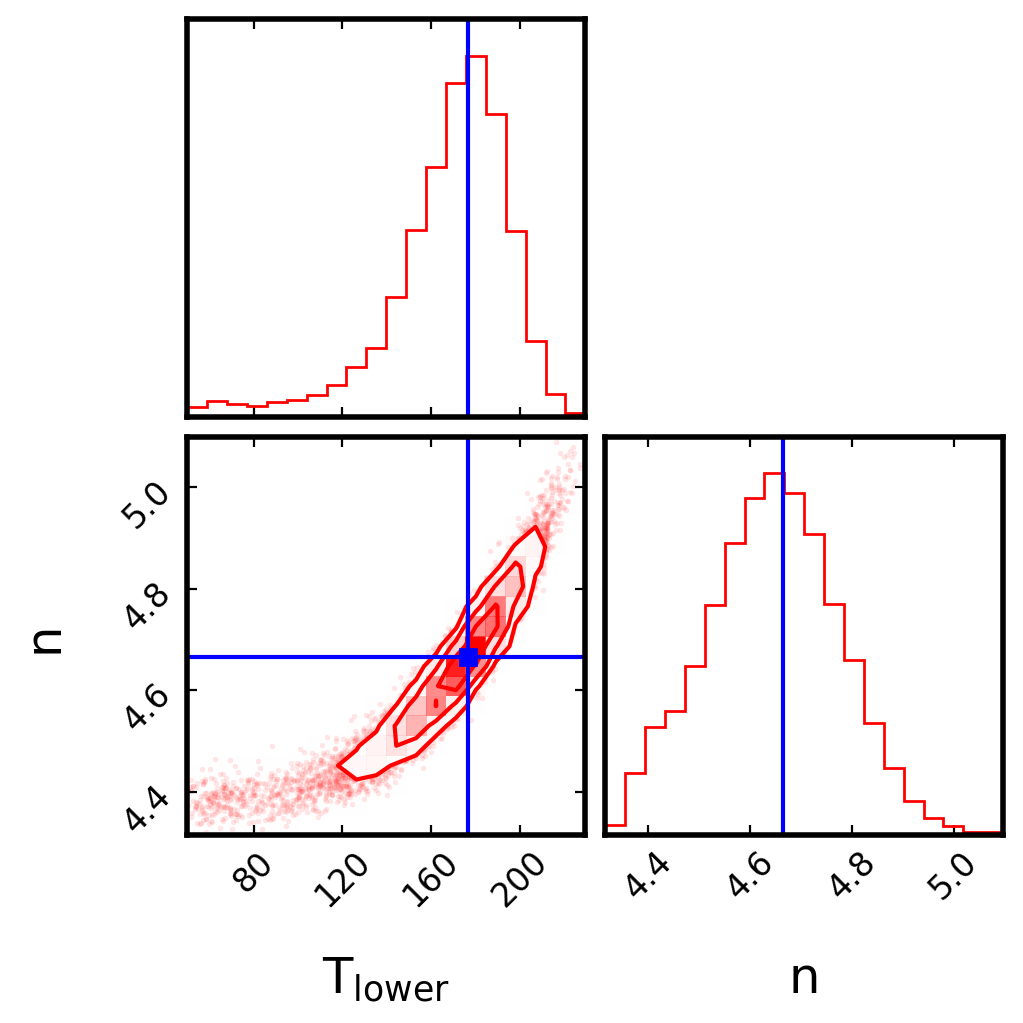}
\includegraphics[width=0.32\linewidth]{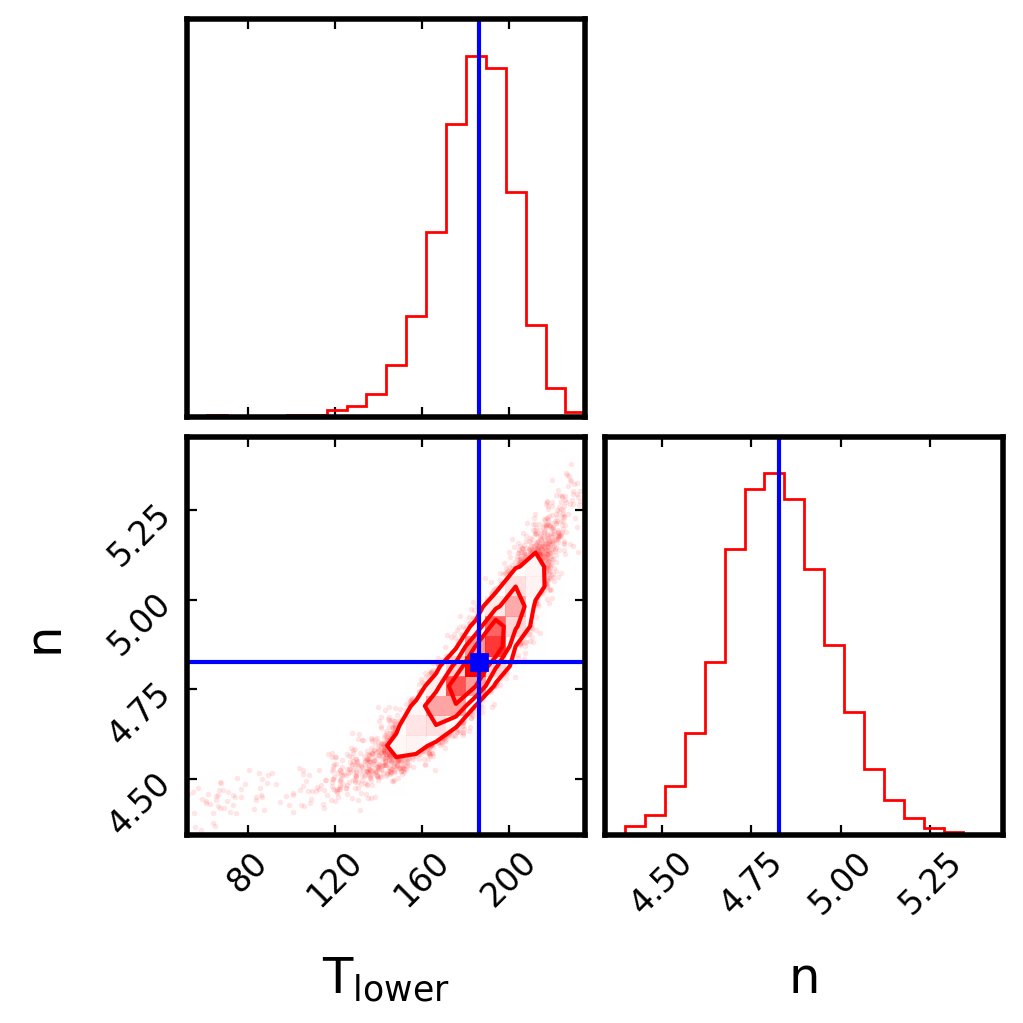} 
\includegraphics[width=0.32\linewidth]{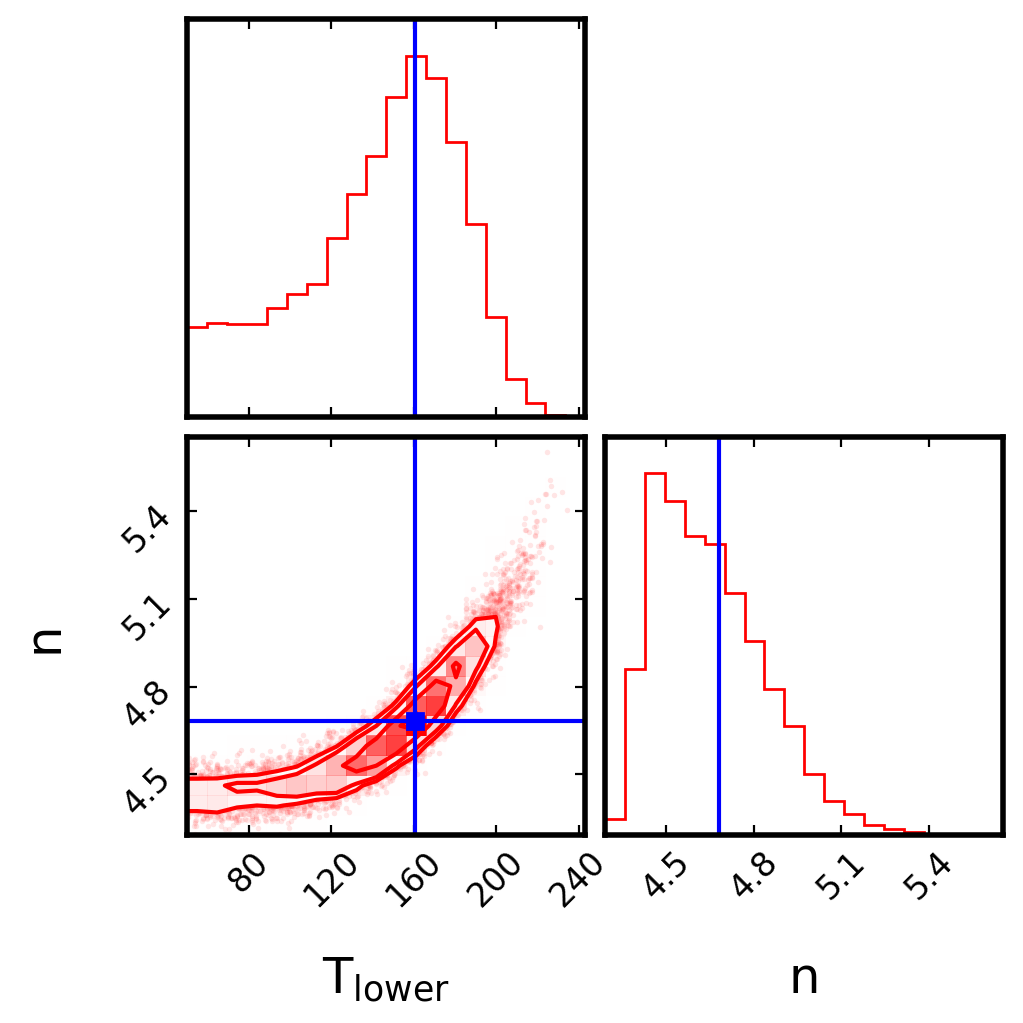} \\
}
\vbox{
}
\vspace{-\baselineskip}
\caption{Corner plots for the (\tl, $n$) fits of the \htwo\ population diagrams with \tu\,=\,2000\,K.
For each region, the upper and right panels show the PDF,
and the lower left panel the co-dependence of the two parameters. 
It can be seen that \tl\ and $n$ are correlated, with an extension
toward flatness particularly pronounced in JWST-SE-1 and VLA-NW-A.  }
\label{fig:corner}
\end{figure*}

\begin{table*}
\caption{Best-fit results for \htwo\ population diagrams with \tu\,=\,2000\,K\label{tab:popfits}}
\begin{center}
\begin{tabular}{lcccccccrrr}
\hline
\hline
\multicolumn{1}{c}{Region} &
\multicolumn{1}{c}{Number} &
\multicolumn{1}{c}{Max} &
\multicolumn{1}{c}{\tl} &
\multicolumn{1}{c}{$n$} &
\multicolumn{1}{c}{OPR} &
\multicolumn{1}{c}{log$_{10}$} &
\multicolumn{1}{c}{log$_{10}$} \\
& \multicolumn{1}{c}{points} &
\multicolumn{1}{c}{$J_\mathrm{up}$} &
\multicolumn{1}{c}{(K)} &
&& \multicolumn{1}{c}{(\Ntot/cm$^{-2}$)$^{a}$} &
\multicolumn{1}{c}{(\htwo/\msun)$^{b}$} &
\multicolumn{1}{c}{AIC$^{c}$} &
\multicolumn{1}{c}{BIC$^{d}$} &
\multicolumn{1}{c}{$\chi^2$} \\
\multicolumn{1}{c}{(1)} &
\multicolumn{1}{c}{(2)} &
\multicolumn{1}{c}{(3)} &
\multicolumn{1}{c}{(4)} &
\multicolumn{1}{c}{(5)} &
\multicolumn{1}{c}{(6)} &
\multicolumn{1}{c}{(7)} &
\multicolumn{1}{c}{(8)} &
\multicolumn{1}{c}{(9)} &
\multicolumn{1}{c}{(10)} &
\multicolumn{1}{c}{(11)} \\
\hline
\hline
\multicolumn{10}{c}{OPR\,=\,3}\\
\hline
JWST-SE-1 & 6 & 9 & 144$^{+48}_{-35}$ & 4.48$^{+0.08}_{-0.15}$ & $-$ & 18.79 $\,\pm\,$ 0.45 & 3.01 $\,\pm\,$ 0.45 & 4.97 & 4.55 & 7.05 \\
JWST-SE-2 & 6 & 9 & 183$^{+15}_{-14}$ & 4.79$^{+0.12}_{-0.13}$ & $-$ & 18.59 $\,\pm\,$ 0.09 & 2.81 $\,\pm\,$ 0.09 & 6.34 & 5.93 & 8.87 \\
JWST-SE-3 & 5 & 7 & 171$^{+15}_{-13}$ & 4.67$^{+0.09}_{-0.10}$ & $-$ & 18.70 $\,\pm\,$ 0.10 & 2.92 $\,\pm\,$ 0.10 & 6.53 & 5.75 & 8.29 \\
SE & 7 & 10 & 177$^{+27}_{-19}$ & 4.67$^{+0.13}_{-0.13}$ & $-$ & 18.60 $\,\pm\,$ 0.26 & 2.82 $\,\pm\,$  0.26& 13.32 & 13.21 & 26.50 \\
VLA-SE & 6 & 9 & 186$^{+19}_{-16}$ & 4.83$^{+0.14}_{-0.15}$ & $-$ & 18.49 $\,\pm\,$ 0.17 & 2.70 $\,\pm\,$ 0.17 & 1.93 & 1.51 & 4.25 \\
VLA-NW-A & 5 & 7 & 160$^{+51}_{-30}$ & 4.68$^{+0.16}_{-0.22}$ & $-$ & 18.57 $\,\pm\,$ 0.43 & 2.79 $\,\pm\,$ 0.43 & 3.48 & 2.70 & 4.51 \\
\hline
\multicolumn{10}{c}{Fitting OPR}\\
\hline
JWST-SE-1 & 6 & 9 & 173$^{+48}_{-29}$ & 4.61$^{+0.17}_{-0.19}$ & 3.54$^{+0.38}_{-0.45}$ & 18.57 $\,\pm\,$ 0.36 & 2.79 $\,\pm\,$ 0.36 & 4.74 & 4.11 & 4.86 \\
JWST-SE-2 & 6 & 9 & 219$^{+19}_{-17}$ & 5.09$^{+0.19}_{-0.21}$ & 3.55$^{+0.25}_{-0.26}$ & 18.41 $\,\pm\,$ 0.08 & 2.63 $\,\pm\,$ 0.08 & 3.74 & 3.12 & 4.12 \\
JWST-SE-3 & 5 & 7 & 123$^{+39}_{-35}$ & 4.49$^{+0.04}_{-0.10}$ & 2.49$^{+0.13}_{-0.16}$ & 19.11 $\,\pm\,$ 0.49 & 3.33 $\,\pm\,$ 0.49 & $-0.15$ & $-1.33$ & 1.46 \\
SE & 7 & 10 & 151$^{+54}_{-36}$ & 4.55$^{+0.09}_{-0.16}$ & 2.60$^{+0.23}_{-0.27}$ & 18.79 $\,\pm\,$ 0.47 & 3.01$\,\pm\,$ 0.47 & 14.83 & 14.66 & 24.70 \\
VLA-SE & 6 & 9 & 198$^{+28}_{-21}$ & 4.91$^{+0.20}_{-0.20}$ & 3.20$^{+0.32}_{-0.32}$ & 18.42 $\,\pm\,$ 0.25 & 2.64 $\,\pm\,$ 0.25 & 3.23 & 2.61 & 3.78 \\
VLA-NW-A & 5 & 7 & 208$^{+34}_{-25}$ & 5.06$^{+0.31}_{-0.34}$ & 3.80$^{+0.44}_{-0.47}$ & 18.29 $\,\pm\,$ 0.25 & 2.50 $\,\pm\,$ 0.25 & $-4.85$ & $-6.02$ & 0.57 \\
\hline
\hline
\end{tabular}
\end{center}
\begin{flushleft}
$^{a}$\,Calculated from Eq. \eqref{eqn:ntot}. \\
$^{b}$\,Calculated from Eq. \eqref{eqn:h2mass}. \\
$^{c}$\,Calculated according to \texttt{SciPy/optimize}: AIC$\,=\,N \ln(\chi^2/N) + 2 N_\mathrm{fit}$ where $N_\mathrm{fit}$ is
the number of fitted parameters, and $N$ is the number of data points.  \\
$^{d}$\,Calculated according to \texttt{SciPy/optimize}: BIC$\,=\,N \ln(\chi^2/N) + \ln(N) N_\mathrm{fit}$.  \\
\end{flushleft}
\end{table*}


\subsection{Fitting procedures \label{sec:fitting}}

The fits were applied only to those apertures where at least 5 \htwo\ transitions were detected
with S/N$\,\geq\,3$: JWST-SE-1, JWST-SE-2, JWST-SE-3, SE, VLA-SE, and VLA-NW-A.
All five regions in the SE were detected in 5 or more \htwo\ transitions, 
while, toward the NW, this was true only for VLA-NW-A. 
Notably, the aperture centered on the CO(2--1) detection did not satisfy this criterion.\footnote{This may partially be due to
a small fraction of the aperture falling outside the Channel 1 FoV (see Fig. \ref{fig:apertures}).}
Because of the low extinction in \izw\ (see Paper\,I), 
we have not applied any extinction correction to the \htwo\ lines before fitting.

We estimated the best-fit parameters \tl\ and $n$ by minimizing
$\chi^2$, or equivalently, maximizing the likelihood $L \propto \exp^{-\chi^2/2}$,
assuming Gaussian errors.
$\chi^2$ is calculated using the model given in Eq. \eqref{eqn:niT}, 
and the data and their uncertainties in Tables \ref{tab:nwflux}
and \ref{tab:seflux} (Appendix \ref{sec:flux}) plugged into Eq. \eqref{eqn:flux}.
To do this, we used a direct minimization with the Python package \texttt{SciPy/minimize},
and explored the parameter space to estimate the uncertainties and mutual correlations of the
best-fit parameters with the 
Markov Chain Monte Carlo (MCMC) package \texttt{emcee} \citep{foreman13}.
This package is a Python implementation of the affine invariant MCMC ensemble sampler 
designed for parameter estimation by \citet{goodman10}.
Uniform priors are adopted for both \tl\ and $n$; in particular, we confined
\tl\ to be $\geq\,50$\,K.

The integrated intensities of the S(1) lines in \izw\ are among the faintest ever measured;
they range from the brightest in JWST-SE-3, $1.0\times10^{-9}$ \wms,
to the faintest in the CO2--1 aperture, $8.1\times10^{-11}$\,\wms,
a factor of 8 below 
the S(1) intensity of $6.3\times10^{-10}$ \wms\ measured in Leo\,P, a nearby faint 
dwarf irregular also at $\sim$3\%\,\zsun\ \citep{telford24},
but powered by just one O star \citep{evans19}.\footnote{Their 
photometric aperture has a diameter roughly half of ours, 0\farcs67.}

\subsection{Results with fixed OPR \label{sec:fixedopr}}

Here we report on the fits with \tu\,=\,2000\,K;
similar results for \tu\,=\,3500\,K are given in Appendix \ref{sec:tupper}, where
we also compare the best-fit parameters from the two sets of fits.
The results 
are shown in Fig. \ref{fig:pop} as blue curves where 
$N_j/g_j$
normalized to S(1) ($N_3/g_3$) is plotted against $E_i$.
The best-fit parameters with the MCMC $1\sigma$ uncertainties are given in Table \ref{tab:popfits}.
Figure \ref{fig:corner} shows the probability distribution function (PDF) of the
fitted parameters in the upper and right panels for each region, and the mutual dependence
of \tl\ and $n$ in the lower left panels.
As the nominal best-fit value, we have used the initial minimization,
rather than the mode or median of the posterior PDF, which however, usually coincide 
(the exceptions are $n$ for
JWST-SE-1 and VLA-NW-A, where the mode of the PDF is skewed toward low values of $n$).
The uncertainties correspond to the 16\% and 84\% percentile of the posterior PDF distribution.
The uncertainties on \Ntot\ 
and \htwo\ mass
have been calculated by sampling randomly 1000 times the joint posterior PDF distribution of the fitted
parameters, computing log(\Ntot) for each sample, and taking the standard deviation of the result.

We have assessed the quality of the two sets of fits with different \tu,
according to the Akaike Information Criterion \citep[AIC,][]{akaike74} and the
Bayesian Information Criterion (BIC, see Table \ref{tab:popfits}, and Table \ref{tab:popfits3500} in Appendix \ref{sec:tupper}),
and find that there is no clear statistical difference between the two sets of fits. 
In three of the regions, the AIC would point to \tu\,=2000\,K as being a better fit,
and the other three to \tu\,=\,3500\,K. 

Figure \ref{fig:corner} shows 
significant covariance of the parameters \tl\ and $n$ (with \tu\,=\,2000\,K) for each of the regions in \izw, 
unlike the majority of the galaxies analyzed by \citet[][see their Figs. 3 and 6]{togi16},\footnote{\citet{togi16} used \spit/IRS data
for which S(0) is available, which may influence this finding.} 
which showed little covariance.
In their case, 
$n$ is practically independent of \tl, especially toward lower \tl.
Similarly, in JWST-SE-1 and VLA-NW-A, the trend between \tl\ and $n$ flattens at low \tl, which suggests that
under such conditions, the best-fit parameters are not easy to determine,
unless methods like the MCMC used here are employed.
Because \tl\ is well determined for \izw, we speculate that the bulk of \izw's warm \htwo\ content 
is traced by the MIRI rotational lines. 


Just as in more massive galaxies 
\citep[e.g., M\,83, NGC\,5728, NGC\,7469:][respectively]{hernandez23,davies24,lai22},
the MIRI \htwo\ population diagrams of \izw\ are well approximated by a continuous
temperature distribution.
However, the massive galaxies have steeper power-law indices
($n\,=4.9-6.3$) and warmer \tl\ for the integral (\tl\,$\sim$\,250-300\,K),
compared to the results shown in Table \ref{tab:popfits}.
The generally shallower indices $n < 5$ for \izw\ 
\citep[and in the nucleus and `hot spot' in NGC\,5728,][]{davies24}
imply a greater fraction of warm \htwo\ that emits at higher temperatures.

The fitted warm \htwo\ column densities (see Table \ref{tab:popfits}) are more than an
order of magnitude lower than \Ntot\,$\sim\,10^{20}-10^{21}$\,\cmtwo, typical of spiral 
disks \citep[e.g.,][]{schinnerer24}.
They are also much lower than the implied \htwo\ columns from resolved CO molecular clouds at pc scales
in two low-metallicity dwarf galaxies, WLM and Sextans\,B \citep[][respectively]{rubio15,shi20};
the CO measurements typically probe the highest \htwo\ column within the cloud
\citep[e.g.,][]{bolatto13}.
As reported in Table \ref{tab:popfits}, 
the implied \htwo\ mass within the 120\,pc apertures 
range from $\sim\,300$\,\msun\ to $\sim\,2000$\,\msun.
The highest of these are close to the range of virial cloud masses 
of $10^3 - 10^4$\,\msun\ found in WLM ($\sim\,$13\%\,\zsun) and Sextans\,B ($\sim$\,7\%\,\zsun) by 
\citet[][]{rubio15,shi20}, respectively.

The results from the population diagrams
imply that the molecular gas in \izw\ is sufficiently dense to 
thermalize the S(5) line,
with a critical density of $\sim 10^5$\,\cmthree\ \citep[e.g.,][]{draine11}, 
denser than the ionized gas with electron densities $\sim 10^2$\,\cmthree\ 
judging from the optical \citep{izotov99}\footnote{Densities of the ionized gas
are discussed in detail in Paper\,III.} and the thermal radio continuum \citep{hunt05}.
The detections of the high-$J$ lines 
suggest that the warm molecular clouds in \izw\  
are characterized by relatively high densities, typical of the cooler clouds probed by CO.
We quantify this below in Sect. \ref{sec:nonLTE}.
The lowest \htwo\ rotational lines (e.g., S(1)) mainly originate in 
cooler gas (T$\,\la\,200$\,K), while higher-$J$ lines need 
warmer temperatures (T$\,\ga\,400$\,K) and densities high enough (\nh$\,\ga\,10^5$\,\cmthree) to
thermalize the levels.
As discussed below in Sect. \ref{sec:excitation},
\htwo\ is excited by collisions and UV pumping \citep[e.g.,][]{draine11}, and
the trade-off between excitation and self-shielding
can be assessed through \htwo\ excitation diagrams like those in Fig. \ref{fig:pop}.
The total \htwo\ column densities, \Ntot, are sufficiently high to provide significant levels
of self-shielding. 

\begin{figure*}[t!]
\vbox{
\hspace{-1cm}\textbf{JWST-SE-1}\hspace{3.9cm}\textbf{JWST-SE-2}\hspace{4.0cm}\textbf{JWST-SE-3}\\ 
\centering
\includegraphics[width=0.32\linewidth]{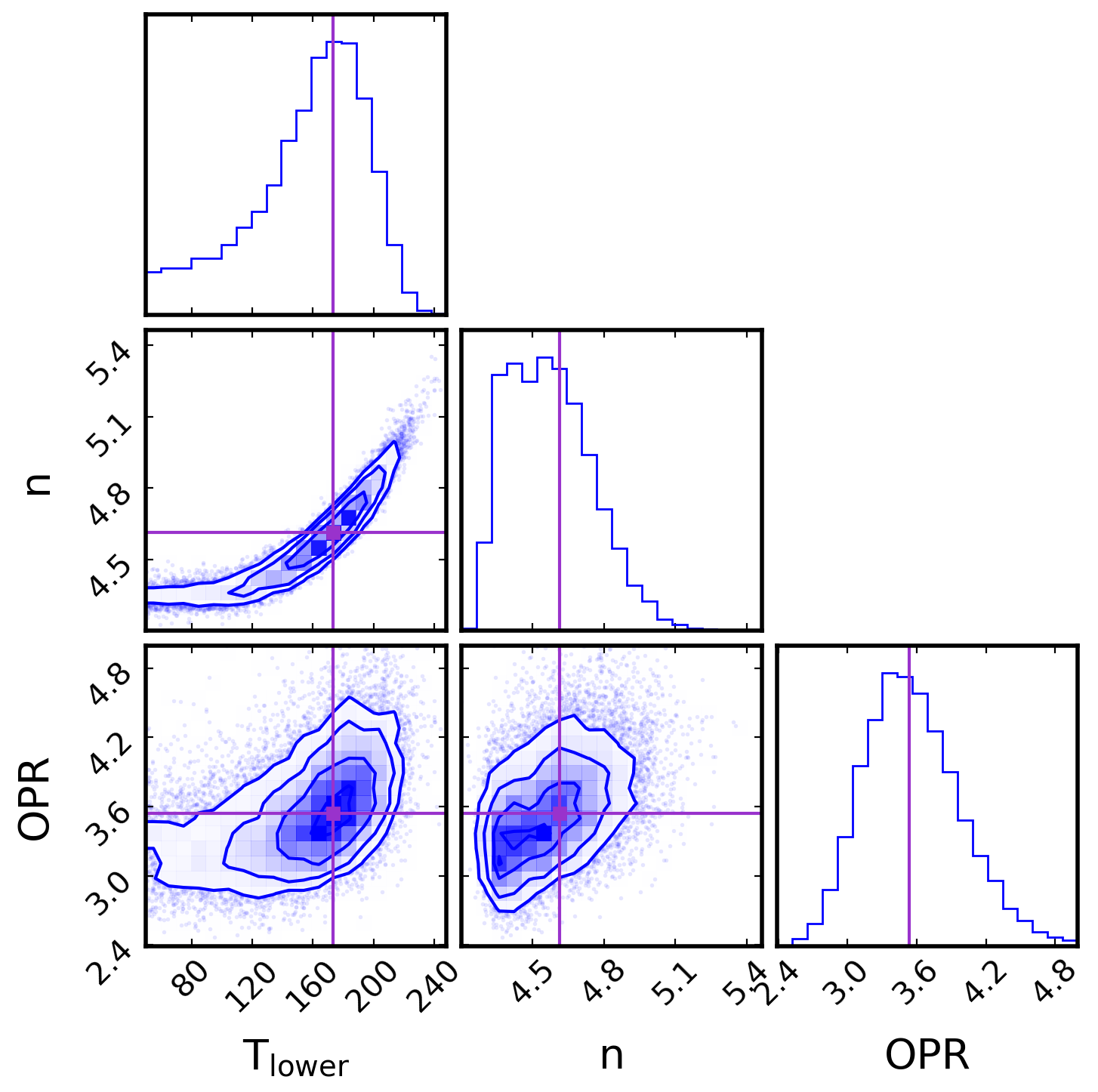}
\includegraphics[width=0.32\linewidth]{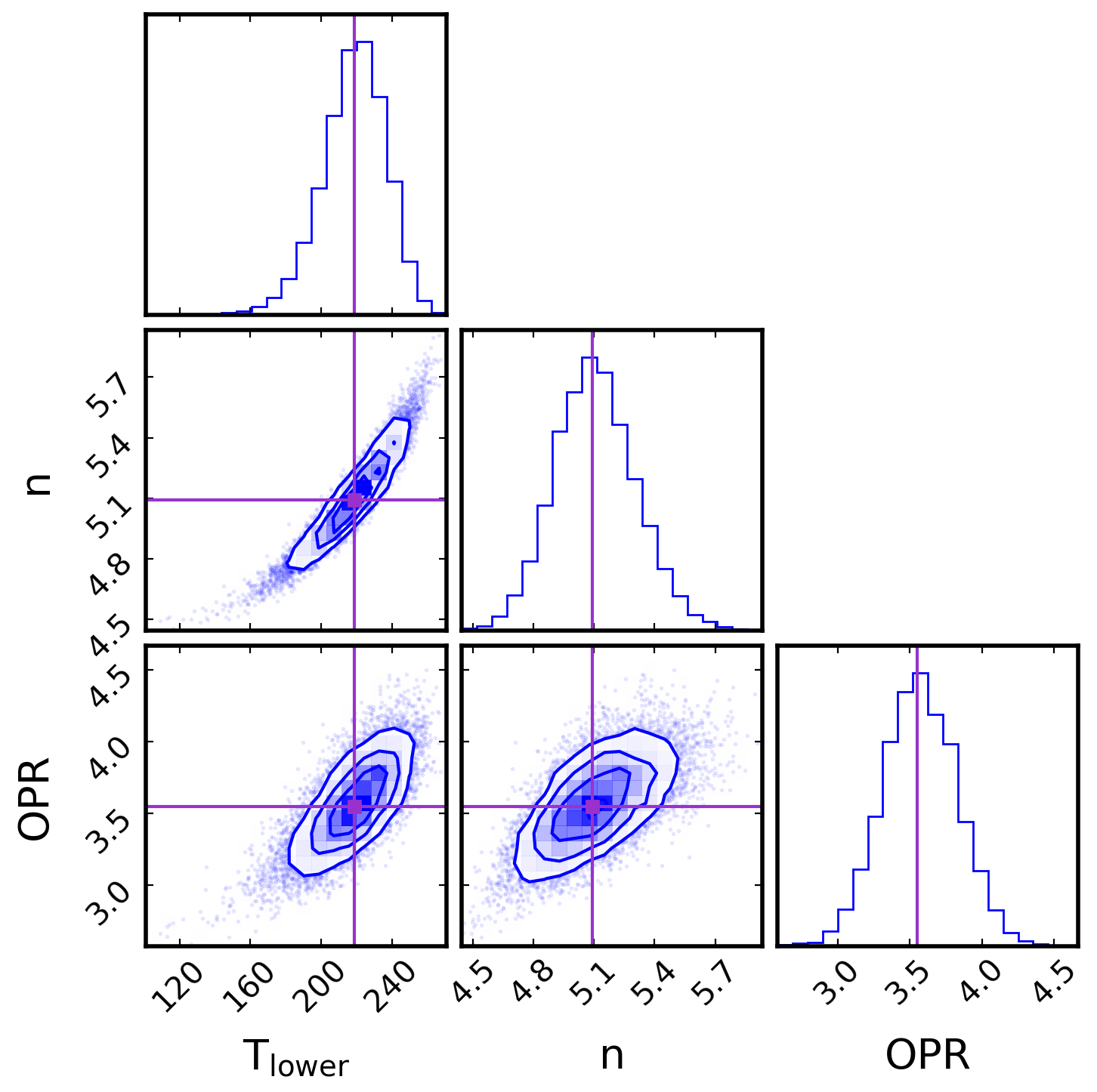} 
\includegraphics[width=0.32\linewidth]{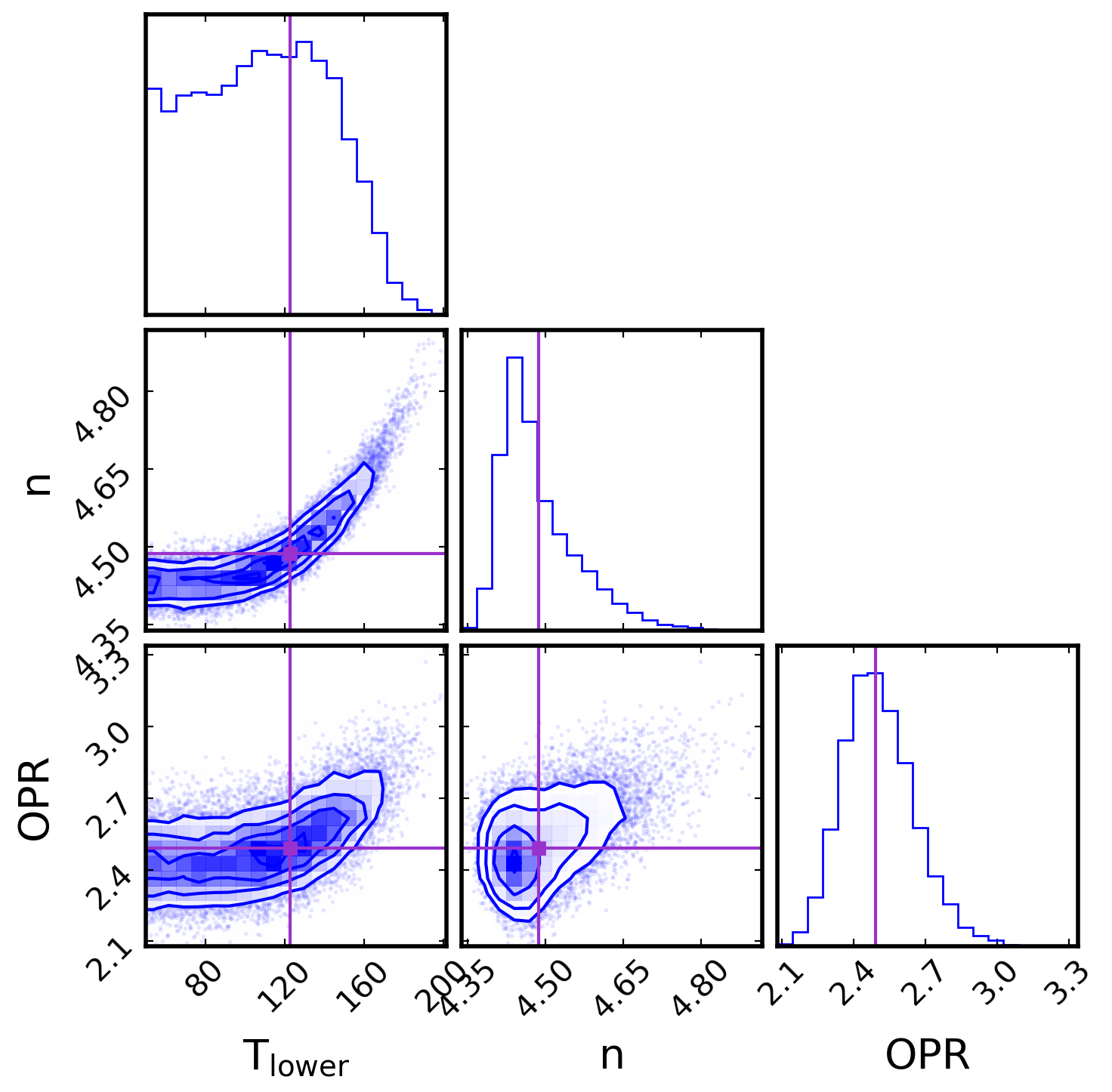}\\
}
\vbox{
\hspace{-0.3cm}\textbf{SE}\hspace{4.8cm}\textbf{VLA-SE}\hspace{4.4cm}\textbf{VLA-NW-A}\\ 
\centering
\includegraphics[width=0.32\linewidth]{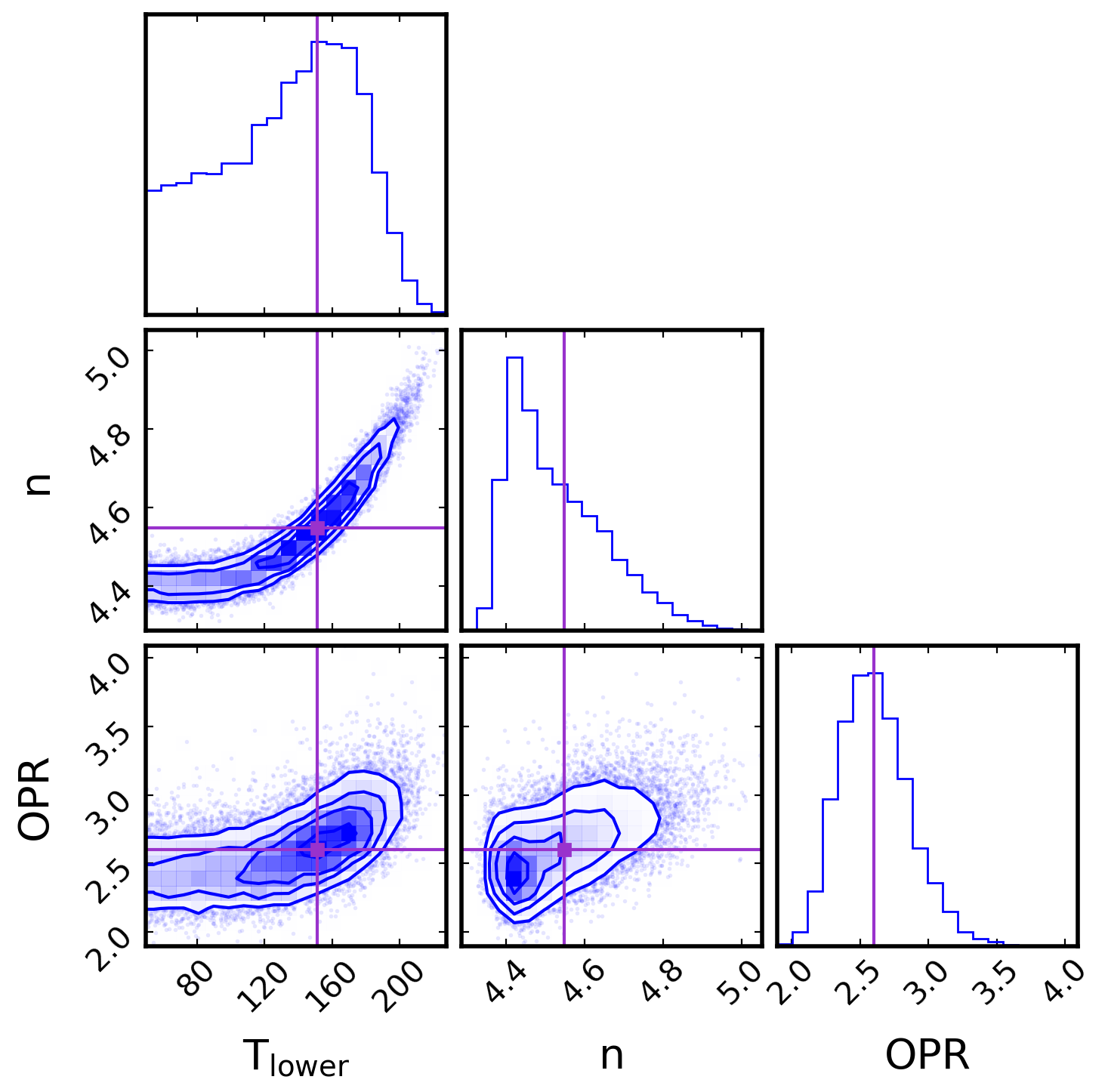}
\includegraphics[width=0.32\linewidth]{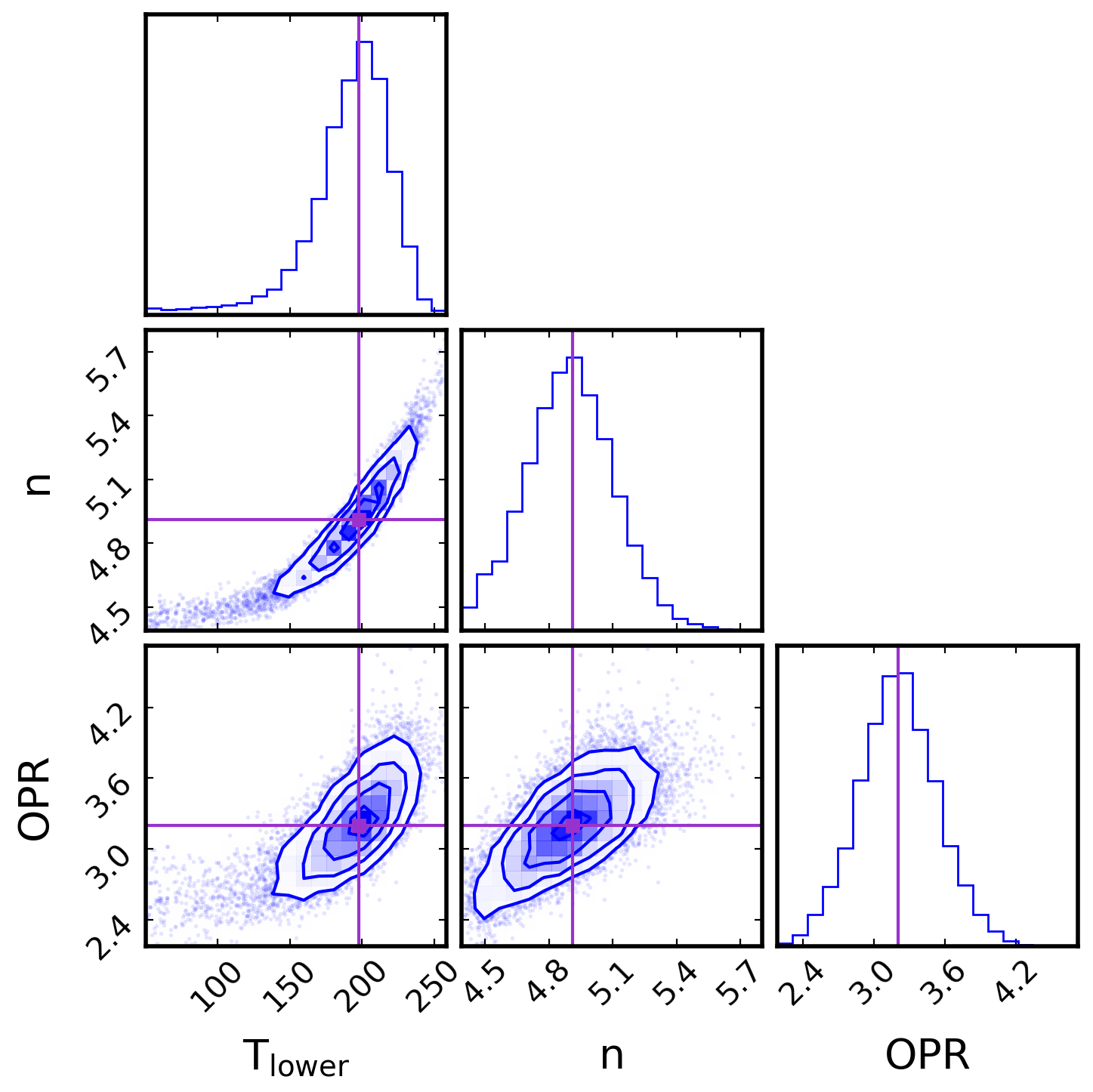} 
\includegraphics[width=0.32\linewidth]{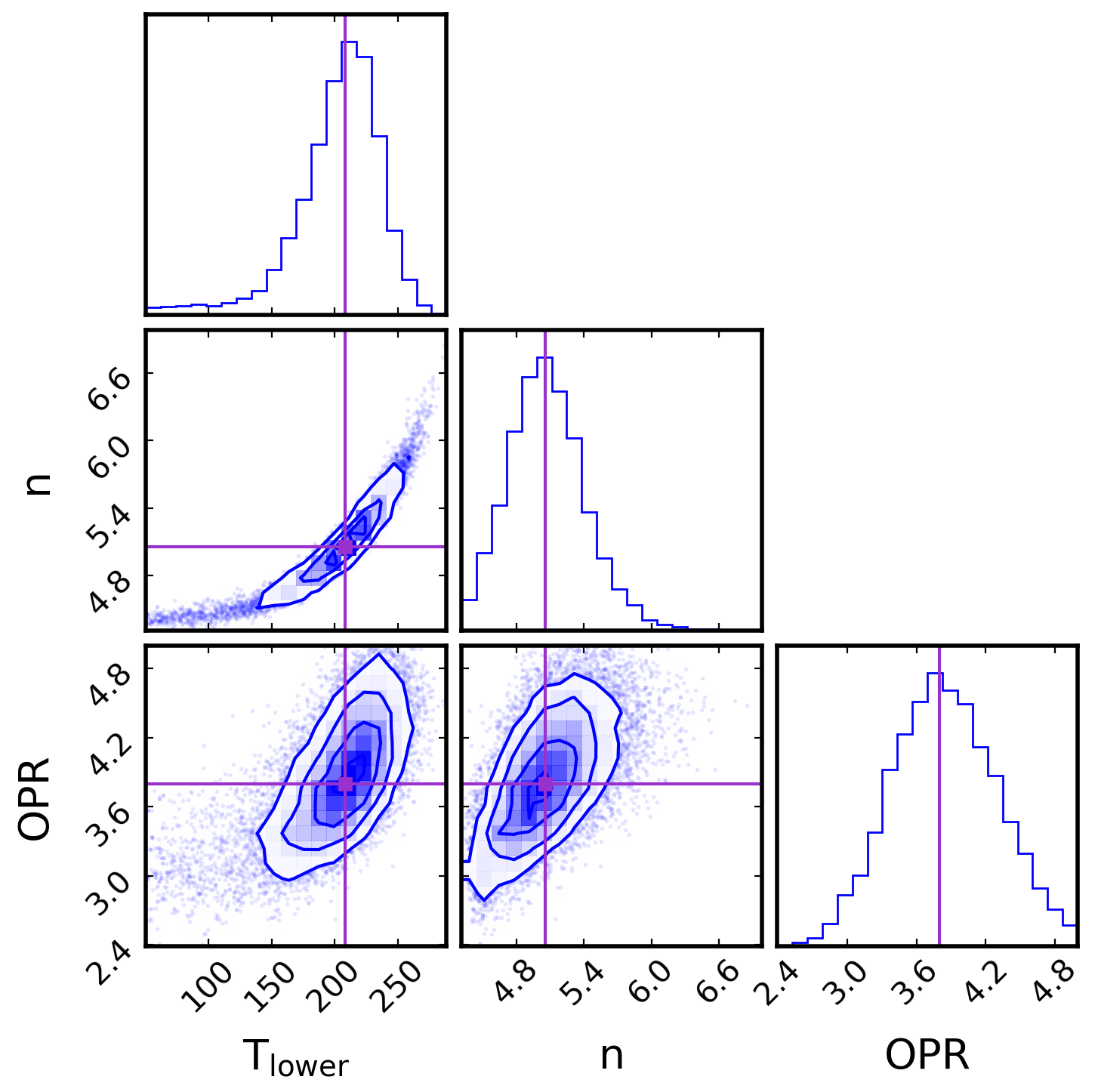} \\
}
\vbox{
}
\caption{Corner plots for the (\tl, $n$, OPR) fits of the \htwo\ population diagrams with \tu\,=\,2000\,K.
As in Fig. \ref{fig:corner}, for each region, the top and right-most panels show the PDF,
and the lower corner panels the co-dependence of the three parameters. 
As in Fig. \ref{fig:corner}, \tl\ and $n$ are correlated, but here
the OPR also is correlated with \tl\ and $n$, with a tendency toward flatness
(lower left sub-panels) in JWST-SE-3 and SE;
these are also the regions where OPR $<\,3$.
}
\label{fig:corner_opr}
\end{figure*}

\subsection{Fitting the OPR \label{sec:opr}}

It can be seen from Fig. \ref{fig:pop} that some of the high-$J$ lines are not well fit
by our empirical model.
This could be due to non-LTE processes exciting the lines, 
such as UV pumping,
which we will discuss in Sects. \ref{sec:nonLTE} and \ref{sec:excitation},
but also to our assumption about a fixed OPR\,=\,3.
Thus, we have extended the LTE model to include a third variable, the OPR.
This reduces the constraints on the fits by introducing an additional variable,
but, with at least five \htwo\ transitions considered, it is feasible.

As for the previous fits, uniform MCMC priors are adopted for \tl, $n$, and OPR, with 
\tl\ constrained to be $\geq\,50$\,K;
we also repeated the fitting procedure for both \tu\,=\,2000\,K and \tu\,=\,3500\,K.
The new fits including the OPR (with \tu\,=\,2000\,K) are shown in Fig. \ref{fig:pop} as red curves,
and the best-fit parameters are reported in the bottom portion of Table \ref{tab:popfits}.
Figure \ref{fig:corner_opr} gives the corner plots from the MCMC sampling.
As before, we have taken the best-fit values to be those that come from the $\chi^2$
minimization, and, in general, they coincide with the mode and median of the posterior PDFs.
This is not the case for $n$ in the SE aperture, where the PDF extends to lower values.
When fitting also the OPR, the uncertainties on \tl\ are generally larger than when OPR is fixed.
Nevertheless, comparing the AIC of the fits with and without fitting the OPR suggests
that the variable OPR fits are generally 
the more statistically valid representation of the data;
in particular, the fits where OPR$\,>\,3$ are always superior from this perspective.

\begin{figure*}[t!]
\centering
\includegraphics[height=0.35\linewidth]{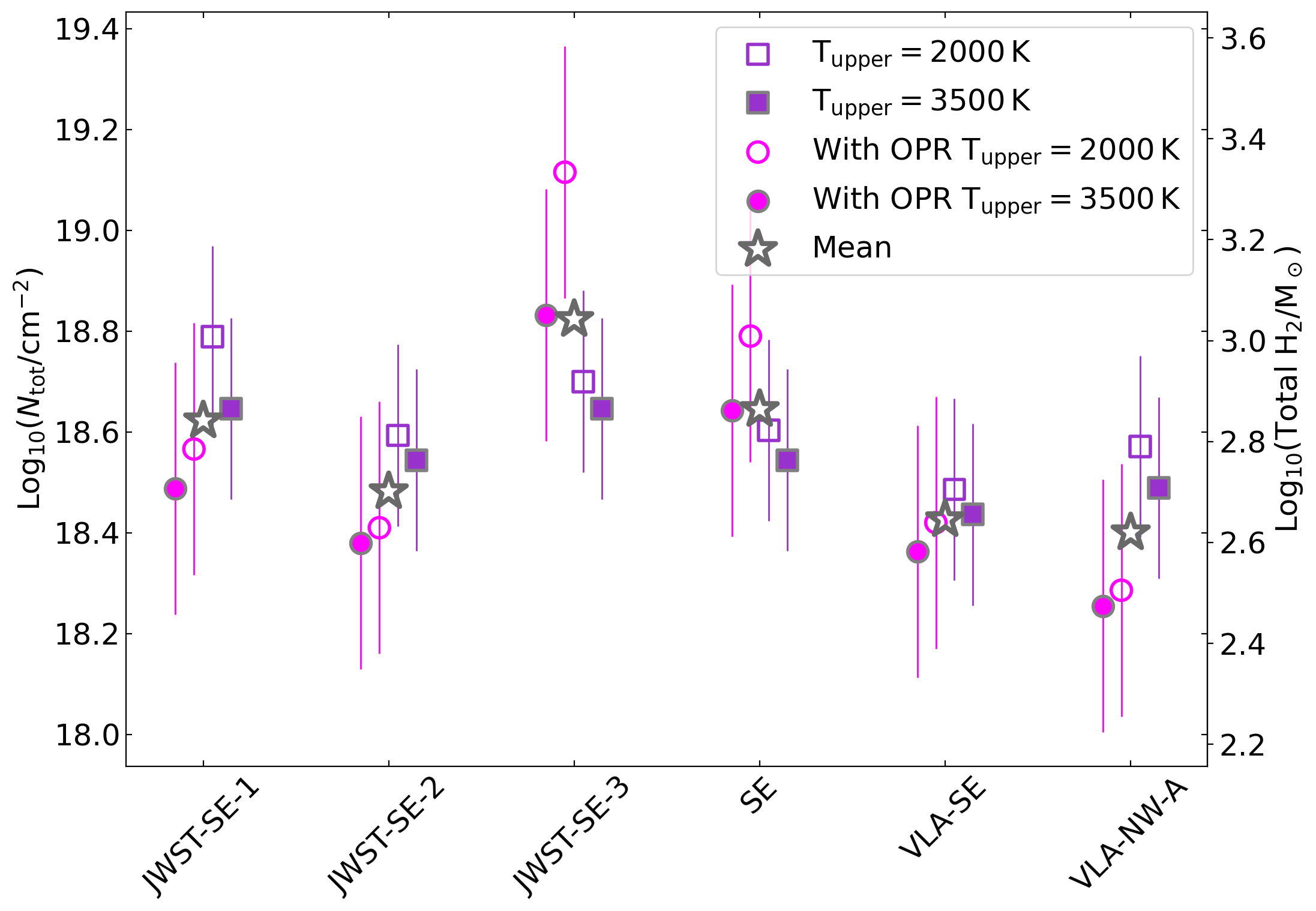}
\hspace{0.03\linewidth}
\includegraphics[height=0.35\linewidth]{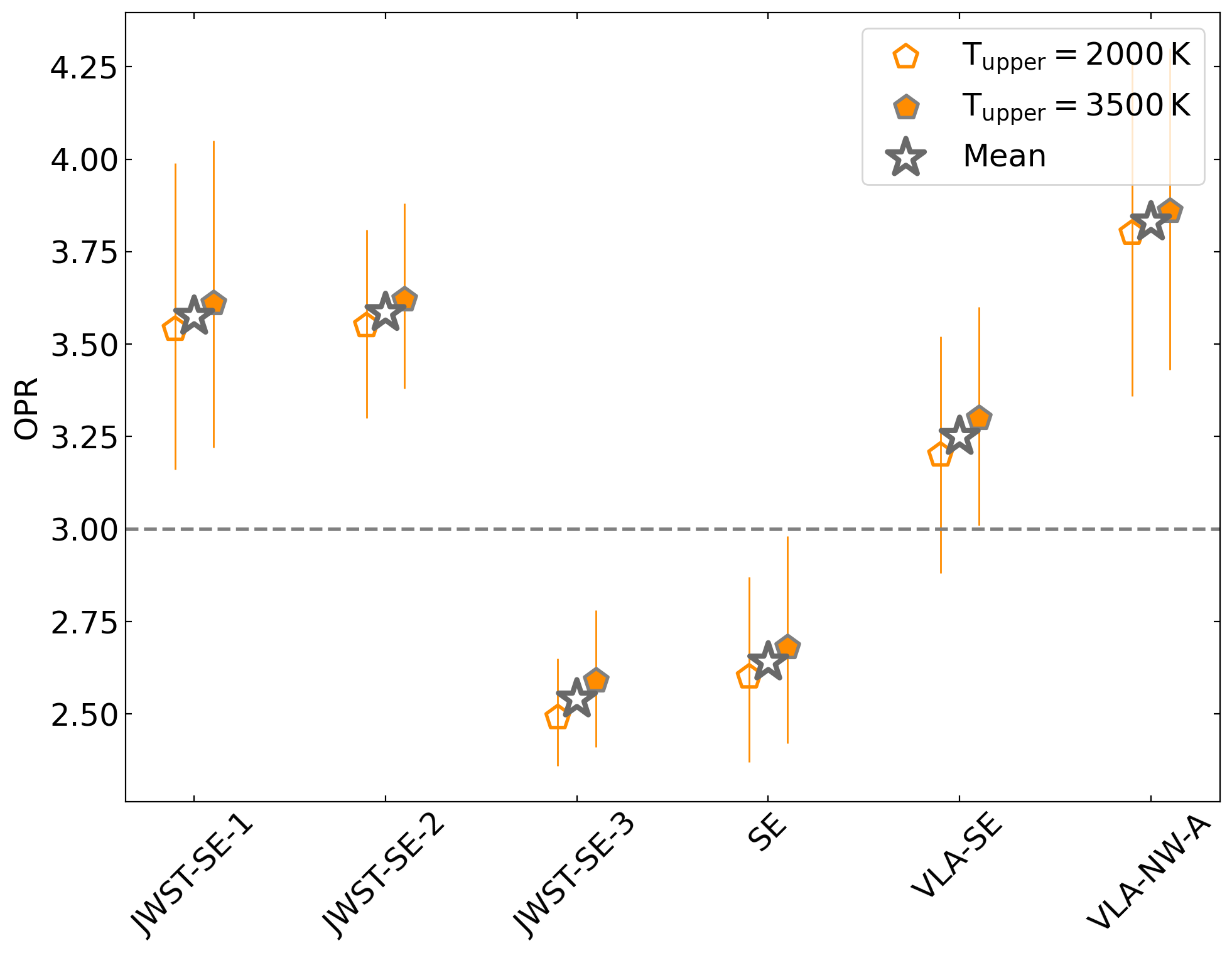}\\
\caption{Left panel: Comparisons of the \htwo\ column \Ntot\ and \htwo\ total mass
inferred from fits at different \tu, and also with and without fitting the OPR.
The left-hand ordinate scale shows \htwo\ column density, and the right \htwo\ total mass
within the apertures.
Right panel: Comparison of OPR inferred from fits with different \tu.
The open star shows the mean for each region. 
In the right panel, the horizontal dashed line indicates the maximum LTE value
of OPR\,=\,3. 
}
\label{fig:compare_tupper}
\end{figure*}

Appendix \ref{sec:tupper} reports the results of the \htwo\ population diagram fits
with \tu\,=\,3500\,K, with and without fitting the OPR.
Higher \tu\ gives higher values of \tl\ and $n$, thus slightly
lower values of \Ntot\ and \htwo\ mass within the aperture. 
The left panel of Fig. \ref{fig:compare_tupper} illustrates the comparison of the estimates of 
\Ntot\ and \htwo\ mass, 
considering the two \tu\ values and with/without fitting the OPR;
the right panel shows the comparison with \tu\ of best-fit OPR.
Both \Ntot\ and \htwo\ mass are well constrained, showing relatively little variation
among the four sets of fits.
The value of the fitted OPR also does not vary significantly with \tu;
the root-mean-square difference of the best-fit OPR for the two values of \tu\ is only 0.08, thus
reinforcing the significance of the OPR values $>3$ and $<3$ resulting from our analysis.
For more details, see Appendix \ref{sec:tupper}.

From Fig. \ref{fig:pop}, it is evident that although the lowest \htwo\ transitions from S(1) to
S(5) are well fit by this simple model,
the higher $J$ transitions, S(7) in JWST-SE-1 and SE (as well as S(8) in SE), tend not to be.
We attribute this to an increasing contribution of non-LTE processes such
as UV pumping to the higher $J$ levels (Sect. \ref{sec:excitation}), 
or simply to the failure of the simple power-law model (Eq. \ref{eqn:temp})
to describe the actual temperature distribution.
More high-$J$ rotational lines or lines from vibrationally-excited levels would be needed to substantiate this hypothesis.

\subsubsection{Best-fit OPR $>$ 3 \label{sec:oprgt3}}

The corner plots in Fig. \ref{fig:corner_opr} show that for three regions, JWST-SE-1, JWST-SE-2, and VLA-NW-A, 
the models significantly favor a best-fit OPR $>$ 3.
These three apertures are spatially independent, so we are sampling similar physical conditions
in different regions of the galaxy.
Although the possibility of OPR $>$ 3 was predicted theoretically by \citet{draine96} and \citet{sternberg99},
this is the first time that an OPR $> 3$ has been measured in the ISM of any star-forming region.

The \htwo\ OPR is a variable arising from a complex interplay of competing physical processes.
The true OPR in a PDR depends on temperature, collisions of \htwo\ with protons and hydrogen atoms,
selective photodissociation of ortho- and para-\htwo, the \htwo\ formation rates,
and the efficiency of grain-surface reactions \citep[e.g.,][]{sternberg99,bron16}.
In star-forming galaxies with metallicities comparable to Solar, \htwo\ formation is dominated
by catalysis on grain surfaces; this process is expected to create \htwo\ molecules
with an OPR$\,\approx\,$3. 
There is also \htwo\ production in the gas phase via $H^- + H \rightarrow H_2 + e^-$, 
but in chemically evolved galaxies,
this process is very sub-dominant compared to grain catalysis.
However, in very metal-poor systems, gas-phase production will be the dominant channel.
Because the gas-phase reaction is relatively indifferent to the \htwo\ proton spins,
this process is also expected to produce \htwo\ with an OPR$\,\approx\,$3.

A higher OPR$\,>\,3$ could arise through effective self-shielding,  
since the dominant process for \htwo\ destruction is photodissociation.
If OPR $> 1$, ortho-\htwo\ self-shields more effectively than para-\htwo. 
Preferential destruction of para-\htwo\ acts to increase OPR. 
In a stationary PDR, photodissociation and \htwo\ formation are balanced, 
and ortho $\rightarrow$ para conversion tends to be fast enough to keep OPR $\la 3$. 
However, when an increase in the FUV flux causes the photodissociation rate to exceed the local \htwo\ formation rate, 
a photodissociation front will propagate rapidly into the \htwo\ gas.

\begin{figure}[h!]
\centering
\includegraphics[width=\linewidth]{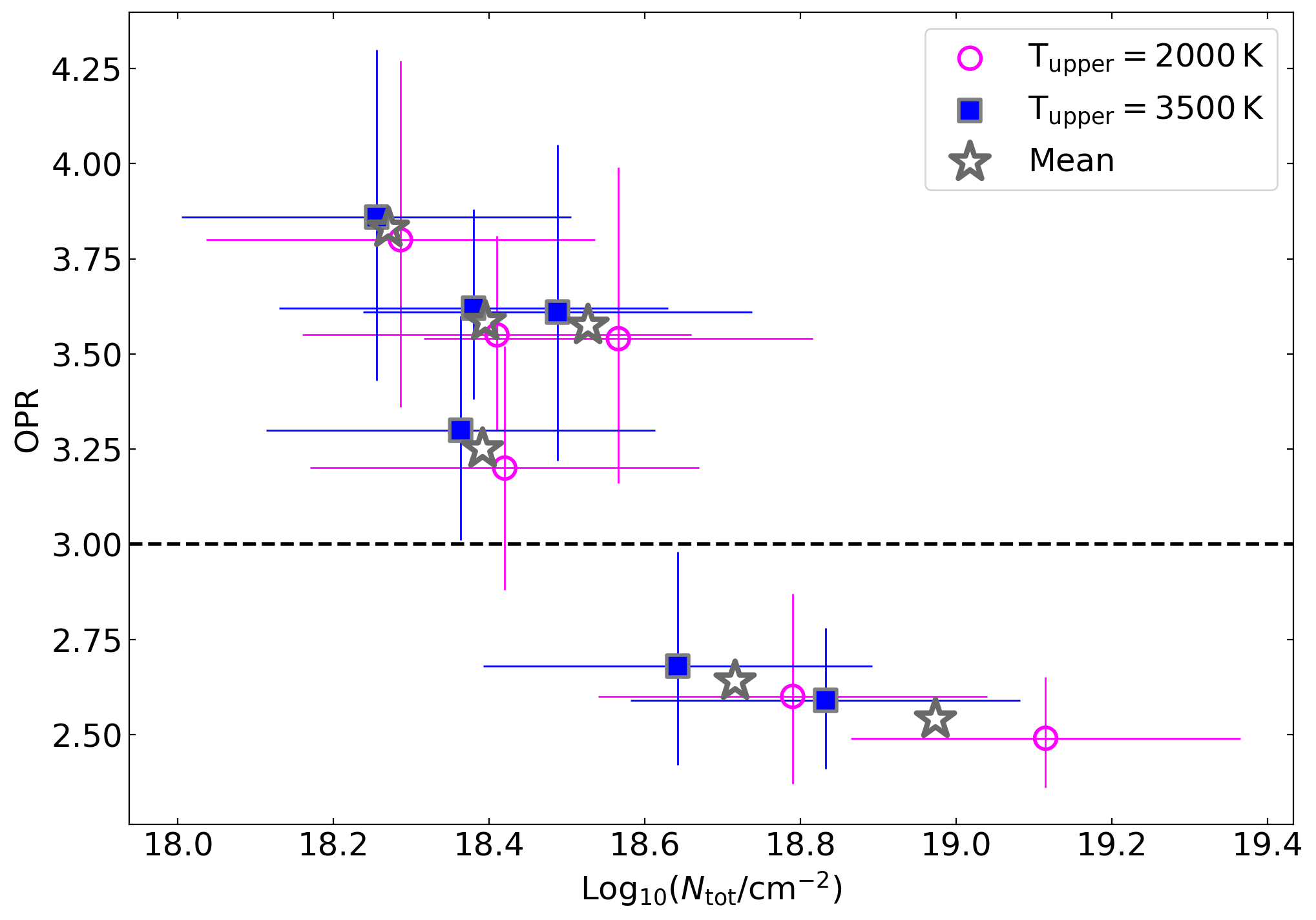}
\caption{OPR plotted against \htwo\ column density \Ntot.
The different values of \tu\ are shown as open (magenta) circles
and filled (blue) squares for \tu\,=\,2000\,K, 3500\,K, respectively.
The mean of the two determinations is shown as an open (gray) star.
The horizontal dashed line corresponds to the LTE maximum OPR value of 3.
}
\label{fig:opr_h2col}
\end{figure}

In a low-metallicity galaxy like \izw, the
highly-suppressed rate for \htwo\ formation on dust grains will
result in rapidly-propagating photodissociation fronts,
leading to an increase in the OPR. 
JWST-SE-1, JWST-SE-2, and VLA-NW-A, with OPR $> 3$, may be regions 
where molecular gas is being irradiated by recently-formed OB stars, 
with preferential destruction of para-\htwo\ causing OPR to increase beyond 3.
In such a case, we might expect lower total \htwo\ column densities where OPR $> 3$.
Figure \ref{fig:opr_h2col} shows the OPR for each region plotted against total \htwo\
column \Ntot.
Indeed, except for VLA-SE which overlaps significantly with SE and JWST-SE-2,
the trend of increasing OPR with decreasing \Ntot\ is essentially monotonic,
consistent with OPR $> 3$ resulting from selective photodissociation.

\subsubsection{Best-fit OPR $\la$ 3 \label{sec:oprlt3}}

In VLA-SE, the best-fit OPR is 3 within the uncertainties, and
in JWST-SE-3 and SE, the best-fit OPR is significantly $<$ 3.
As shown in Fig. \ref{fig:apertures}, these apertures partially
overlap, so the conditions sampled by the apertures are not altogether independent.
The variations of OPR among the different regions indicate that in some regions (JWST-SE-3 and SE), 
the ortho $\rightarrow$ para conversion in the gas (or on grain
surfaces) overcomes the OPR-increasing effects of preferential destruction of para-\htwo.
We will explore this point further below.

\begin{figure*}[h!]
\includegraphics[width=0.32\linewidth]{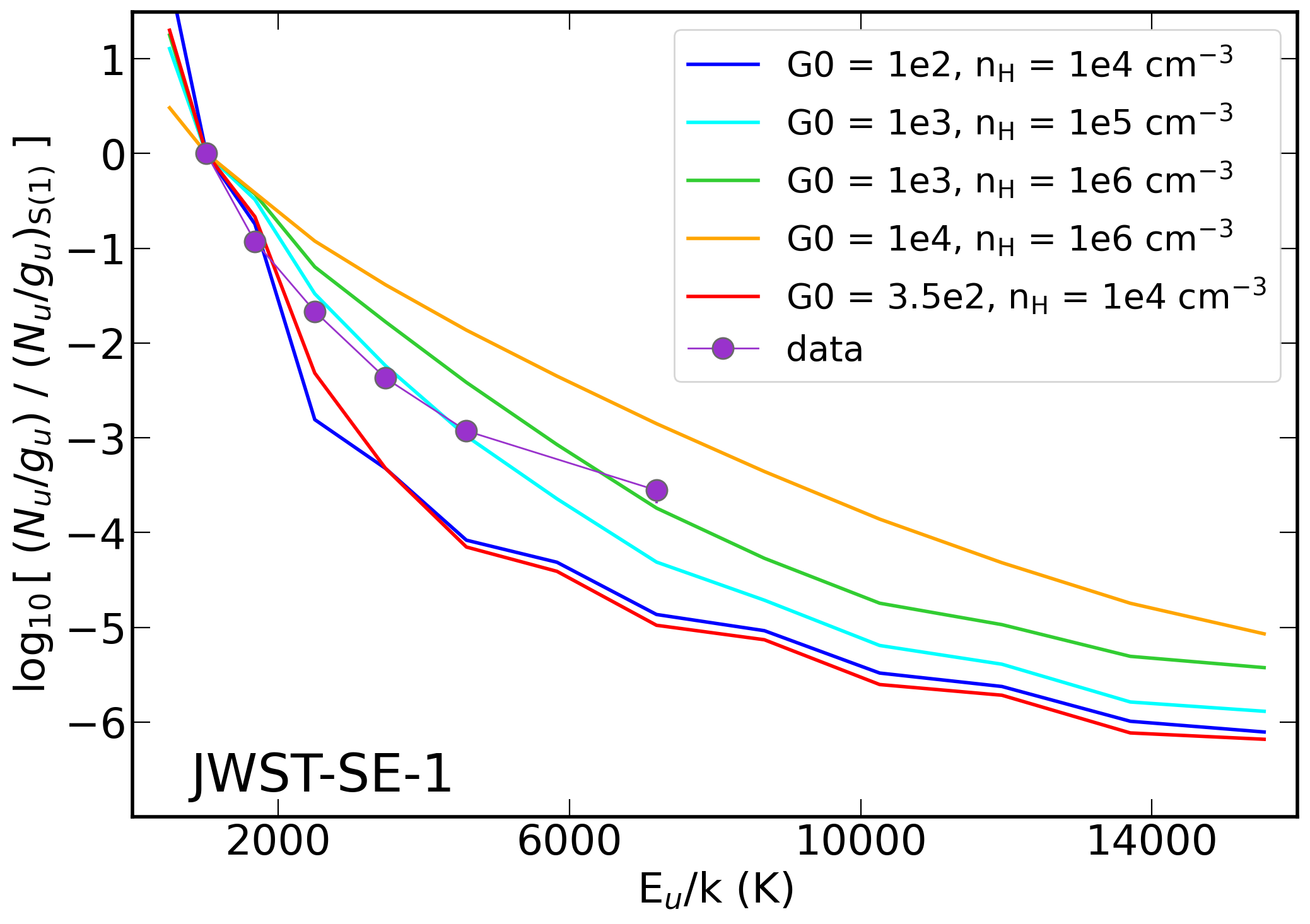}
\includegraphics[width=0.32\linewidth]{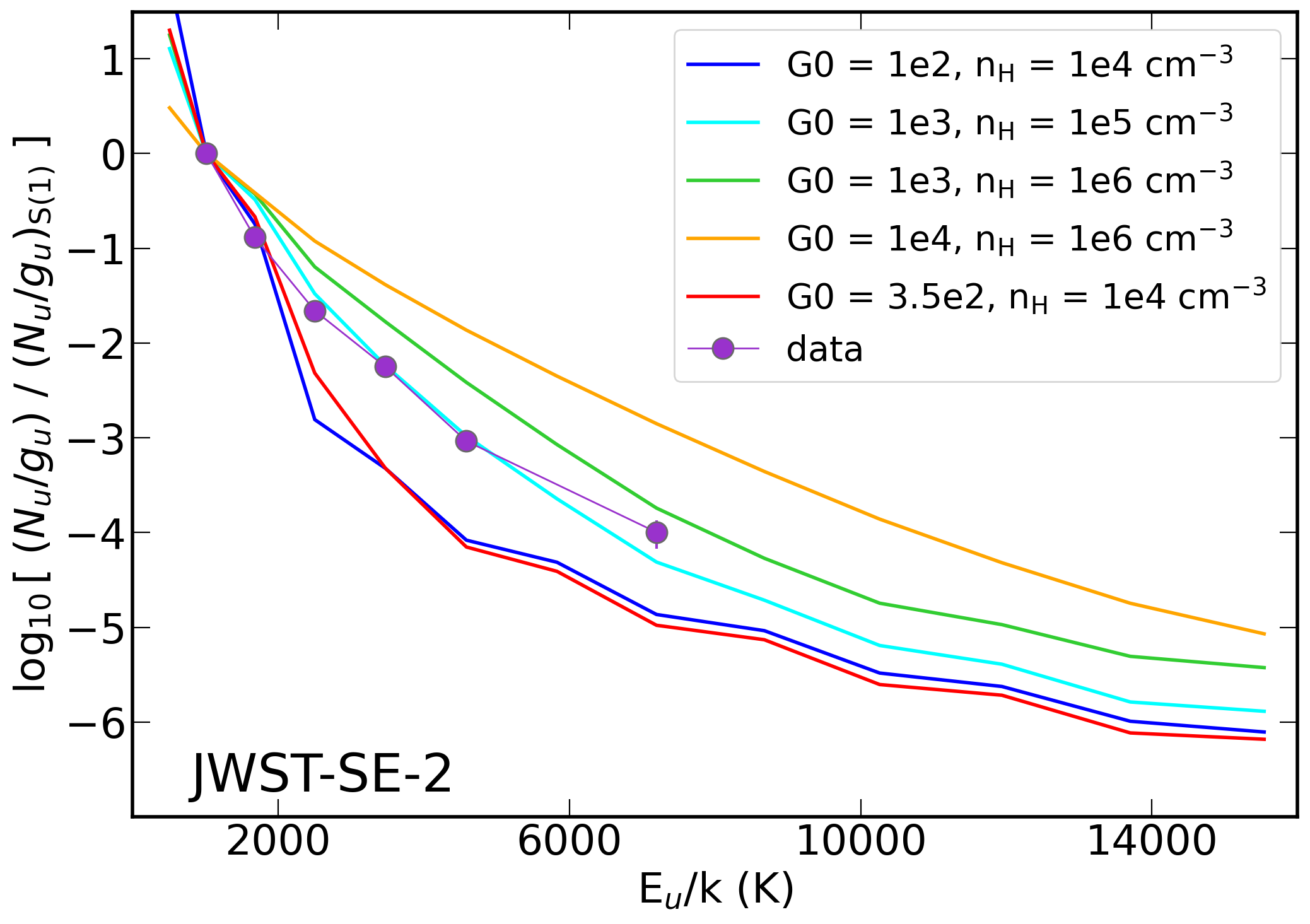}
\includegraphics[width=0.32\linewidth]{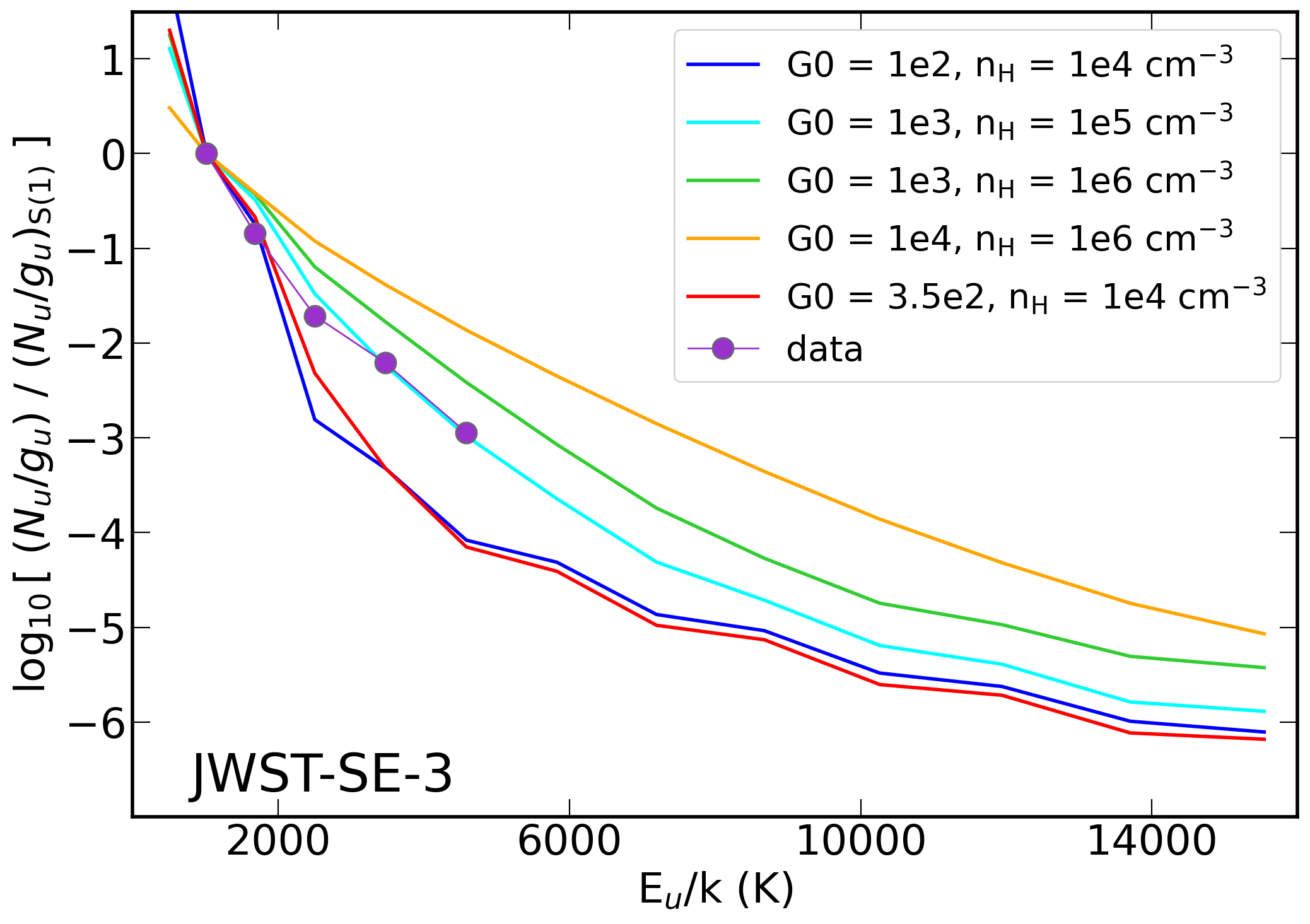} \\
\includegraphics[width=0.32\linewidth]{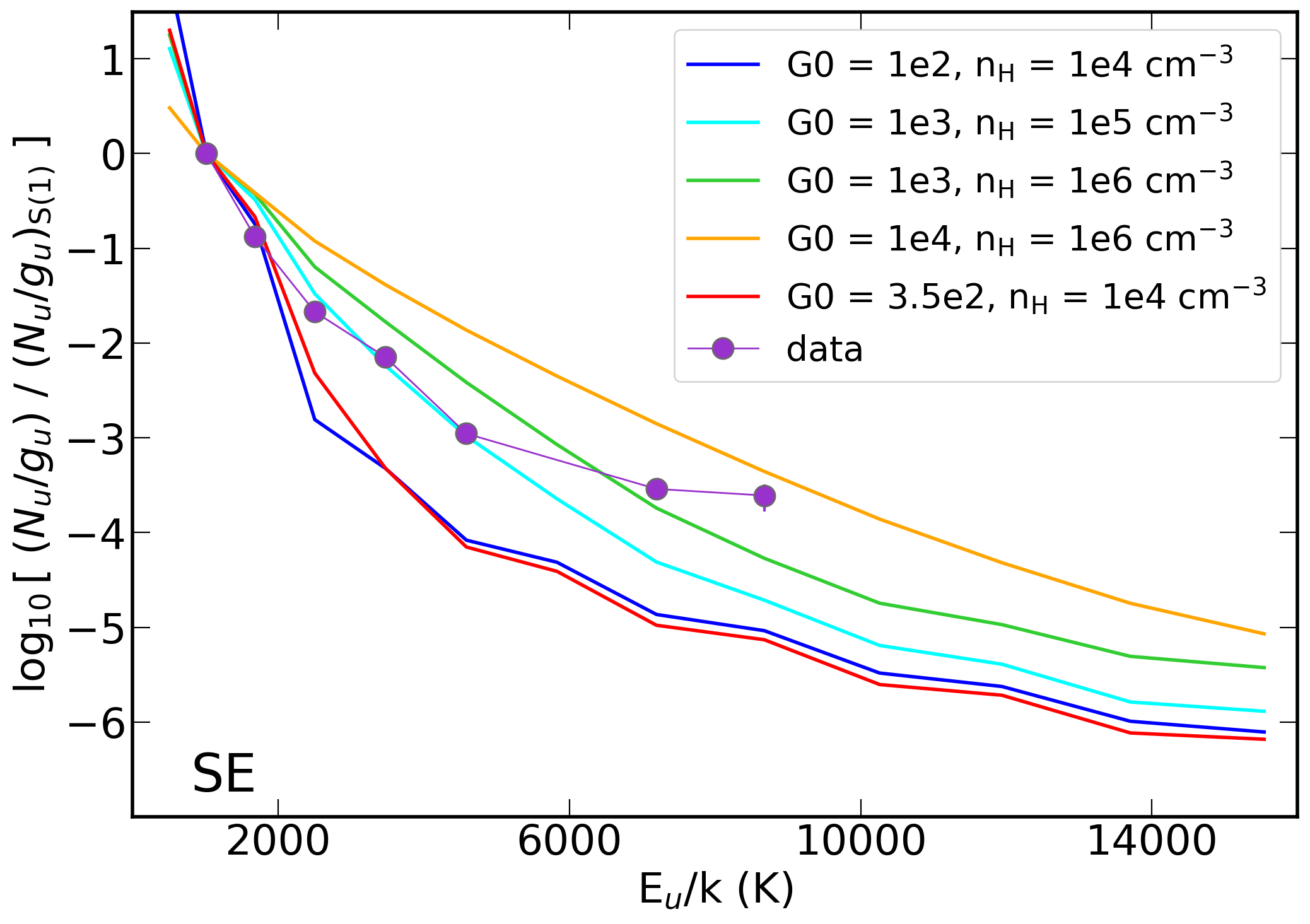}
\includegraphics[width=0.32\linewidth]{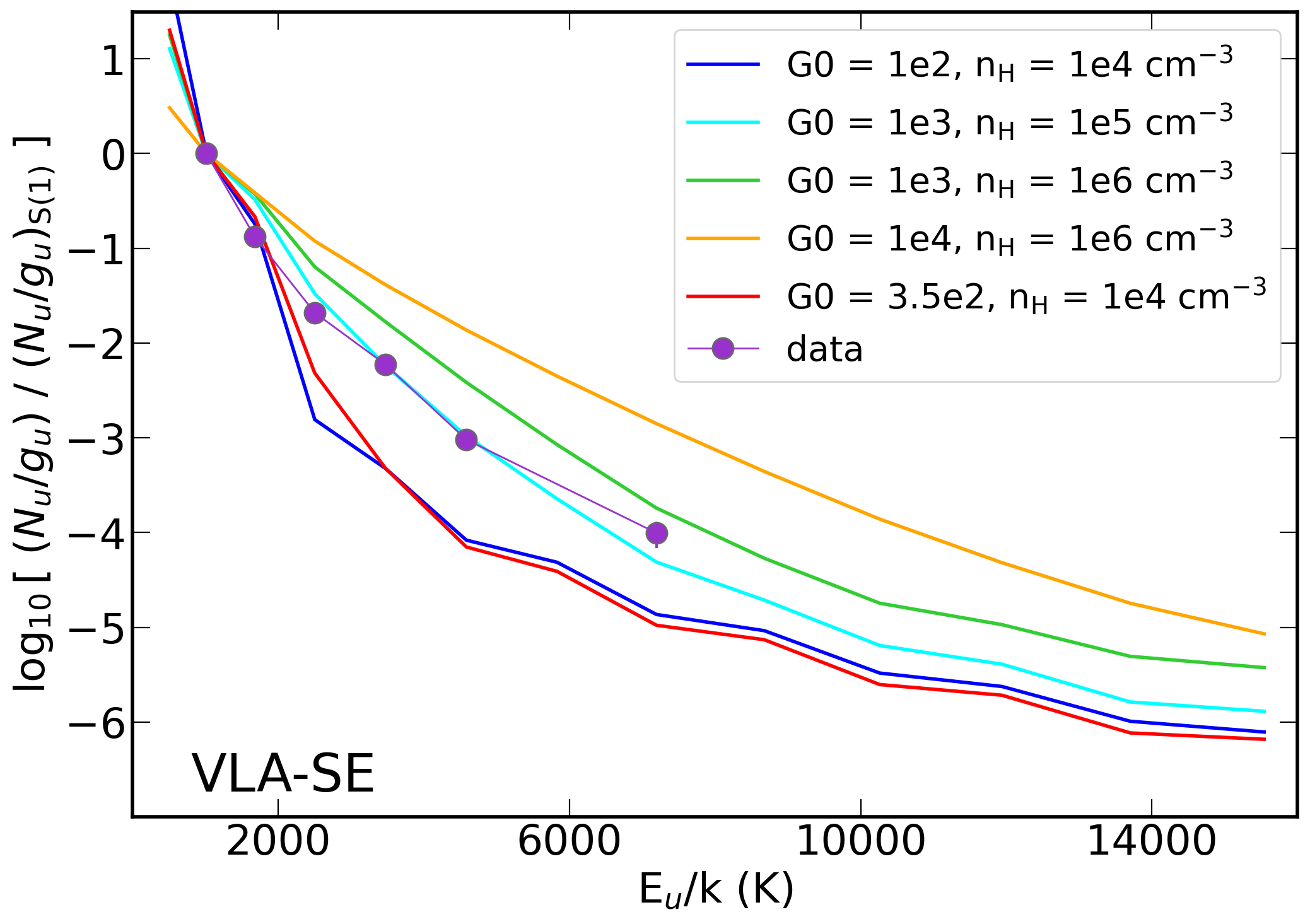}
\includegraphics[width=0.32\linewidth]{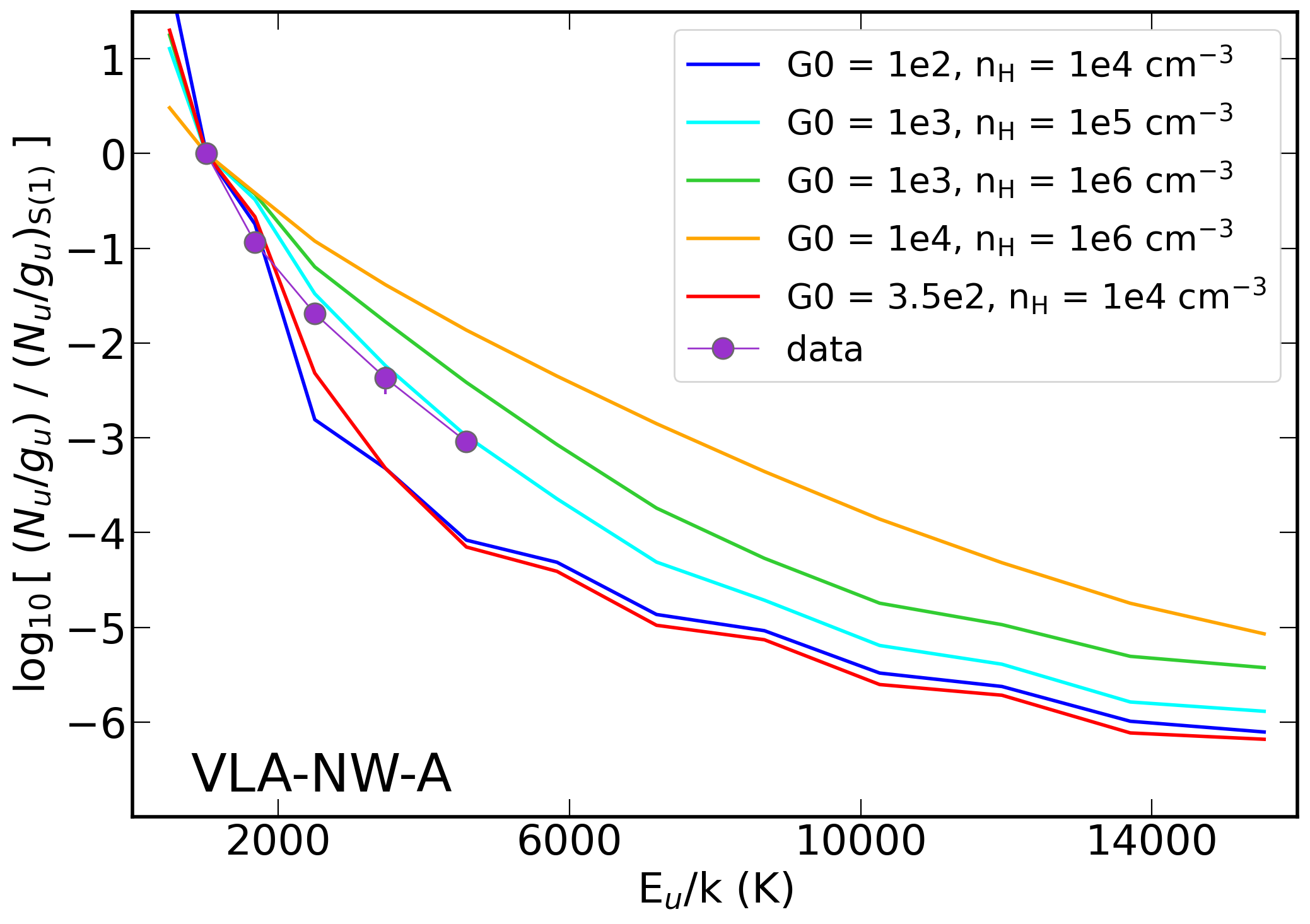} \\
\caption{Comparison of \izw\ \htwo\ population diagrams with Meudon PDR-7 isochoric
models. The visual extinction \av\ was fixed to have a maximum of unity;
only the models with \gnot/\nh\,$\leq\,0.05$ are shown since they
best approximate the observed data.
Further constraints are discussed in the text.
}
\label{fig:meudon}
\end{figure*}

\begin{figure*}[h!]
\includegraphics[width=0.32\linewidth]{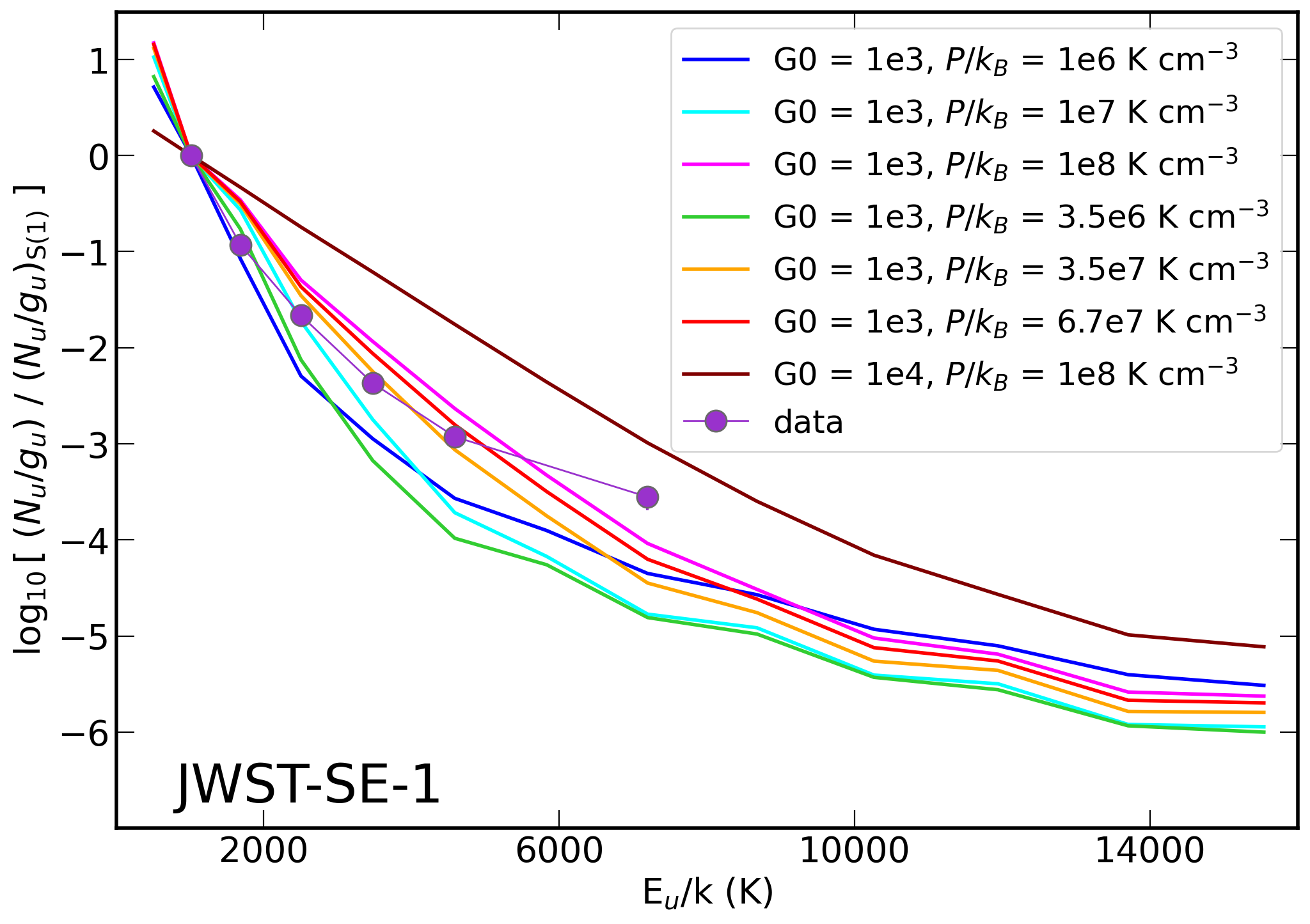}
\includegraphics[width=0.32\linewidth]{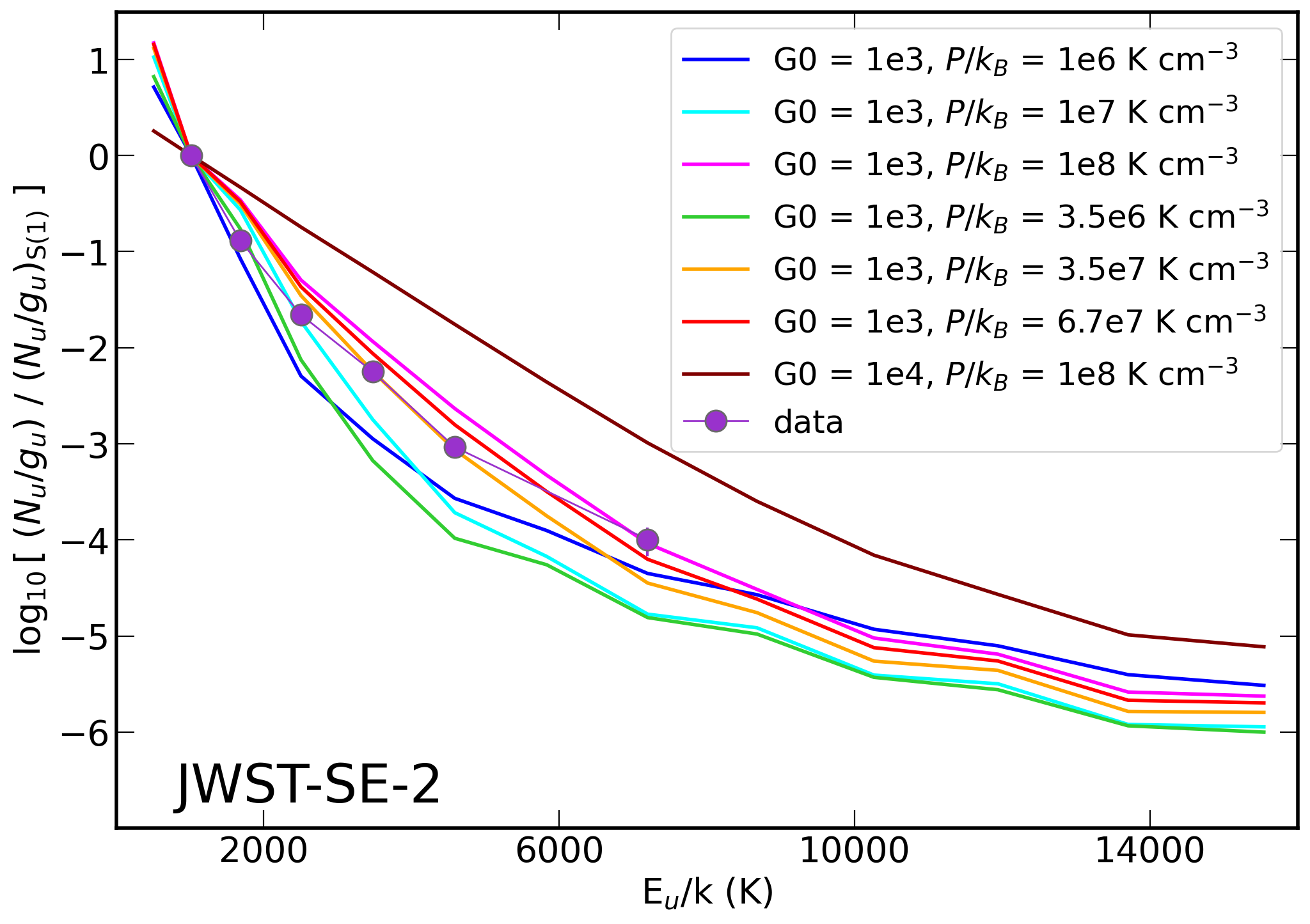}
\includegraphics[width=0.32\linewidth]{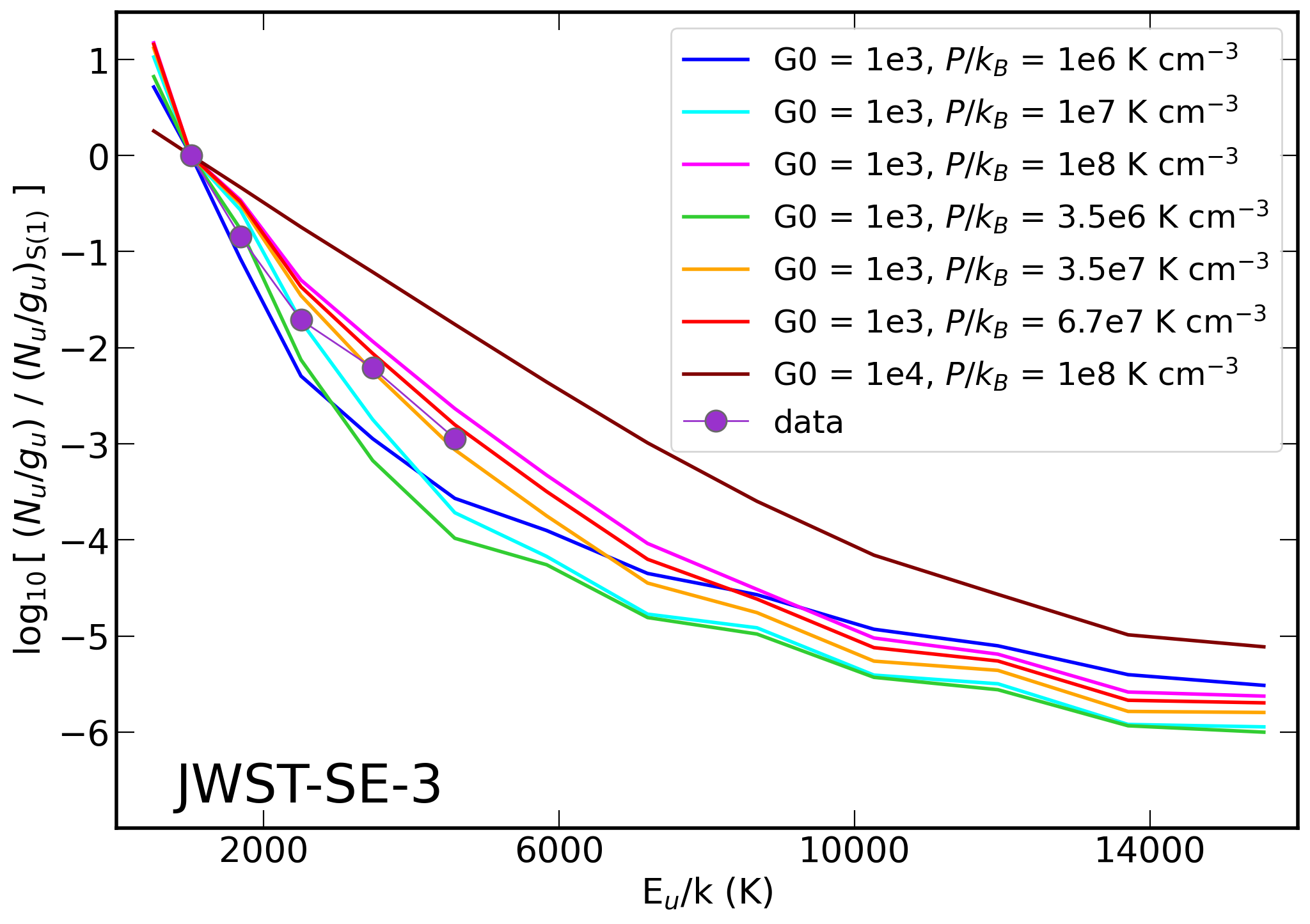} \\
\includegraphics[width=0.32\linewidth]{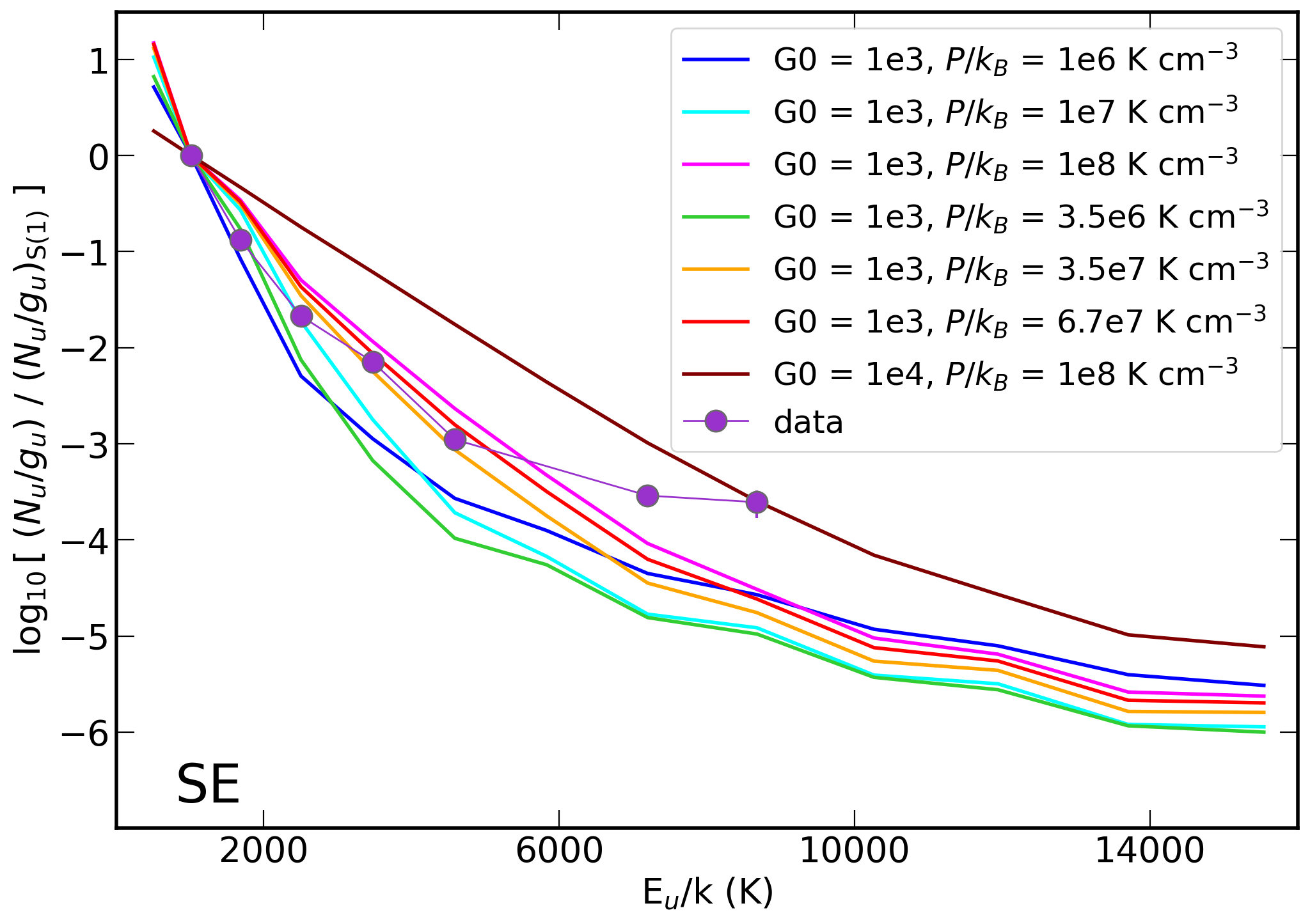}
\includegraphics[width=0.32\linewidth]{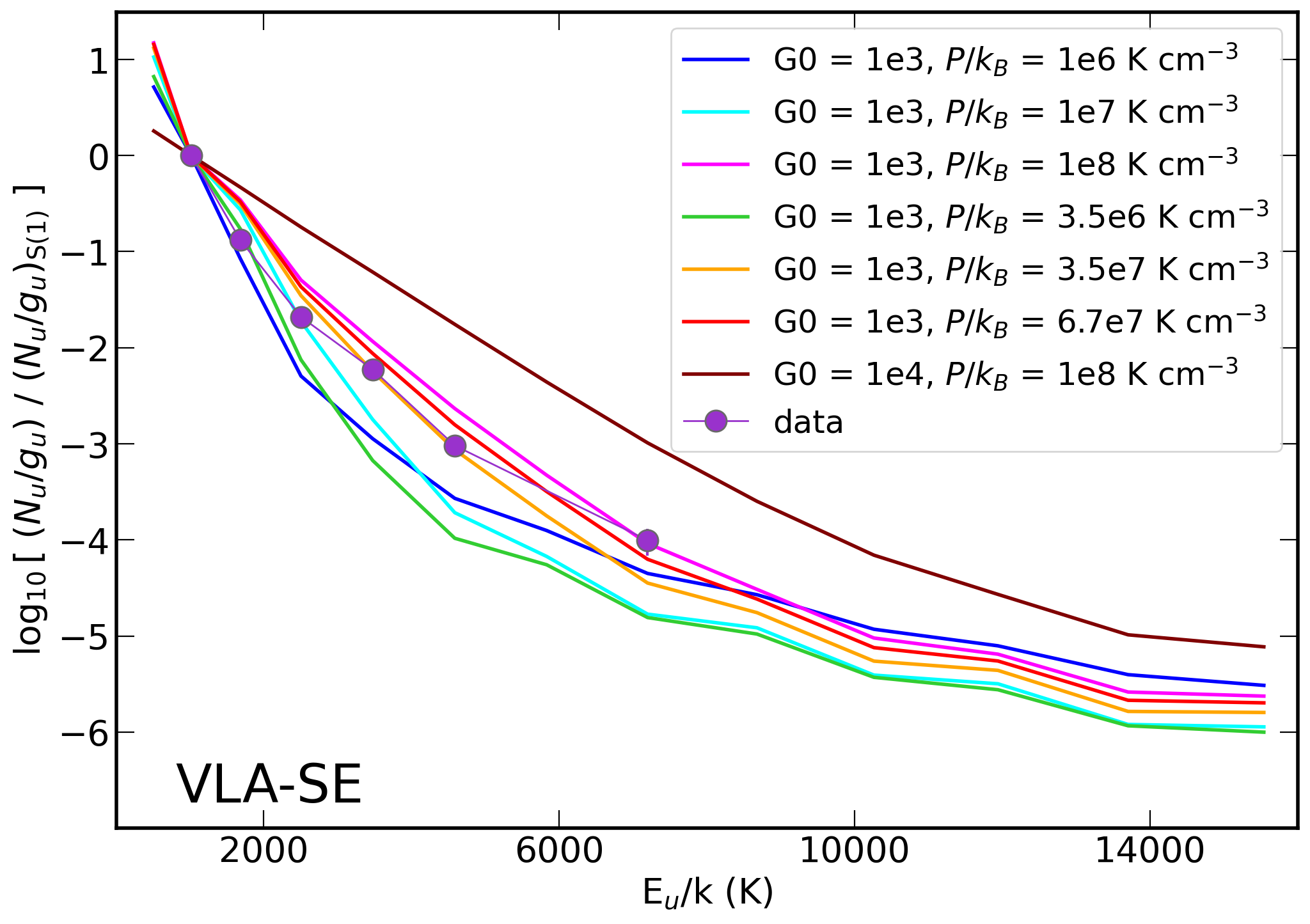}
\includegraphics[width=0.32\linewidth]{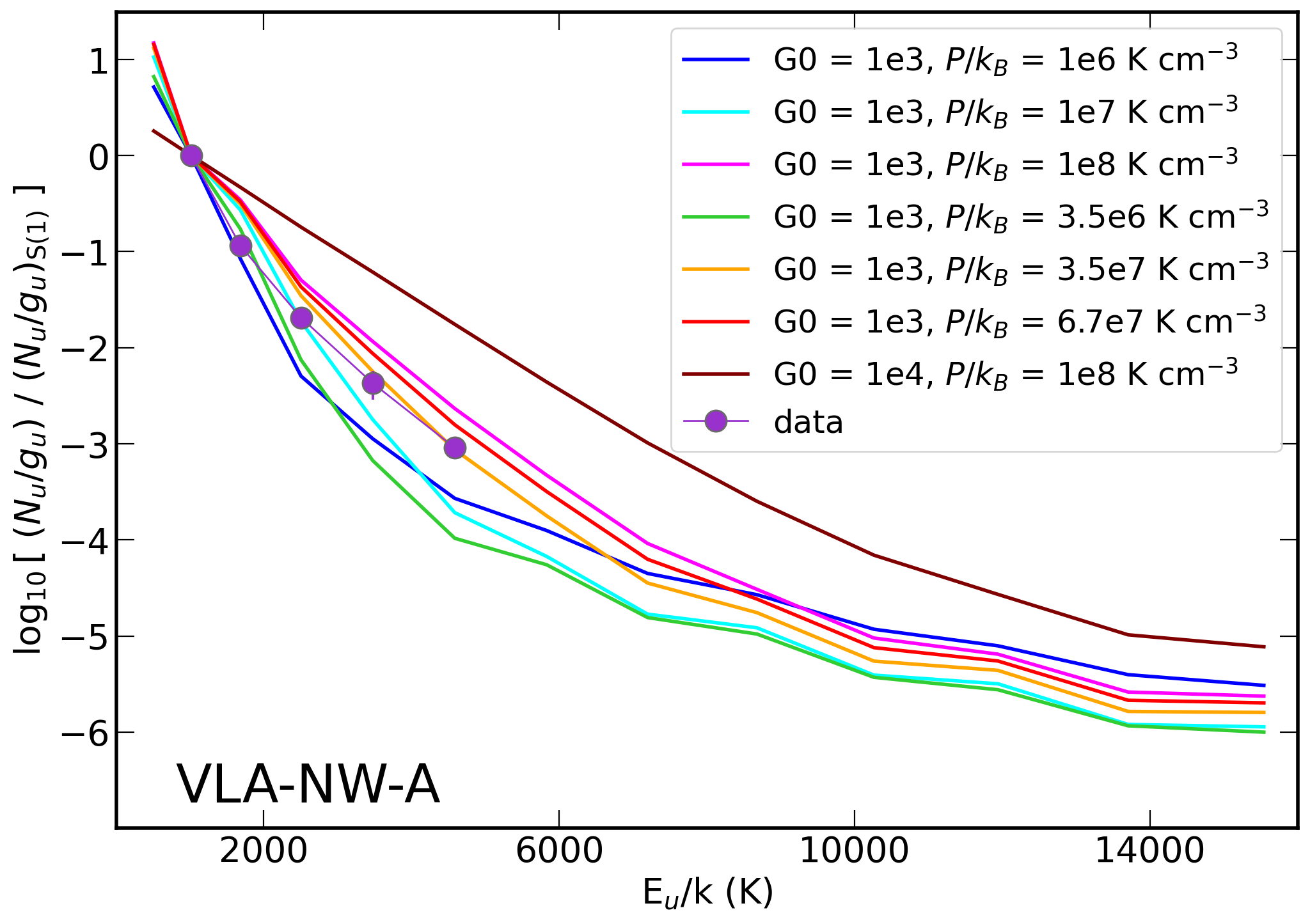} \\
\caption{Comparison of \izw\ \htwo\ population diagrams with Meudon PDR-7 isobaric
models. As for the isochoric models,
the visual extinction \av\ was fixed to have a maximum of unity;
only the models that best approximate the observed data are shown.
Further constraints are discussed in the main text (Sect. \ref{sec:nonLTE}).
}
\label{fig:isobaric}
\end{figure*}

\subsection{Non-LTE excitation of molecular hydrogen \label{sec:nonLTE}}

To better characterize the warm \htwo, we compare the \htwo\ population diagrams with those 
predicted by the Meudon PDR models (PDR-7).\footnote{The Meudon PDR code is part 
of the ISM database, ISMDB, a web-based
fitting tool to fit observations to PDR models \url{https://app.ism.obspm.fr//ismdb/}.}
The UV pumping photophysics in the Meudon model, 
as well as rate coefficients for collisional excitation and
deexcitation of the \htwo, should be appropriate for \izw.
However, the Meudon PDR models assume dust abundances, and coolant (e.g., C$^+$) abundances, 
appropriate for the Milky Way; thus, applicability of the Meudon models to \izw\ is
questionable. 
In addition, the Meudon models assume stationary PDRs, with thermal and chemical balance at each point. 
The ionization-dissociation fronts in \izw\ could be propagating rapidly enough that the 
stationary approximation may not be valid. 
Thus, any conclusions drawn from comparison of \izw\ with the models must be regarded as highly tentative.

We first consider isochoric (constant density) models \citep{lepetit06,lebourlot12};
we varied  \gnot, the FUV radiation field intensity, from $10^2$ to $10^5$ 
(the maximum value provided in the models), and the H nucleon density, \nh,
from $10^2$ to $10^6$\,\cmthree.
Because of the low extinction in \izw\ (Paper\,I), we 
limited \av\ to a maximum of unity; the viewing angle was taken to be face-on, 
with inclination\,=\,0$^\circ$.
No fits have been attempted, but we succeed in constraining \nh\ and \gnot.


The results are shown in Fig. \ref{fig:meudon} where only the models that come closest
to approximating the observed \htwo\ are plotted.
We find that \gnot/\nh\,$\la$\,0.05\,cm$^{3}$; 
otherwise the gas is too warm.
\gnot$\,\sim\,10^2$ makes the fall-off at low $J$ too steep, since in that case, the gas is too cold.
The two best approximations are \gnot\,=\,$10^3$ with either \nh\,=\,$10^5$\,\cmthree,
or \nh\,=\,$10^6$\,\cmthree.
However, \nh\,=\,$10^6$\,\cmthree\ 
(green, orange curves) 
gives high-$J$ level populations that are too high.
Thus, we conclude that \gnot$\sim 10^3$ and \nh\,$\sim 10^5$\,\cmthree\ in the
warm \htwo\ in \izw.
These \nh\ values are toward the upper range of those found in the Orion Bar
based on a similar comparison with the Meudon PDR models by \citet{peeters24}.

We have also compared our \htwo\ observations with the Meudon isobaric (constant pressure) models;
the results are shown in Fig. \ref{fig:isobaric}.
They confirm the \gnot\ value inferred from the isochoric models, and suggest
relatively high pressure;
the best approximation to the \htwo\ population diagrams is a pressure
$P/k_B\,=\,5\times10^{7}$\,K\,cm$^{-3}$ (lying roughly between the plotted curves) and \gnot$\,\sim\,10^3$.
The high pressures inferred from the PDR models are consistent with those
thought to be needed for globular cluster formation
\citep[$P/k_B\,\ga\,10^6$\,cm$^{-3}$\,K, e.g.,][]{elmegreen97,kruijssen15}.
In Paper\,I, we ascribed the nature of the 14\,\micron\ continuum sources
to young stellar clusters (YSCs), 
possibly similar to the compact 21\,\micron-selected sources identified in nearby
galaxies by \citet{hassani23}.
If JWST-SE-1, 2, and 3 are truly YSCs, then such a high pressure
would be necessary to set the stage for their formation.
It could also mean that there may be more undiscovered YSCs in \izw,
since there is no clear evidence for higher pressure around the continuum
sources compared to the other regions.

The values of \gnot\,$\sim 10^3$ we infer for \izw\ are
comparable to those found in the N\,13 PDR in the Small Magellanic Cloud 
\citep[SMC,][]{clark25}, although the pressure inferred is roughly
six times higher ($P/k_B\,\sim\,5\times10^7$\,K\,cm$^{-3}$ vs
$P/k_B\,\sim\,8\times10^6$\,K\,cm$^{-3}$ for N13 in the SMC).
The Meudon PDR models are for gas with $Z \approx $\zsun\ and dust/gas ratios characteristic of the Milky
Way, and are therefore imperfect guides to the PDRs in \izw. 
The much lower metallicity of \izw\ ($Z\,=\,0.03$\,\zsun) compared even to the SMC 
\citep[\zzsun\,$\sim 0.2$,][]{toribio17}
implies reduced rates for dust-catalyzed formation of \htwo, reduced photoelectric heating, and 
reduced cooling by \cii\ and \oi. 
Hence, it would not be surprising if the physical conditions in the PDRs 
responsible for the \htwo\ emission in \izw\ differ considerably from the Meudon models, and the N31 PDR in the SMC. 
In Sect. \ref{sec:dust}, we discuss the IR emission from dust in four regions of \izw\ with strong \htwo\ emission. 
The inferred dust temperatures, $T \sim 45-120$\,K, likely require \gnot$\sim 10^2 - 10^4$ to heat the dust, 
consistent with the \gnot\ values inferred from the PDR models.

Since the Meudon PDR models include ortho-para conversion \citep{bron14,bron16}, as well as \htwo\ 
photodissociation, we have also compared the Meudon OPR
with what we derive from our LTE fits.
Over the entire parameter space sampled by the Meudon models, the OPR is $< 3$:
in the isochoric, constant-density models, the OPR varies from roughly unity with 
(\gnot$\,=\,100$, \nh\,=\,$10^4$\,\cmthree),
to roughly 3 with (\gnot$\,=\,10^4$, \nh\,=\,$10^6$\,\cmthree), and in the 
the isobaric ones, the OPR ranges from 1.5 to 2.7,
over $P/k_B$ from $10^6$ to $10^8$ K\,cm$^{-3}$.

As reported in Sect. \ref{sec:opr},
we find OPR $>\,3$ in three regions, JWST-SE-1, JWST-SE-2, and VLA-NW-A.
We speculate that OPR $> 3$ in \izw\ may be a consequence of preferential photodissociation 
of para-\htwo\ in a propagating photodissociation front (see Fig. \ref{fig:opr_h2col}),
with little reformation of \htwo\ because of the reduced dust abundance.

\subsection{Fraction of \htwo\ in the apertures \label{sec:fraction}}

We have estimated the fraction of \htwo\ falling within the apertures
relative to the total \htwo\ in \izw\ by 
calculating the fraction of the S(1) transition in the extracted spectra
with respect to the spectrum extracted from the full Channel 3 cube
(see Fig. \ref{fig:apertures}).
The spectrum extracted from the Channel 3 cube gives an S(1) flux of roughly $2.7\times10^{-19}$\,W\,m$^{-2}$.
The seven independent (non-overlapping) apertures
(CO2-1, JWST-SE-1, JWST-SE-2, JWST-SE-3, VLA-NW-A, NW, and VLA-NW-C),
together account for $\approx 1.15\times10^{-19}$\,W\,m$^{-2}$,
or $\approx 43$\% of the S(1) flux from the full Channel 3 cube.

The \htwo\ excitation has been seen to vary among the different apertures, 
and there may be even greater variation when considering the fainter regions outside the selected
apertures. Nevertheless, with $\ga 40$\% of the S(1) flux,
we are likely accounting for an appreciable fraction of the
warm \htwo\ in \izw. Of course, it is possible that there are
additional cold \htwo\ clouds that are not exposed to the
heating processes required to populate levels $J_\mathrm{upper}\,=\,3$ and above.

\subsection{Excitation of \htwo\ and UV pumping \label{sec:excitation}}

We have seen that in most regions the observed 
\htwo\ $v=0$ levels in \izw\ -- from $J=3\rightarrow1$ up to $J=9\rightarrow7$ -- can be reproduced by a simple
model with a specified ortho/para ratio OPR, and a specified
power-law distribution of temperatures $dN({\rm H}_2)/dT \propto
T^{-n}$, if the \htwo\ level populations are assumed to be in LTE at
each temperature $T$.  This model predicts that $N_i/g_i$ should
depend {\bf\emph{only}} on the energy $E_i$ of the excited state, with
$N_i/g_i$ monotonically decreasing with increasing $E_i$.

If the rotationally excited \htwo\ is due to collisional excitation,
we expect vibrationally excited levels to follow the same dependence of 
$N_i/g_i$ on $E_i$.
However, if UV pumping is important, 
the vibrationally excited levels will have $N_i/g_i$
falling above the trend for the $v=0$ levels. 
This behavior is seen in
Galactic PDRs 
\citep[e.g.,][]{Burton+Howe+Geballe+Brand_1998,Bertoldi+Draine+Rosenthal+etal_2000,Meyer+Lauroesch+Sofia+etal_2001}.
NIRSpec measurements of emission lines from $v=1$ levels would disclose the role of UV pumping in \izw.

\section{Spectral Signature of Dust at Low Metallicity \label{sec:dust}}

Here, we focus on the dust continuum apparent in the spectra shown in Fig. \ref{fig:spectra}.
For this analysis, the data
cubes have been convolved to $27$\,\micron\ resolution prior to
extraction of the spectrum in the selected regions (see Sect. \ref{sec:convolved}).  
To better reveal the dust continuum emission, 
the detected emission lines (see Paper\,I) in the MIRI-MRS
spectra have been removed.
Then, to improve the signal-to-noise, the line-subtracted spectra were smoothed 
by applying a moving median filter over 71 spectral elements, 
reducing the effective resolution to $R\approx 60$.  


\begin{figure*}
\begin{center}
\includegraphics[angle=0,width=0.35\linewidth]{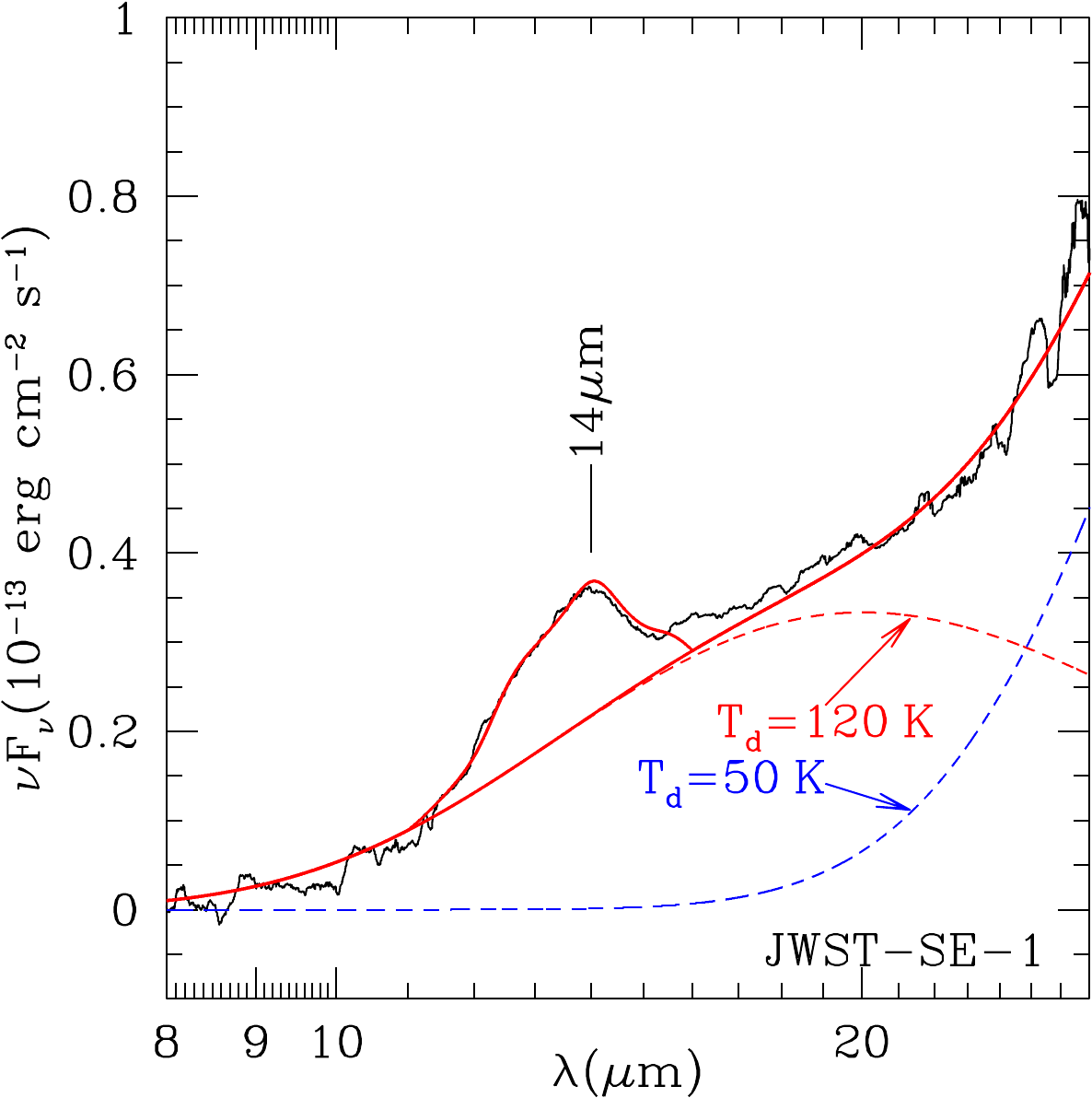}
\hspace{0.05\linewidth}
\includegraphics[angle=0,width=0.35\linewidth]{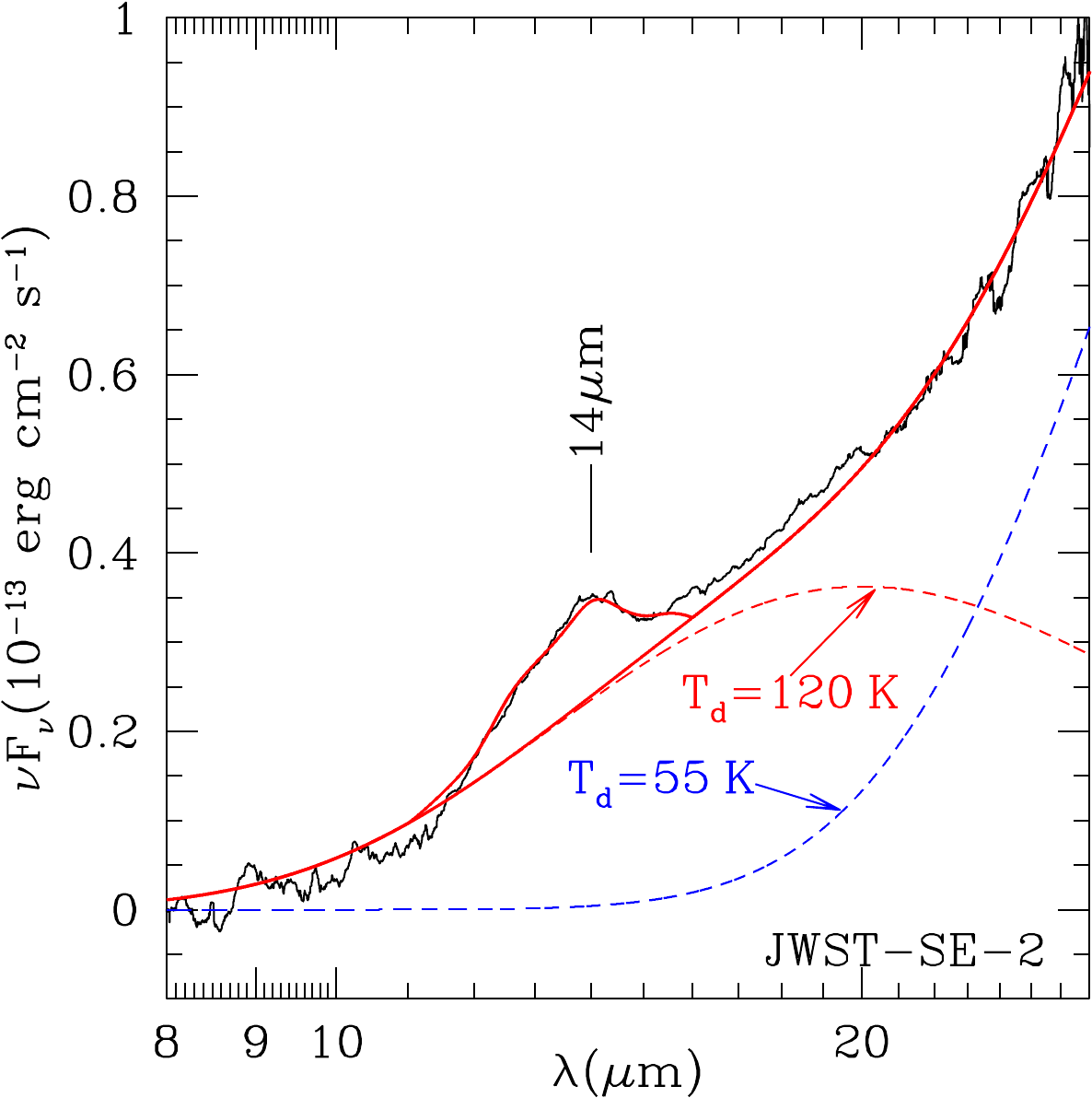} \\
\includegraphics[angle=0,width=0.35\linewidth]{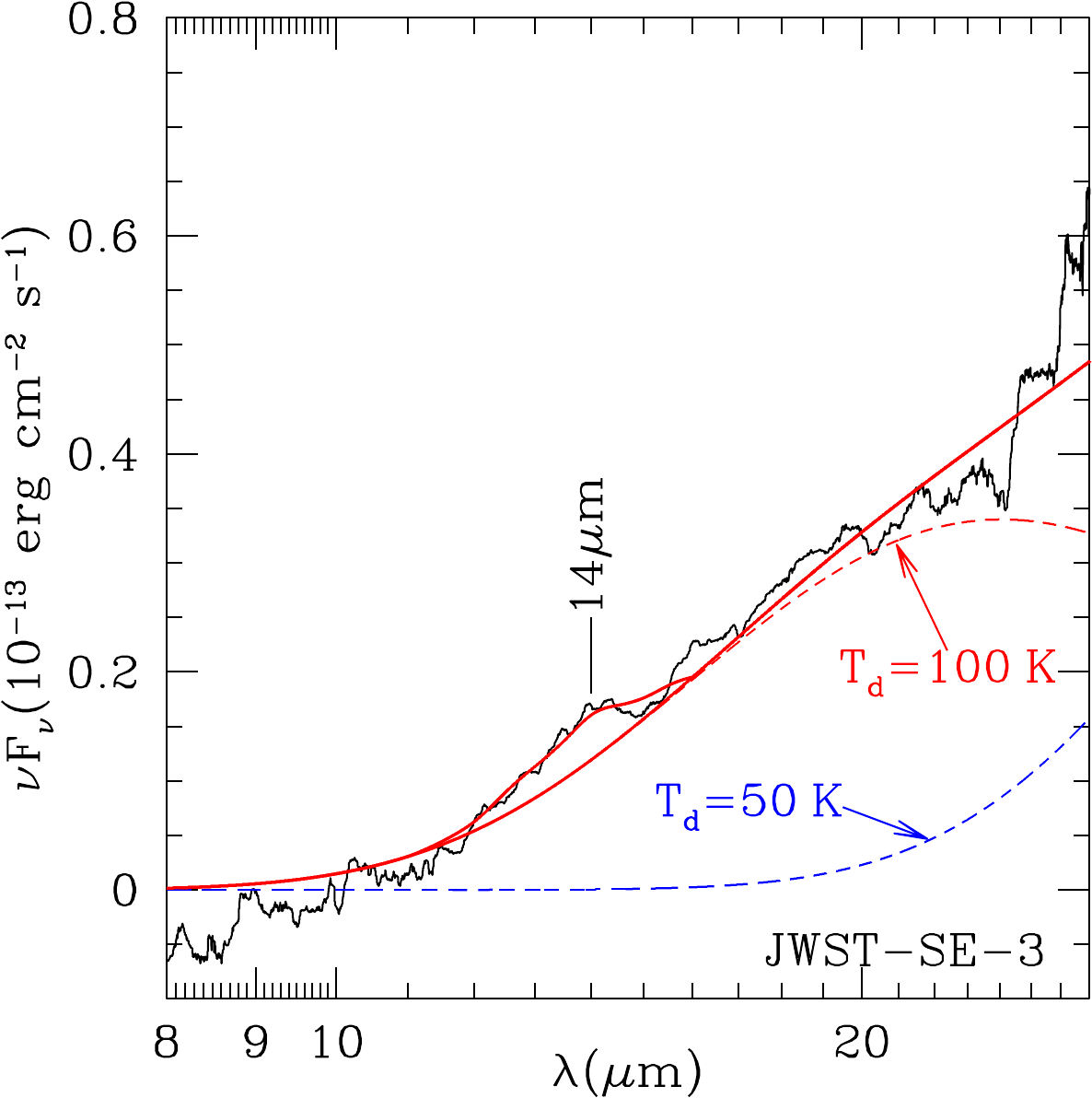}
\hspace{0.05\linewidth}
\includegraphics[angle=0,width=0.35\linewidth]{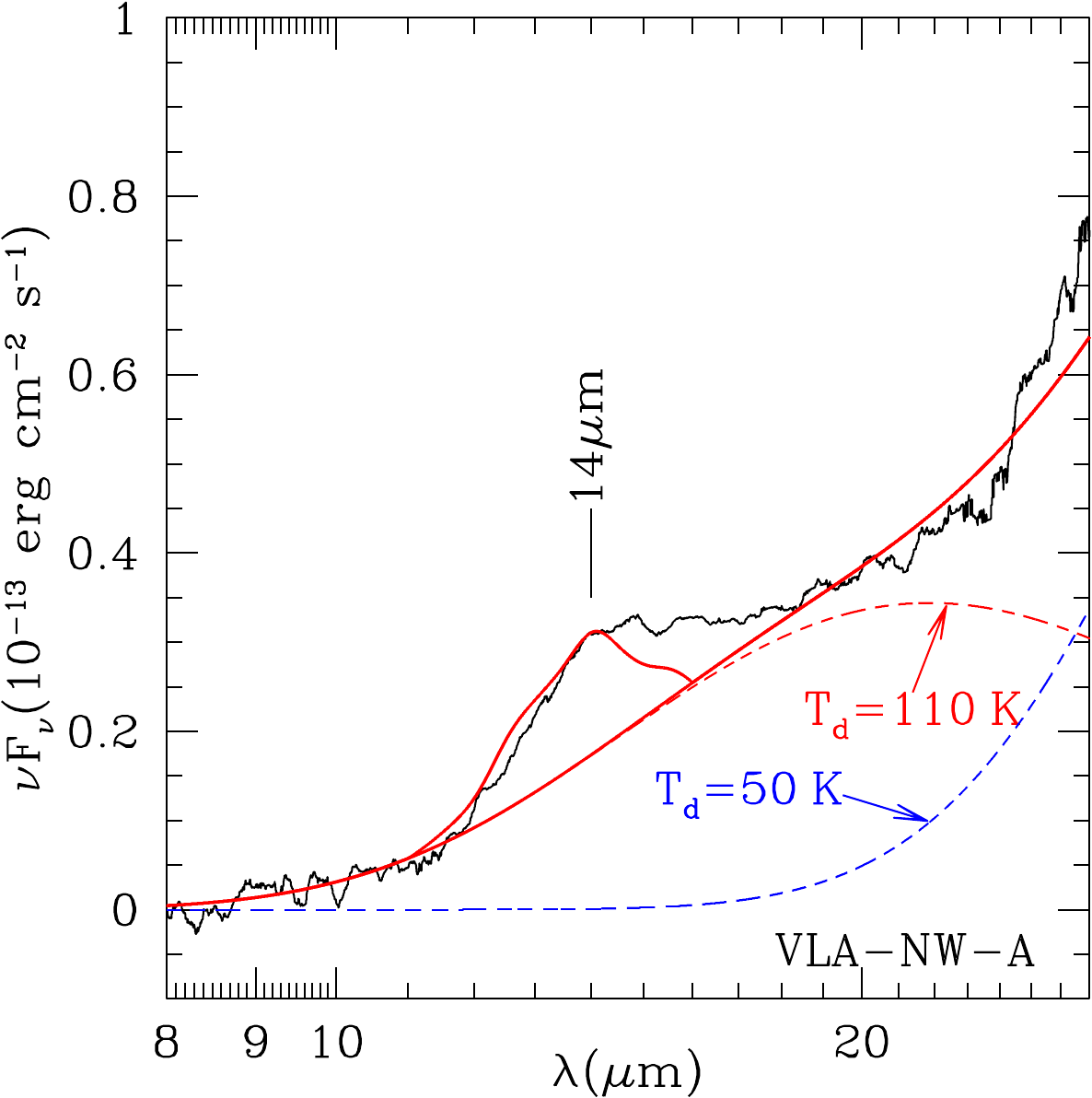}
\caption{\label{fig:4spectra}
Spectra of dust emission from $8-27$\,\micron\ 
in JWST-SE-1, JWST-SE-2, JWST-SE-3, and VLA-NW-A.  The
  three SE regions show emission in the broad feature centered near
  $14$\,\micron\ (see Fig. \ref{fig:opac}), in addition to a featureless
  continuum emission.  VLA-NW-A appears to have similar emission, but
  apparently extending to longer wavelengths.  The spectra are modeled
  with two dust temperatures (see Table \ref{tab:components}).  
Dashed
  curves show the featureless continuum for each of the temperature
  components.}
\end{center}
\end{figure*}

Figure \ref{fig:4spectra} shows the $8-27$\,\micron\ dust emission for
four selected $0.65\arcsec$ radius regions in Fig. \ref{fig:apertures}.  
These apertures are spatially independent, and correspond to the 14-\micron\
continuum sources introduced in Paper\,I.
The strongest emission is found in the three SE regions, but VLA-NW-A also shows
significant emission from the dust continuum.  
Much of the observed $8-27$\,\micron\ emission can be reproduced by a
power-law opacity $\kappa_c \propto \lambda^{-2}$, and dust
temperatures in the range $45-120{\rm K}$ as shown in Fig. \ref{fig:4spectra}.

\subsection{Emission feature near $14\mu$m \label{sec:dustfeature}}

\begin{figure}[h!]
\begin{center}
\includegraphics[angle=0,width=\linewidth]{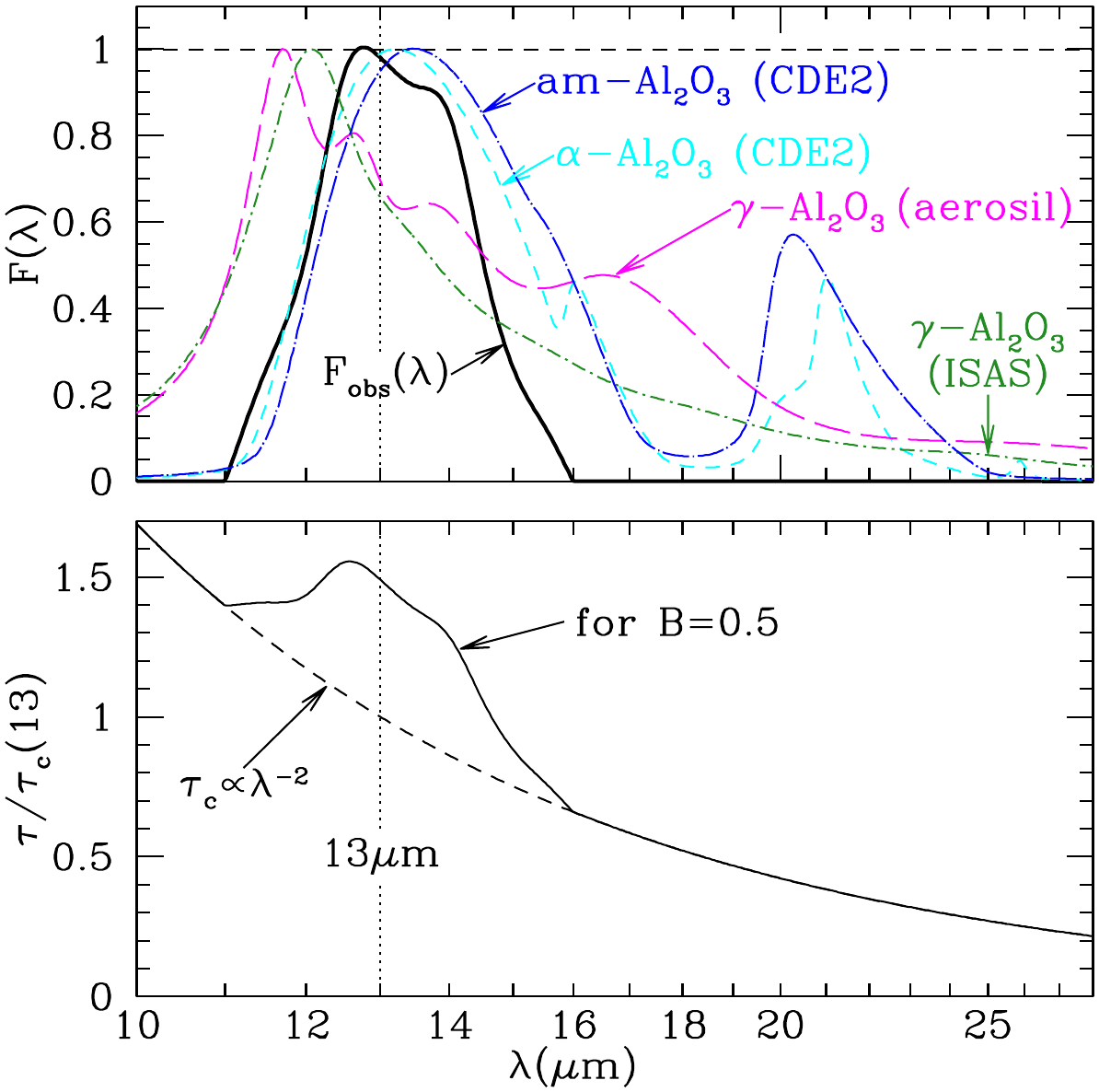}
\caption{\label{fig:opac}
Upper panel: $F_{\rm obs}(\lambda)$ (solid curve) is the normalized opacity profile required to reproduce the observed $14\micron$ emission feature. Broken curves are opacities either measured calculated for various samples of Al$_2$O$_3$ (see text). 
Lower panel: Model opacity for feature strength parameter $B=0.5$ (see text and Table \ref{tab:components}).
   }
\end{center}
\end{figure}

In addition to the $\lambda>10$\,\micron\ continuum emission, the four
spectra in Fig. \ref{fig:4spectra} exhibit an emission feature near
$\sim$$14$\,\micron.  
To model the observed feature, we use an empirical profile 
$F_{\rm obs}(\lambda)$ added
to the $\kappa_c \propto \lambda^{-2}$ continuum opacity:
\begin{equation}
\label{eq:model}
\kappa(\lambda)=
\kappa_c(13\micron)
\left[
\left(\frac{13\micron}{\lambda}\right)^2
+
B F_{\rm obs}(\lambda)\right]
\end{equation}
\noindent
The normalized opacity profile $F_{\rm obs}(\lambda)$ -- peaking near
13$\mu$m -- is shown in Fig. \ref{fig:opac}.  The overall strength
of the profile relative to the continuum is determined by parameter
$B$ in Eq.\ \eqref{eq:model}.  The four spectra in Fig.
\ref{fig:4spectra} are fitted with values of $B$ 
in the range $0.35-0.8$ (see Table \ref{tab:components}).
Because the opacity peaks near 13\,\micron, 
the 14\,\micron\ emission feature will also be referred to as the 13\,\micron\ opacity feature.

The identity of the $14$\,\micron\ feature is unknown.  
Silicates (e.g., MgSiO$_3$ or Fe$_x$Mg$_{2-x}$SiO$_4$) provide a substantial fraction of the grain mass in the Milky Way, but the
absence of an emission feature at $20$\,\micron\ in the observed spectra 
(see Fig. \ref{fig:4spectra}) rules out silicates as a major dust constituent in \izw. 
Iron oxides have infrared features at $17.1$\,\micron\ for hematite
$\alpha$-Fe$_2$O$_3$ \citep{Wang+Muramatsu+Sugimoto_1998},
$15.0$\,\micron\ for maghemite \hbox{$\gamma$-Fe$_2$O$_3$}
\citep{Draine+Hensley_2013}, and $15.7$\,\micron\ for magnetite
Fe$_3$O$_4$ \citep{Draine+Hensley_2013}, and $\sim$$20-25$\,\micron\ for
w\"{u}stite FeO \citep{Koike+Matsuno+Chihara_2017}. 
None of these iron oxides match the \izw\ spectra.

In the Milky Way, 
some dusty outflows from oxygen-rich evolved stars show a ``$13$\,\micron\ feature''
\citep{Posch+Kerschbaum+Mutschke+etal_1999,Sloan+Kraemer+Goebel+Price_2003}, 
but the feature is much narrower
\citep[FWHM\,$\approx0.55$\,\micron,][]{Posch+Kerschbaum+Mutschke+etal_1999}
than the \izw\ feature
(FWHM\,$\approx 2.5\micron$).
Spinel MgAl$_2$O$_4$ has been suggested
\citep{Posch+Kerschbaum+Mutschke+etal_1999,Zeidler+Posch+Mutschke_2013}.
Spinels have an accompanying feature at $16.8$\,\micron\ which is also present
in the spectra of oxygen-rich asymptotic giant branch (AGB) stars.
However, there is no evidence of a $16.8$\,\micron\ feature in the \izw\ spectra; 
hence spinels seem unlikely to be responsible for the emission in \izw.

Alumina (Al$_2$O$_3$) has many crystalline phases.
Presolar Al$_2$O$_3$ grains, condensed in the outflow from AGB stars, 
have been found in meteorites \citep[e.g.,][]{Takigawa+Stroud+Nittler+etal_2018}.
\citet{Sloan+Kraemer+Goebel+Price_2003} suggest that the
$13$\,\micron\ feature in oxygen-rich evolved stars arises from corundum 
($\alpha$-Al$_2$O$_3$).  
With strong infrared resonances, the opacity
of Al$_2$O$_3$ particles in the Rayleigh limit is sensitive to
particle shape.  
The opacity calculated for ellipsoids with the ``CDE2'' 
distribution of axial ratios
\citep{Ossenkopf+Henning+Mathis_1992,
       Fabian+Henning+Jager+etal_2001},\footnote{%
We assume that the $\alpha$-Al$_2$O$_3$ c-axis is equally likely to be
parallel to any of the principal axes of the ellipsoids.}
using anisotropic optical constants from \citet{Barker_1963}, is
shown in Fig. \ref{fig:opac}. 
Also shown is the opacity calculated
for ellipsoids with the CDE2 shape distribution, but using optical constants determined for amorphous Al$_2$O$_3$ 
\citep{Giacomazzi+Shcheblanov+Povarnityn+etal_2023}. 
$\alpha$-Al$_2$O$_3$ and amorphous Al$_2$O$_3$ both have a strong opacity peak near 20\,\micron\ that is inconsistent with
the \izw\ spectra in Fig. \ref{fig:4spectra}.



\begin{table*}[t!]
\begin{center}
\caption{Dust properties in selected regions \label{tab:components}}
\begin{tabular}{ccccccccc}
\hline
\hline
\multicolumn{1}{c}{Region}  & 
\multicolumn{1}{c}{B$^a$} & 
\multicolumn{1}{c}{$T_{\rm d,hot}$} & 
\multicolumn{1}{c}{$T_{\rm d,warm}$} & 
\multicolumn{1}{c}{$M_{\rm d,hot}^b$} & 
\multicolumn{1}{c}{$M_{\rm d,warm}^b$} &
\multicolumn{1}{c}{$M_{\rm HI}$} & 
\multicolumn{1}{c}{$M_d/M_{\rm HI}$} & 
\multicolumn{1}{c}{$f_{\rm dust}$} \\
& & \multicolumn{1}{c}{(K)} & 
\multicolumn{1}{c}{(K)} & 
\multicolumn{1}{c}{($M_\odot$)} & 
\multicolumn{1}{c}{($M_\odot$)} & ($M_\odot$) \\
\hline
\\
JWST-SE-1 & 0.7 & 120 & 50 & 0.07 & 59 & $7.1\times10^5$ & $8\times10^{-5}$ & 0.3 \\
JWST-SE-2 & 0.45& 120 & 55 & 0.07 & 32 & $7.6\times10^5$ & $4\times10^{-5}$ & 0.14\\
JWST-SE-3 & 0.35& 100 & 50 & 0.20 & 20 & $9.9\times10^5$ & $2\times10^{-5}$ & 0.07\\
VLA-NW-A & 0.8 & 110 & 50 & 0.12 & 44 & $2.0\times10^5$ & $2\times10^{-4}$ & 0.7 \\
\\
\hline
\hline 
\end{tabular}
\end{center}
$^a$~Parameter in Eq.\ (\ref{eq:model}) determining strength of the $13.5\micron$ feature.\\
$^b$~Assuming $\kappa_c(\lambda)=10^3{\rm \,cm}^2{\rm\, g}^{-1} (\lambda/13\micron)^{-2}$, and $D=18.2$\,Mpc.\\
\end{table*}

Figure \ref{fig:opac} also reports measured opacities
\citep{Koike+Kaito+Yamamoto+etal_1995} for commercially-available
particles of $\gamma$-Al$_2$O$_3$ (``aerosil''), as well as
for predominantly $\gamma$-Al$_2$O$_3$ particles 
produced by combustion of solid rocket fuel (``ISAS'').
The $\gamma$-Al$_2$O$_3$ particles have opacity peaking at
$12$\,\micron, inconsistent with \izw.  In addition, the ``aerosil''
particles have a broad absorption peak at $\sim 17$\,\micron,
again inconsistent with the \izw\ spectra.

Al$_2$O$_3$ is predicted to form in SN ejecta \citep{schneider24}.  
However, none of the \al\ variants considered provide a good match to the observed spectra. 
$\alpha$-Al$_2$O$_3$ and amorphous Al$_2$O$_3$ both have a strong peak near 13\,\micron, 
but are ruled out by a secondary peak near 20\,\micron\ that is absent in
the observed spectra. 
Samples of predominantly $\gamma$-Al$_2$O$_3$ (the ``aerosil'' and ``ISAS'' curves in Fig. \ref{fig:opac}) 
lack the 20\,\micron\ feature, but do not provide a good match to $F_{\rm obs}$.
Because silicates, iron oxides, and spinels appear to be inconsistent with the observed spectra, 
we seek another refractory material composed of reasonably abundant elements to provide the broad 13\,\micron\ opacity peak 
needed to explain the 14\,\micron\ emission feature. 
While none of the examples of \al\ studied provide a good match to the observed spectral shape, all
forms of \al\ exhibit strong absorption somewhere in the $12-15$\,\micron\ range, 
with dielectric functions that make the opacity profile quite sensitive to particle shape. 
Some forms of \al\ (e.g., "ISAS" $\gamma$-\al) lack strong emission features near 20\,\micron. 
We tentatively identify the 14\,\micron\ feature as being dust to some form of \al-based material.

The spectrum of VLA-NW-A also appears to show a possible emission
excess on the red shoulder of the $14$\,\micron\ emission feature (see Fig.
\ref{fig:4spectra}).  If real (i.e., not due to noise or inaccurate
background subtraction), this implies either a change in the profile of
the $14\micron$ feature, or an additional emission component at
$\sim 15$\,\micron.

\subsection{Dust Mass \label{sec:dustmass}}

A $\kappa_c\propto\lambda^{-2}$ opacity could plausibly arise from
metallic grains or amorphous carbon grains.  
Iron grains were proposed to explain the
featureless mid-infrared emission from the winds from low-metallicity
AGB stars in globular clusters
\citep{McDonald+Sloan+Zijlstra+etal_2010,McDonald+Boyer+vanLoon+Zijlstra_2011}.
For purposes of estimating the mass of the continuum-emitting dust, we
adopt a provisional opacity $\kappa_c(13\micron)=1\times10^3 {\rm
  \,cm}^2 \,{\rm g}^{-1}$.  This is much larger than estimates for the
opacity of metallic Fe 
grains,\footnote{$a=0.01\micron$ spheres composed of bcc Fe have $\kappa(\lambda)
\approx 12 (\lambda/10\micron)^{-2} {\rm \,cm}^2 \,{\rm g}^{-1}$ at
$T\,=\,300{\rm K}$ \citep[][Appendix B]{Draine+Hensley_2013}, with
$\kappa$ decreasing as $T$ is decreased.  However, if the electron
scattering time is greatly decreased by impurities (e.g., Ni, Mg, Al,
C, O) or defects, the opacity could be substantially increased.  Our
provisional opacity $\kappa_c = 10^3(\lambda/13\micron)^{-2}$ should be
regarded as very uncertain.}
but is comparable to estimates for, say, amorphous
carbon 
\citep{Rouleau+Martin_1991,
       Zubko+Mennella+Colangeli+Bussoletti_1996,
       Jaeger+Mutschke+Henning_1998a}.

The $10-27$\,\micron\ spectra in Fig. \ref{fig:4spectra} require at
least two dust temperatures -- ``hot'' dust with $T\approx 100$\,K, and
``warm'' dust with $T\approx 50$\,K.  Masses of the hot and warm dust
components are given in Table \ref{tab:components}.  The bulk of the
dust mass is in the lower temperature component.  
Dust masses are for the provisional opacity $\kappa_c=1\times10^3(13\mu{\rm m}/\lambda)^{-2}\,{\rm cm}^2\,{\rm g}^{-1}$; 
if the actual opacity is higher or lower, the dust masses
will be lower or higher.


We have estimated the \hi\ mass \mhi\ in the JWST-SE-1, JWST-SE-2, JWST-SE-3, and VLA-NW-A apertures using \Nhi\ from the
2\arcsec\ resolution \hi\ map of \citet{lelli12}; 
the values are given in Table \ref{tab:components}. 
The corresponding \Nhi\ column densities range from $2\times10^{21}$\,cm$^{-2}$ in VLA-NW-A to
$\sim 10^{22}$\,cm$^{-2}$ in the SE apertures.
The ratio of refractory grain mass to H mass 
$M_{\rm d}/M_{\rm H} = f_{\rm dust}\times 0.01(Z/Z_\odot)=3\times10^{-4}f_{\rm dust}$, 
where $f_{\rm dust}<1$ is the fraction of the refractory elements that are contained in the dust,
and $0.01$ corresponds to the assumed value of the dust-to-H nucleon mass ratio at Solar metallicity
\citep{aniano20}.
For the dust mass estimates in
Table \ref{tab:components}, $f_{\rm dust}$ ranges from $0.07$ in JWST-SE-3 to $0.7$ in VLA-NW-A.
While the Milky Way has $f_{\rm dust}\approx 1$, low-metallicity galaxies are usually found to 
have low global values of $f_{\rm dust}$,
with $f_{\rm dust} \la 0.01$ for $Z/Z_\odot \la 0.05$ \citep[e.g.,][]{remy14}. 
However, in galaxies with metallicities as low as \zzsun\,$\sim 0.1$,
regions of high \hi\ column density $\sim 10^{22}$\,cm$^{-2}$ can have $f_\mathrm{dust}$ $\la 1$
\citep{haman24}.
It is therefore perhaps not surprising to find
$f_{\rm dust}$ as large as 0.3 and 0.7 in JWST-SE-1 and VLA-NW-A. 

Because our spectroscopy stops at $\sim 27$\,\micron, the JWST
observations are insensitive to cooler dust.  Using photometry from
Spitzer and Herschel, \citet{fisher14}
estimated a total dust mass $M_{\rm dust}\approx 900_{-500}^{+900}$\,\msun,
while \citet{hunt14} estimated $M_{\rm dust}\approx
340\,\pm\,10$\,\msun.  Our similar estimate here from only a limited area
arises from use of a lower (but uncertain) opacity for metallic grains at $\sim 25$\,\micron.  
While VLA-NW-A is much less bright at $11.3$\,\micron\ than JWST-SE-1 and JWST-SE-2 (see Fig. \ref{fig:apertures}), 
the dust mass seems to be comparable in the three sources; 
the 11.3\,\micron\ emission is weak only because the 
``hot" dust is slightly cooler than in JWST-SE-1 and JWST-SE-2.

\begin{figure*}[t!]
\includegraphics[width=0.33\linewidth]{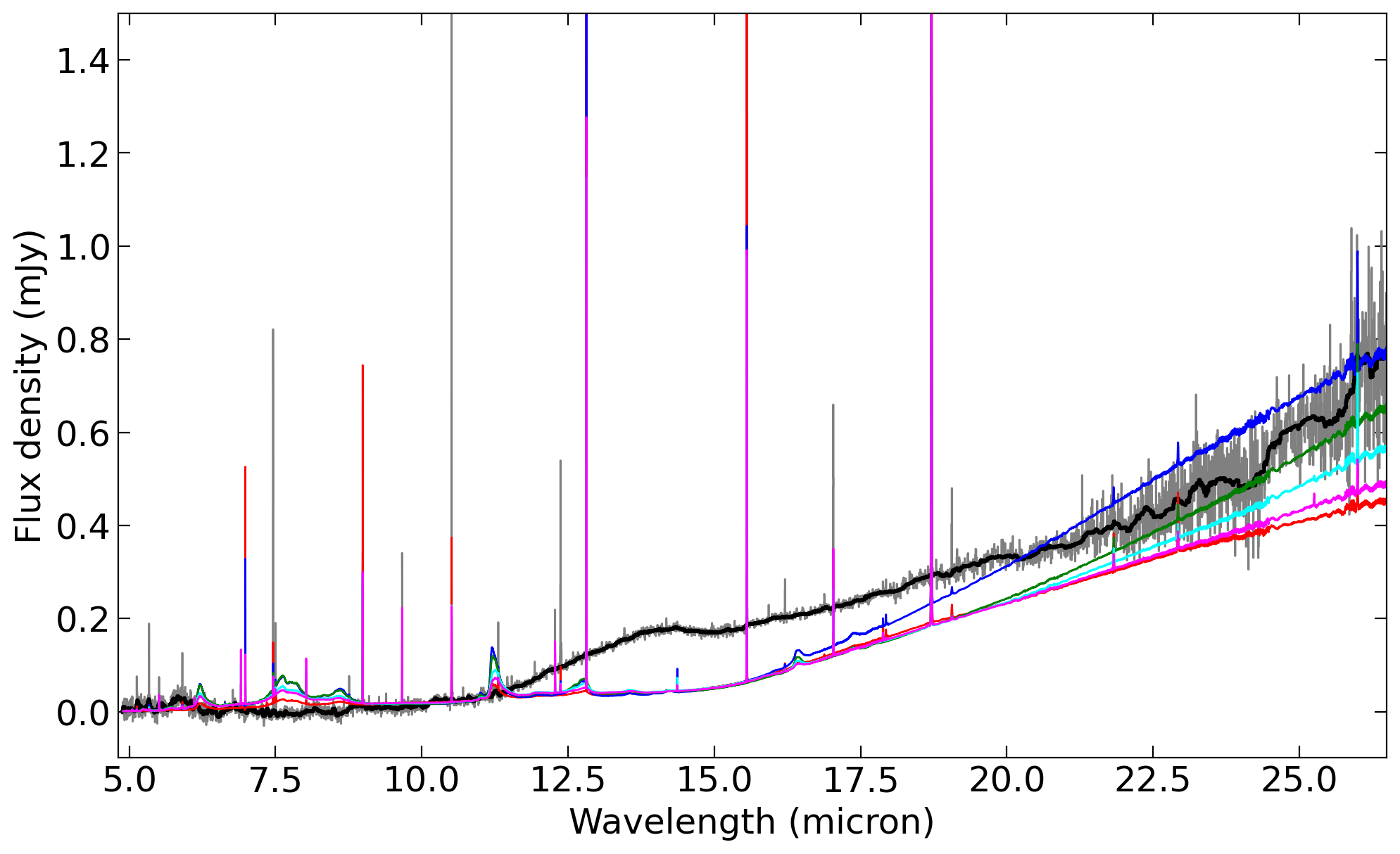}
\includegraphics[width=0.33\linewidth]{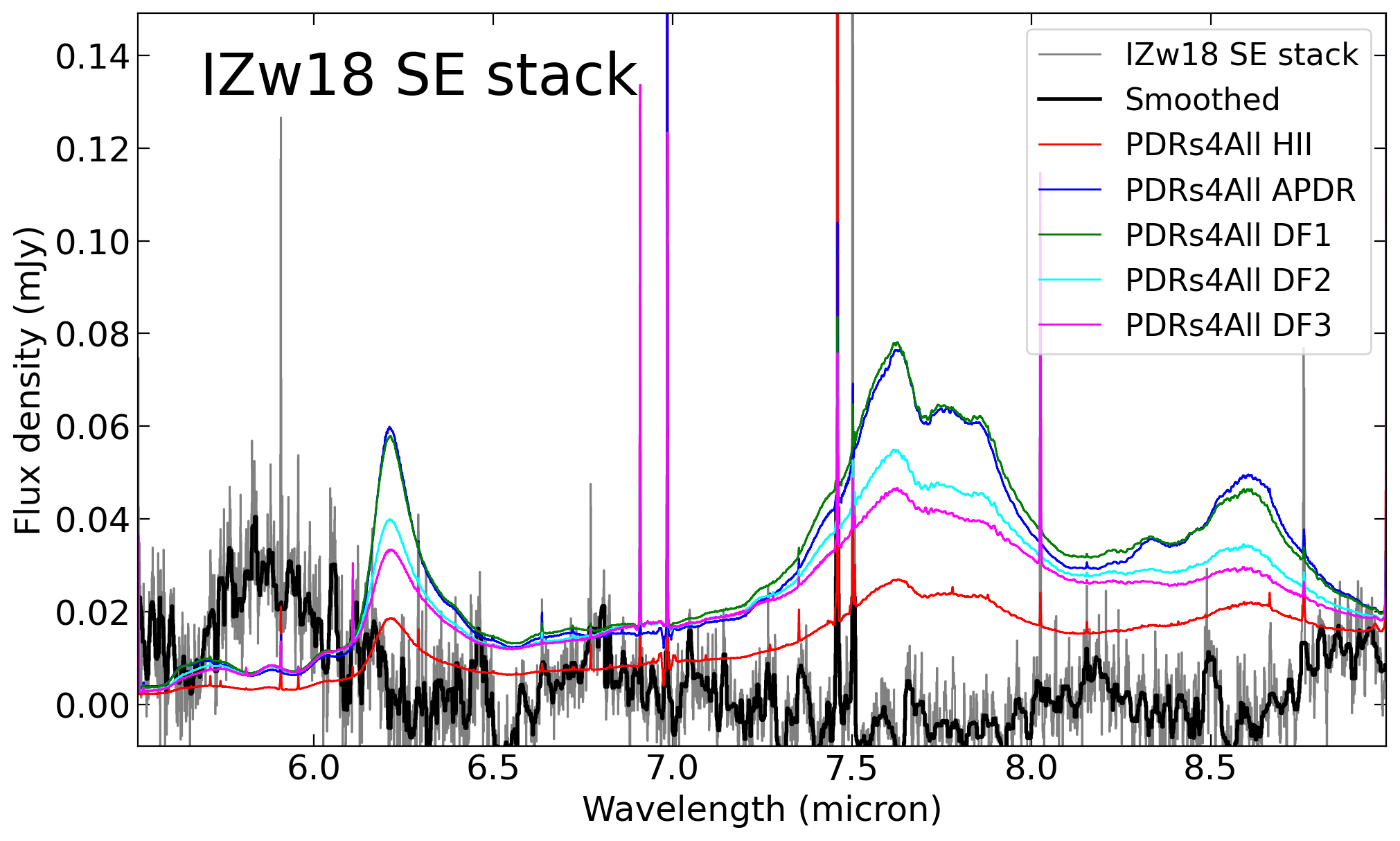}
\includegraphics[width=0.33\linewidth]{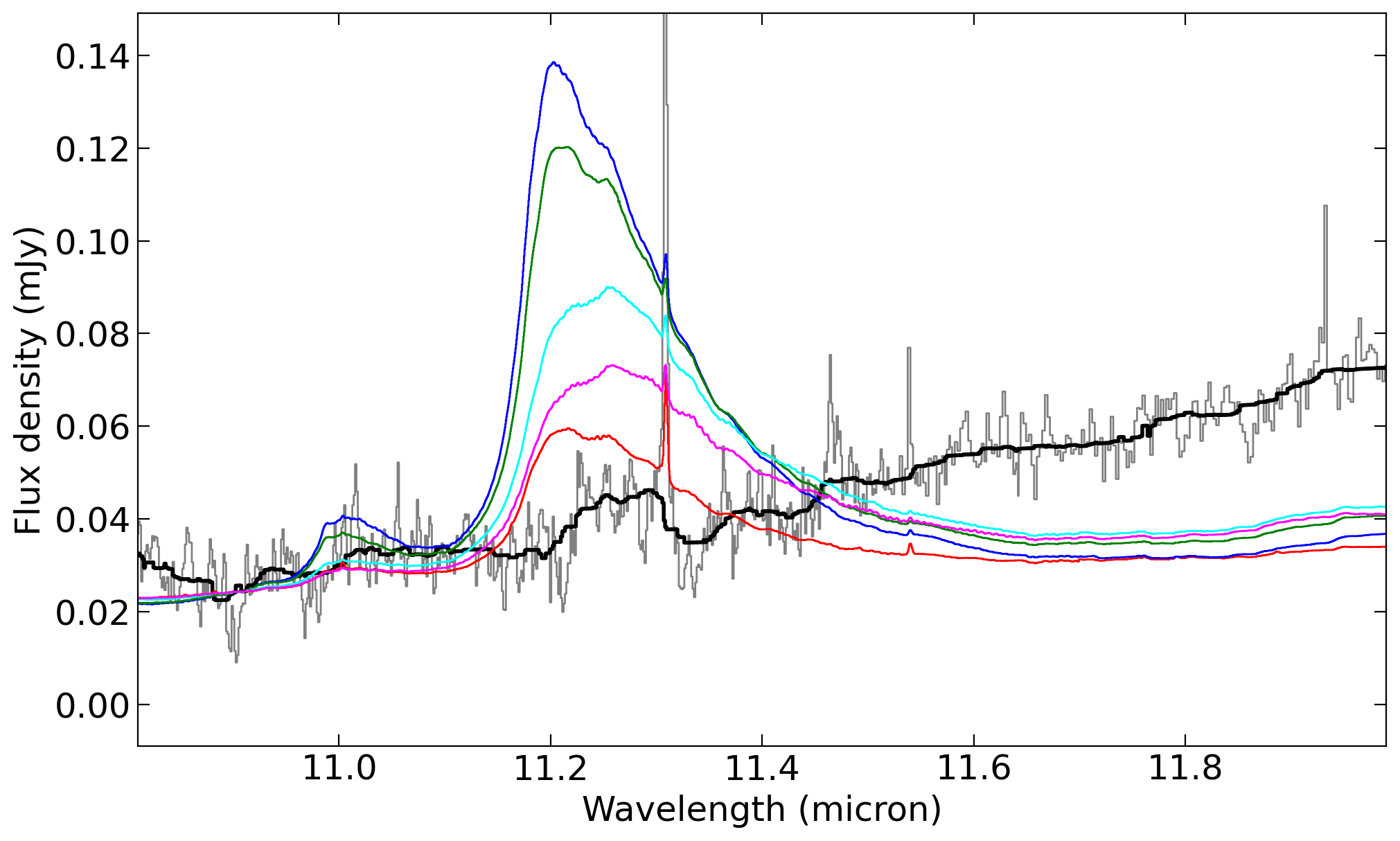} \\
\caption{\izw\ stacked spectra from the SE compared with templates from PDRs4All \citep{chown24}.
The left panel shows the full wavelength coverage; 
the middle panels show a zoom-in around $\sim 7$\,\micron, and  
the right around 11\,\micron.
PDR templates are normalized in all panels to the mean at 10.9\,\micron;
all flux density units are mJy for \izw, and for the PDRs arbitrary, rescaled to the normalization.
The PDRs4All templates are given by colored curves as in the legend, and the \izw\
stacked spectrum is in gray, with the smoothed data as heavy black curves.
It can be seen that the \izw\ spectra have a ``bump'' at $\sim 14$\,\micron,
implying warmer dust than the PDRs4All Orion Bar templates with possibly a different dust composition.
Moreover, there is no discernible PAH emission between 5 and 8\,\micron, nor clearly at $\sim\,11.2$\,\micron.
}
\label{fig:pah}
\end{figure*}

The opacity of the material responsible for the 14$\micron$ emission feature
is unknown.  
Crystalline $\alpha-$\al\ and amorphous \al\ particles with the CDE2 shape distribution 
have $\kappa(13$\,\micron)$\approx 7300$ and $8200$\,cm$^2$\,g$^{-1}$, respectively. 
For $\kappa(13$\,\micron)$\approx 8000$\,cm$^2$\,g$^{-1}$, the mass of \al\ material is
\begin{equation}
M_{{\rm Al}_2{\rm O}_3} \approx
B \frac{\kappa_c(13\micron)}{\kappa_{{\rm Al}_2{\rm O}_3}(13\micron)} M_c
\approx \frac{B}{8} M_c
~~.
\end{equation}
where $M_c$ is the mass of the material responsible for the continuum
emission (assuming $\kappa_c(13$\,\micron$)\approx 10^3{\rm\,cm}^2\,{\rm g}^{-1}$).  
With $B\approx 0.5$, the Al$_2$O$_3$ would account for only
$\sim 7\%$ of the total dust mass.

%

It seems likely that dust chemistry in low-metallicity galaxies at
high redshifts will resemble what is happening in \izw: injection
of elements from evolved stars and supernovae into a nearly dustless
ISM permeated with harsh ultraviolet radiation,
heating the interstellar dust to $T\gtrsim50$\,K.  Dust formed in the
stellar outflows will undergo both growth and destruction in the
interstellar medium.  The end-product present in \izw\ -- 
metallic or carbonaceous material plus an unidentified compound (possibly \al) producing the broad emission feature 
near $14$\,\micron\ -- may be a guide to dust physics
in young galaxies at $z > 10$.

\subsection{Searching for PAH emission in \izw\ \label{sec:pahs}}

It is well known that in metal-poor environments, PAH emission tends to be suppressed 
\citep[e.g.,][]{engelbracht05,madden06,hunt10}.
Nevertheless, given the high S/N of our MIRI spectra of \izw, we have looked for PAHs
by comparing \izw\ to the templates of the Orion Bar from PDRs4All \citep[][]{chown24}.
During the line search described in Sect. \ref{sec:spectra}, we used the wavelengths
given in \citet{chown24} as initial guesses for PAH features;
only one, in JWST-SE-1 at $\sim 11\,$\,\micron, has S/N$\,\ga 5$.

Because we would expect PAH emission together with \htwo\ in PDRs, to increase S/N,
we have averaged the spectra in the five regions to the SE in \izw, where we find most of the \htwo.
Figure \ref{fig:pah} shows this mean SE spectrum compared to the
PDRs4All Orion Bar templates by \citet{chown24}. 
The latter have been normalized to the \izw\ spectra at $\sim 10.9$\,\micron.
The left panel illustrates the entire MIRI spectral range, while the middle panel
zooms in around $\sim 7$\,\micron; the right panel shows a close-up around the 11.2\,\micron\
PAH feature, thought to be a carrier of large, neutral PAHs
\citep[e.g.,][]{draine21,rigopoulou21}.
Because the apertures are not independent (see Fig. \ref{fig:apertures}),
this is not a rigorous exercise; the aim is a qualitative enhancement of any PAH emission features.

It can be seen from Fig. \ref{fig:pah} that \izw\ does not resemble a typical 
metal-rich \hii\ region or PDR.
Although the continuum of the stacked SE spectrum in the left panel rises toward the red, 
most similar to the Orion Bar \hii\ region, it also has a significant bump at $\sim 14$\,\micron\ 
(see Sec. \ref{sec:dustfeature}).
Even with the stacks, the S/N between 6 and 7\,\micron\ in the middle panel
is insufficient to characterize PAH features with any certainty,
although there is an enhancement in \izw\ around $\sim 5.8-6.0$\,\micron, 
blueward of where a PAH feature is seen in the Orion spectra.
The right panel of Fig. \ref{fig:pah} 
shows an enhancement around 11.25\,\micron,
but we hesitate to consider this a true detection of the nanoparticles associated
with PAHs or other aromatic features.
If these are PAHs, they are radically different from any such emission previously known.

There is no measurable PAH emission at 7.7\,\micron\ in \izw.
Previous work has shown that in low-metallicity dwarfs, like in more metal-enriched galaxies,
the 7.7\,\micron\ feature tends to dominate the total PAH emission carried by each band 
\citep[e.g.,][]{hunt10,sandstrom12,chastenet19}.  
However, the ratio of 7.7\,\micron\ intensity, relative to warm dust grain emission,
is suppressed at low metallicity \citep{chown25}.
PAH emission in metal-poor environments also tends to be preferentially lost in ionized gas 
relative to the diffuse neutral medium, 
even at comparable radiation field intensities \citep[e.g.,][]{chastenet19}.  
Since the apertures we have used for this analysis of \izw\ are 
either directly associated with \hii\ regions or contain copious quantities of
ionized gas (see Paper\,I), this
could be part of the reason for the lack of discernible PAH emission in the MIRI spectra.

Indeed, the lack of PAH emission could be because the PAH particles are 
destroyed by the strong RF, capable of exciting \oiv\ and \nev,
or it could be because they are unable to form in the first place given the paucity of dust.
If only very small PAHs are present, the strongest PAH emission feature might be the 3.3\,\micron\
C-H stretch \citep[e.g.,][]{rigopoulou21,draine21}. 
This feature is associated with neutral PAHs, which may be more robust than ionized
PAHs under extreme conditions \citep[e.g.,][]{garciabernete24,rigopoulou24}.
NIRSpec data would be needed to search for this feature.
While the lack of PAH emission is not particularly surprising, the MIRI spectra
of \izw\ supply strong constraints for dust models at these low metallicities.

\section{Discussion \label{sec:discussion}}


Despite the low dust content in \izw\ \citep[][]{fisher14,hunt14},
there is measurable warm \htwo, in particular in the SE, where all five of
the aperture spectra have significant \htwo\ detections in at least five transitions
(see Table \ref{tab:popfits}).
The warm \htwo\ column densities are more than one order of magnitude below the typical
values in more massive spirals.
The warm \htwo\ masses we measure within 120\,pc diameter regions 
range from $\sim\,300$\,\msun\ to $\sim\,2000$\,\msun,
again roughly an order of magnitude lower than the virial \htwo\ cloud masses
found by \citet{rubio15} and \citet{shi20} from CO measured in galaxies with metallicities
around 7-13\%\,\zsun\ (WLM, Sextans\,B).
As mentioned above, \tl\ is well constrained by our fits, so that we may be
detecting most of the total \htwo\ content with the rotational lines.
The heating (from UV pumping of \htwo, plus photoelectrons from the grains that are present,
some heating by X-rays, and presumably cosmic rays) 
may be able to maintain the warm \htwo\ (i.e., \tl$\,\ga\,$150\,K), 
even in gas that is self-shielded by \Ntot$\approx 10^{18}$\,\cmtwo.  
It is likely that the \htwo\ was cooler before the starburst, 
but once OB stars turned on the FUV may be sufficient to keep the \htwo\ warm. 

An OPR in the $v=0$ \htwo\ levels of $> 3$ is measured in three independent regions.
We  attribute this to more rapid photodissociation of \htwo, relative to its formation rate,
which results in more effective self-shielding in ortho-\htwo\ with respect to its para counterpart.
This raises the OPR above its LTE value;
if photodissociation is sufficiently rapid compared to the rates for ortho-para conversion
in the gas (or on grain surfaces),
selective photodissociation will result in OPR $> 3$.
Because the photodissociation rate rises as \htwo\ column density \Ntot\ decreases,
this is consistent with the observed trend in Fig. \ref{fig:opr_h2col} for OPR to be higher
in regions of lower \Ntot.

If we are truly detecting most of the \htwo\ in \izw\ with MIRI,
then we are faced with the question of whether or not stars can form from this warm gas.
The answer to this is not clear.
Given the lack of cool \htwo, it could be that star formation is occurring in
reservoirs of atomic gas as proposed by \citet{glover12} and \citet{krumholz12}.
However, because the lowest-$J$ \htwo\ transition is not covered by our observations,
they are insensitive to the coolest component of \htwo\ with temperatures $\la 100$\,K;
possibly we are detecting just the
tip of the iceberg of the total \htwo\ mass as the potential
fuel for future star formation in \izw.

The detection of \htwo\ up to 
$J_\mathrm{upper}\,=\,10$ (Table \ref{tab:popfits})
raises the question of gas density.
The comparison of the \htwo\ population diagrams in \izw\ with the isochoric Meudon PDR 
models indicates molecular-gas densities of $\sim 10^5$\,\cmthree,
at least for the higher-$J$ transitions.
Optical and radio observations of the ionized gas in \izw\ \citep{izotov99,hunt05} suggest
that the electron density is fairly low, $\la 100$\,\cmthree\ (see also Paper\,III).
The implication is that within 120\,pc regions in \izw\ there is neutral warm, 
dense gas at $\sim 200$\,K co-existing 
with highly ionized, more tenuous, gas at $\sim$20\,000\,K. 
The highest ionized gas densities will be near ionization fronts. 
If some of these have $n_e\,\ga\,2500\,\mathrm{cm}^{-3}$ and $T\,\approx\,20\,000$\,K
(e.g., Paper\,III), 
and the ionization front has evolved to become D-type, the PDR will have a similar pressure 
$p/k_\mathrm{B}\,\approx\,10^{8}\,\mathrm{K\,cm}^{-3}$. 
The zone in the PDR where $T\,\approx\,10^3$\,K would then have $n\,\approx\,10^5$\,\cmthree, 
sufficient to approximately thermalize the high-$J$ levels.

Possibly the most complex aspect of our results is that highly ionized gas such as \oiv\
and \nev\ found in several of the apertures considered here
(Paper\,I) can coexist with warm \htwo\ within the same 120\,pc region.
We would naively expect that the strong radiation field that excites the gas 
would photodissociate \htwo.
However, there could be a ``sweet spot'' such that the dense, neutral gas emitting \htwo\
lives in an environment that is warm, 
self-shielded from FUV, but not sufficiently hot to collisionally dissociate the molecule
\citep[a similar argument for vibrationally excited \htwo\ can be found in][]{draine90}.
Some of the radiation from the sources responsible
for the observed \nev\ (which requires $h\nu>$\,97\,eV) may penetrate 
into and provide heating of the partially ionized molecular gas.

That molecular clouds can survive in extreme conditions is shown
by the observations of \citet{valdivia25} who find sub-parsec and parsec-sized clumps
of dense molecular gas around the young massive cluster R136, in the Large Magellanic Cloud.
R136 is a 1$-$2\,Myr old compact
star cluster hosting several stars more massive than 150\,\msun\ \citep{crowther10}.
The molecular clumps lie within 2 and 10\,pc in projection from R136,
appearing to be the densest remnants of the molecular reservoir that has been carved out through 
intense radiation.


\section{Summary \label{sec:summary}}

In order to characterize the ISM in an extreme, metal-poor dwarf galaxy,
we obtained \jwst\ MIRI/MRS observations of the main body of \izw.
Spectra have been extracted from the convolved cubes within 11 apertures (Fig. \ref{fig:apertures}), 
each $\sim 120$\,pc in diameter,
placed on regions of interest across the galaxy (Fig. \ref{fig:spectra}).
Here we have focused on the analysis of \htwo\ and the dust continuum, and 
our findings can be summarized as follows:
\begin{itemize}
\renewcommand\labelitemi{--}
\item 
MIRI has detected a series of rotational \htwo\ lines from S(1) to S(8) (S(0) is beyond the sensitivity range of MIRI). 
We fit population diagrams in six of the eleven apertures with five or more significant \htwo\
detections (Figs. \ref{fig:pop}, \ref{fig:corner}),
assuming a continuous temperature distribution.
The best-fit power-law index for the temperature distribution tends to be flatter than
for more massive spirals, implying an excess of warm \htwo\ gas.
Interestingly, while \htwo\ emission is seen, 
our criterion of 5 or more significant \htwo\ detections was not met in the aperture centered
on the CO(2--1) detection.
\item 
In a second series of fits, we left the \htwo\ OPR as a free parameter and found
OPR significantly $\,>\,3$ in three of the six apertures for which we were able to construct population
diagrams (Figs. \ref{fig:pop}, \ref{fig:corner_opr}).
Although OPR$\,>\,3$ in PDRs has been considered a theoretical possibility
\citep[e.g.,][]{draine96,sternberg99}, it has never before been measured in interstellar gas.
\item
Comparing the emission measured in different (sometimes overlapping) apertures, we find
(Fig. \ref{fig:opr_h2col}) that OPR tends to increase with decreasing \htwo\ column density, 
\Ntot, consistent with OPR $> 3$ resulting from selective photodissociation.
\item
Comparison of the \htwo\ population diagrams with Meudon PDR isochoric and isobaric models
suggests a \gnot/\nh\ ratio $\la 0.05$\,cm$^3$ and a H nucleon density of $\sim 10^5$\,\cmthree\
(Figs. \ref{fig:meudon}, \ref{fig:isobaric}).
This dense gas coexists within the same $\sim 120$\,pc region as highly ionized,
more tenuous, gas emitting \oiv\ and \nev. 
These different gas phases appear to be in approximate pressure equilibrium.
Nevertheless, the comparison with the Meudon PDR models must be interpreted
with caution because of the extreme physical conditions in \izw\ not accounted
for in the models.
\item
An unidentified dust emission feature at $\sim 14$\,\micron\ was found in \izw's continuum
spectra in three regions in the SE, and in VLA-NW-A (Fig. \ref{fig:4spectra}).
Silicates, iron oxides, and spinels appear unable to account for the observed feature. The feature may be due to alumina (Al$_2$O$_3$), predicted to be found in dust produced by supernovae 
\citep[e.g.,][]{marassi19,schneider24},
and in outflows from O-rich AGB stars 
\citep{dellagli14,Takigawa+Stroud+Nittler+etal_2018}.
All forms of Al$_2$O$_3$ have strong absorption near 13\,\micron, but the the $10-20$\,\micron\ spectrum
varies considerably from one sample to another. While none of the 
\al\ compositions studied here provides a good match to the observed spectra, 
we tentatively identify the 14\,\micron\ emission in \izw\ as being due to some form of Al$_2$O$_3$.
\item
The dust-to-gas mass ratio in these four regions indicates that a substantial
fraction, from $\sim 0.1 - 0.7$, of the refractory elements is incorporated in dust grains,
although this result is sensitive to the adopted grain opacity.
\item 
We searched for PAH emission in \izw\ by comparing templates
from PDRs4ALL (Fig. \ref{fig:pah}).
There is little evidence for PAH emission in \izw, at least at MIRI wavelengths,
although stacking the SE spectra hints at a very weak, tentative 11.2\,\micron\ feature.
\end{itemize}

Warm \htwo\ emission in one transition, S(1), was also recently found in another dwarf
galaxy at a similar low metallicity, Leo\,P \citep{telford24}, powered
by a single O star \citep{evans19}.
Now in \izw, a dwarf starburst also at 3\% Solar metal abundance, 
we have been able to fit \htwo\ level population diagrams
and derive \htwo\ column densities, parameters of continuous temperature distributions,
and even infer the OPR.
The \htwo\ gas in \izw\ is warm, so it is not clear whether stars can form 
from the \htwo\ we measure here.
Nevertheless,
the presence of \htwo\ even at these low levels of chemical enrichment 
casts doubt on the need to form stars in pure atomic gas in these extreme environments
\citep[e.g.,][]{glover12,krumholz12}.
The constraints provided by our observations on \htwo\ formation and destruction
in a chemically unevolved ISM will help to better model the metal-poor 
dwarf galaxy population being identified by \jwst\ at high redshift.

\begin{acknowledgments}
We are grateful to the anonymous referee for comments that helped improve the paper.
We warmly thank 
Federico Lelli for the \hi\ zero-moment map taken from \citet{lelli12}. 
This work is based on observations made with the 
NASA/ESA/CSA James Webb Space Telescope. The
data were obtained from the Mikulski Archive for Space Telescopes at the Space Telescope Science Institute, 
which is operated by the Association of Universities for Research in Astronomy, Inc., under NASA contract NAS 5-03127 for \jwst. 
These observations are associated with program JWST-3533, under \jwst\ grant number B0227.
RRV is grateful for the support of this program, provided by NASA through a grant from the Space Telescope Science Institute.
SH, BLJ, and M. Mingozzi are thankful for support from the European Space Agency (ESA).
M. Meixner acknowledges that a portion of her research
was carried out at the Jet Propulsion Laboratory, California
Institute of Technology, under a contract with the National Aeronautics and Space Administration 
(80NM0018D0004).
BTD thanks Greg Sloan for very helpful discussions.
LKH gratefully acknowledges Martha Haynes for interesting exchanges,
her patience, and for sharing her understanding of low-metallicity galaxies.
\end{acknowledgments}

%

\vspace{5mm}
\facilities{JWST(MIRI), HST(ACS), Keck(KCWI)}


\software{Astropy \citep{2013A&A...558A..33A,2018AJ....156..123A}, Numpy \citep{harris2020array}, SciPy \citep{2020SciPy-NMeth}, dustmaps \citep{green18},
emcee \citep{foreman13} }

The JWST data presented in this article were obtained from the Mikulski Archive for Space Telescopes (MAST) 
at the Space Telescope Science Institute. 
The specific observations analyzed can be accessed via \dataset[doi: 10.17909/n80x-b534 ]{https://doi.org/10.17909/n80x-b534}.




\appendix


\section{Flux measurements of detected \htwo\ lines in the 1D spectra from the convolved cubes\label{sec:flux}}

\begin{figure}[h!]
\centering
\includegraphics[width=0.22\linewidth]{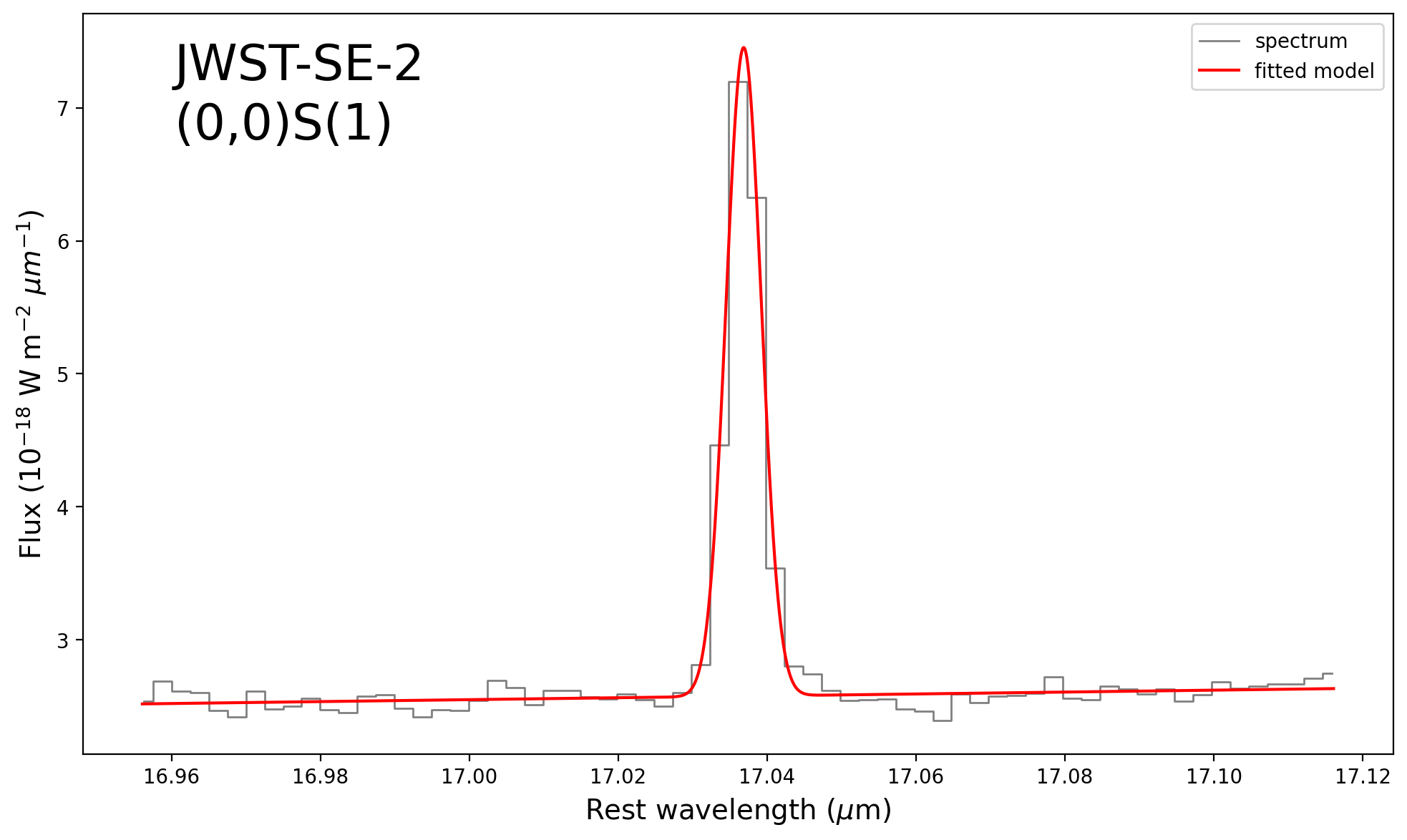} 
\includegraphics[width=0.22\linewidth]{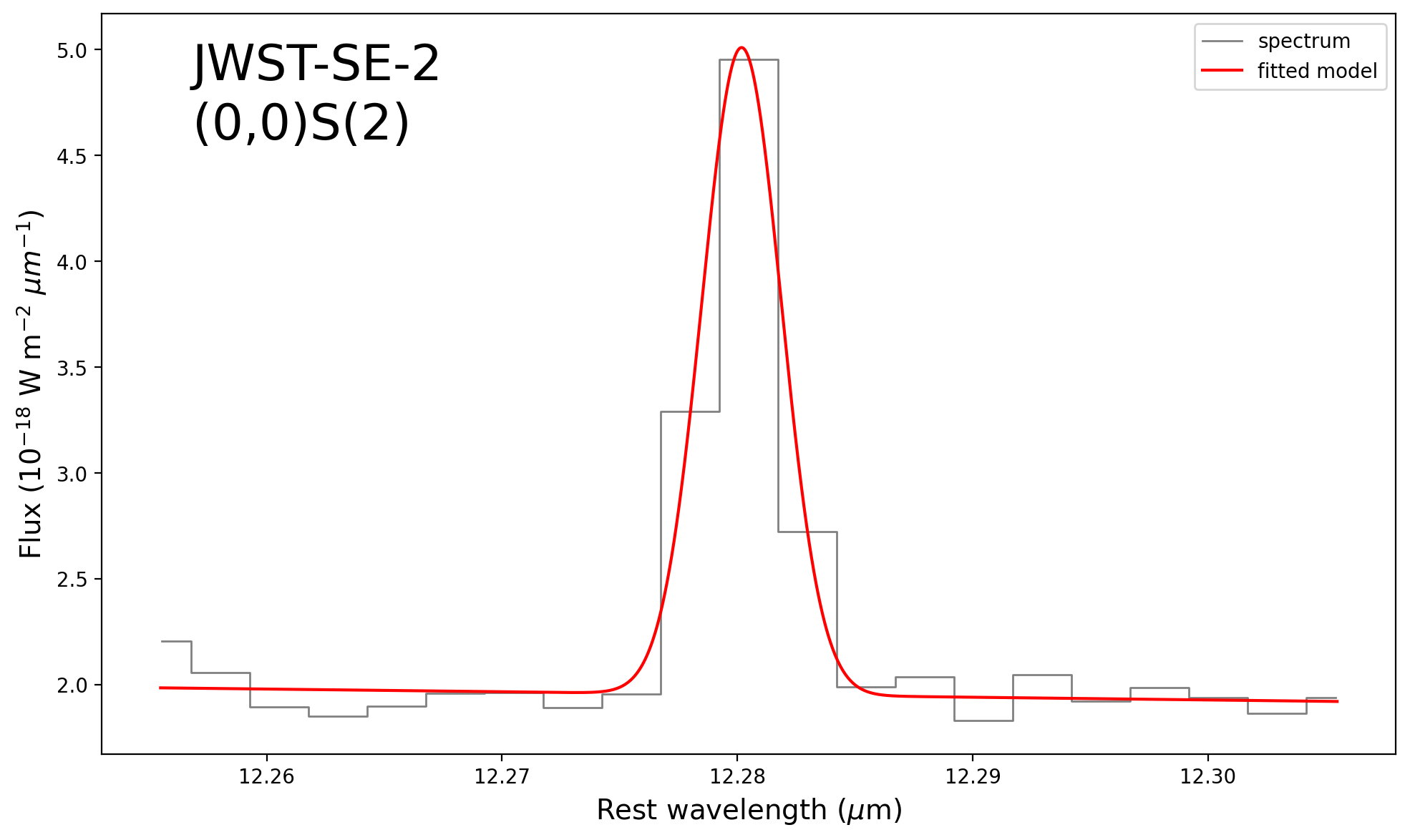}
\includegraphics[width=0.22\linewidth]{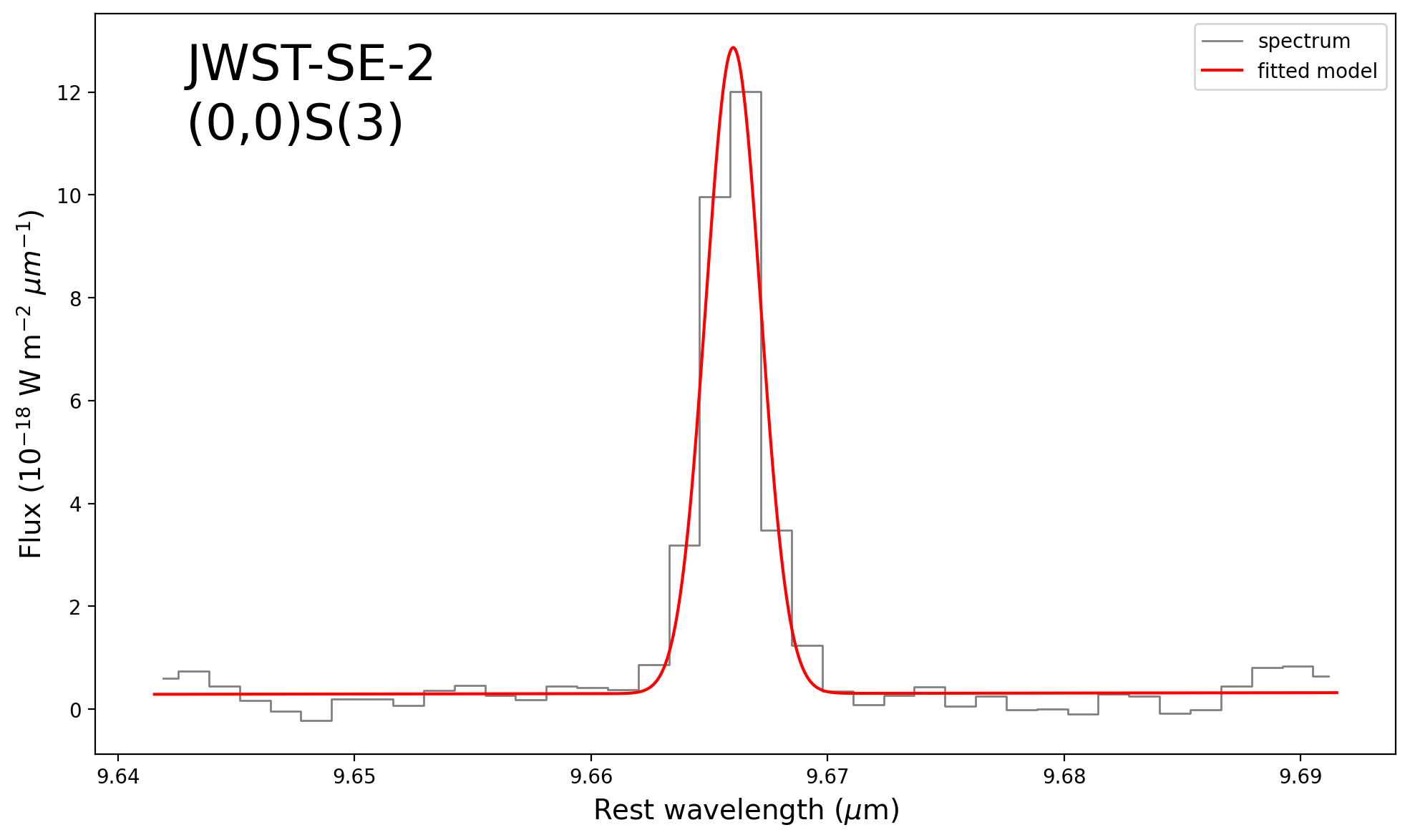}
\includegraphics[width=0.22\linewidth]{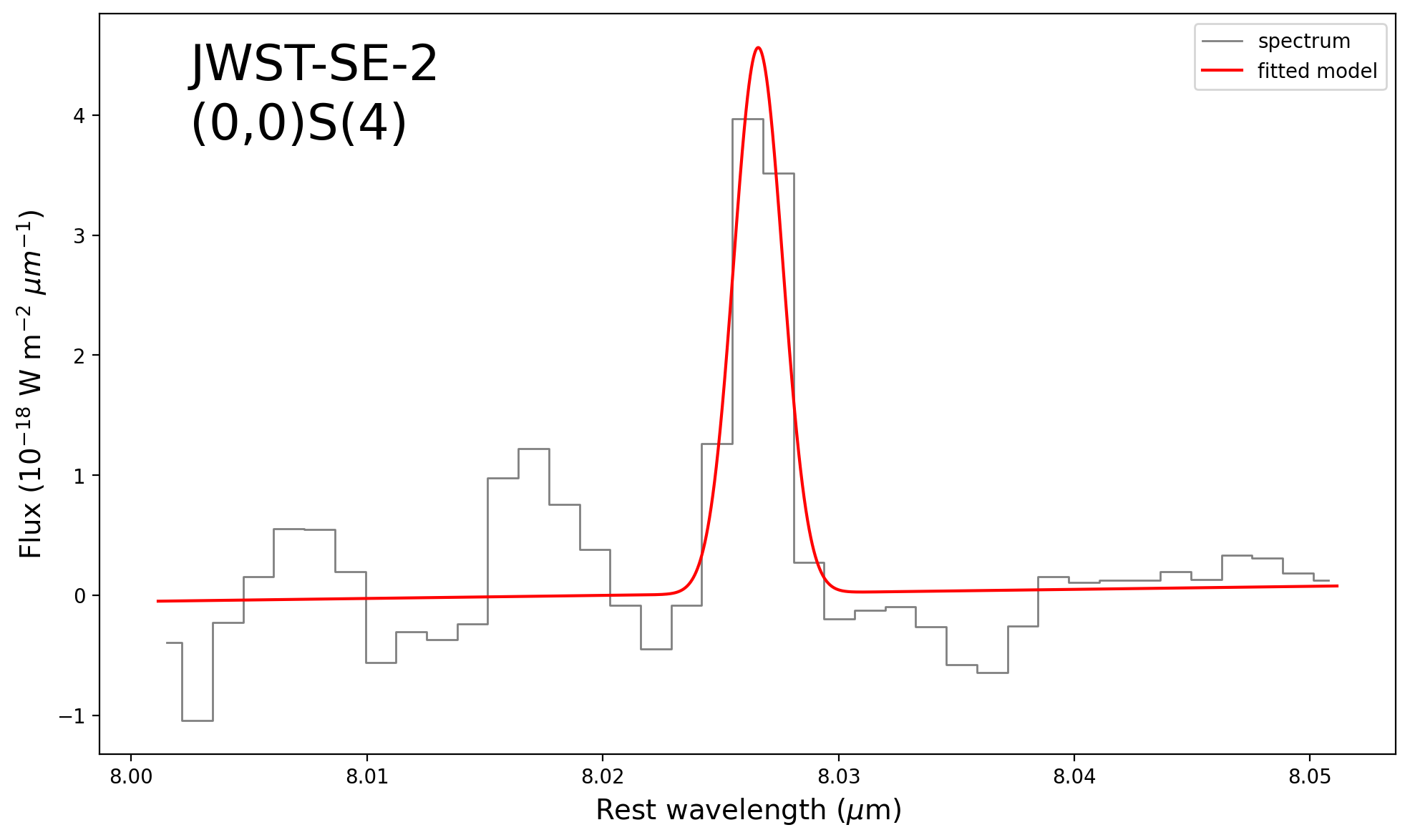} \\ 
\centering
\includegraphics[width=0.22\linewidth]{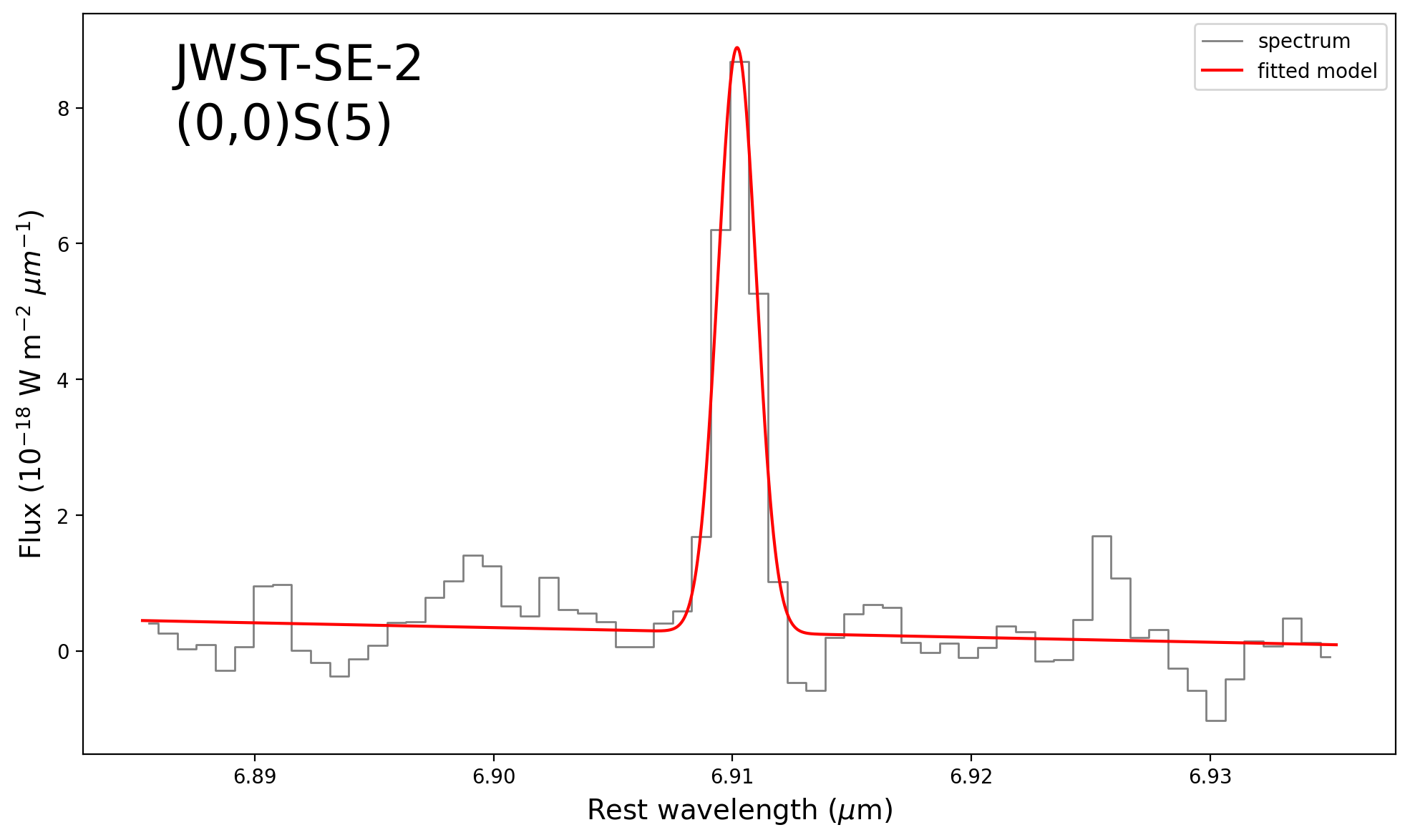}
\includegraphics[width=0.22\linewidth]{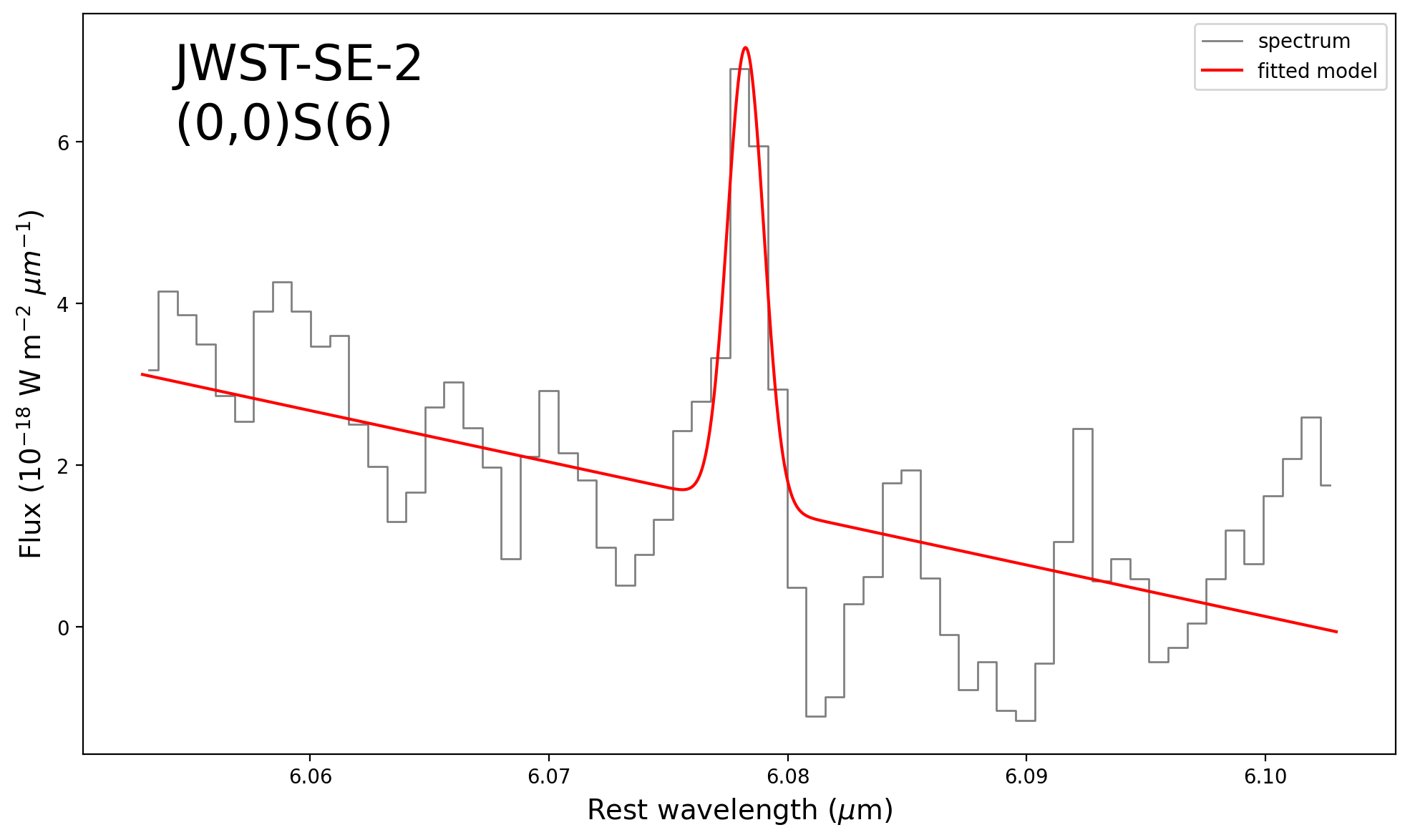}
\includegraphics[width=0.22\linewidth]{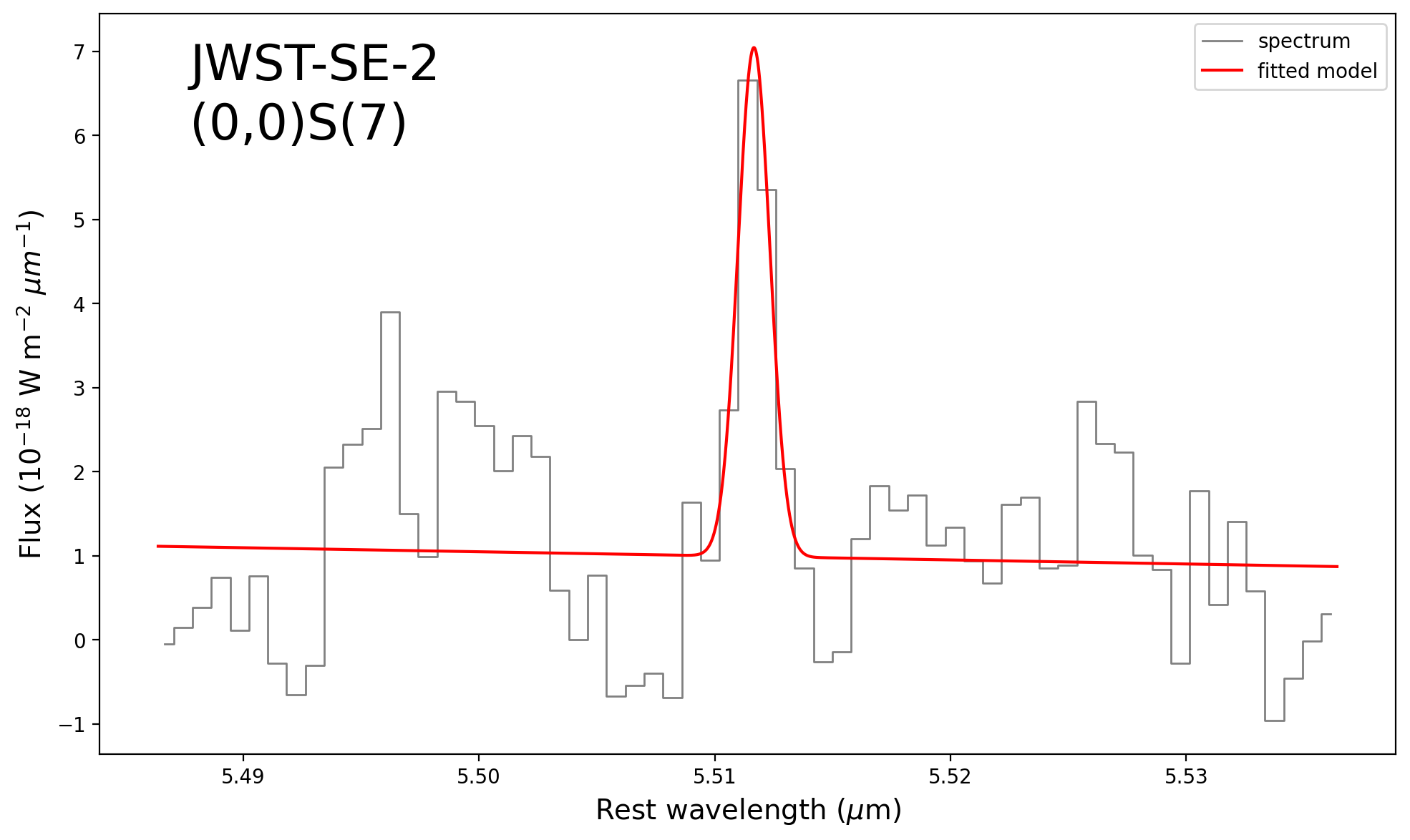}
\caption{Gaussian line fits to the detected \htwo\ lines in a representative region, JWST-SE-2; 
the vertical flux density scale is in units of $10^{-18}$ W\,m\,$^{-2}$\,\micron. 
More details are given in the main text.
}
\label{fig:linefits_h2}
\end{figure}

\textbf{Figure \ref{fig:linefits_h2} shows the fits for
the rotational \htwo\ lines seen in a representative region, JWST-SE-2,
and Tables \ref{tab:nwflux} and \ref{tab:seflux} report the measured
fluxes for all apertures within a radius of 0\farcs65 as described in Sect. \ref{sec:spectra}.
}

\begin{table*}[h!]
\caption{Integrated line fluxes from the convolved cubes in the NW-region 1\farcs3 diameter apertures with S/N$\,\geq\,3$ \label{tab:nwflux}}
\begin{center}
\begin{tabular}{lrcccccc}
\hline
\multicolumn{1}{c}{Line} & \multicolumn{1}{c}{Rest $\lambda$} & NW & VLA-NW-A & VLA-NW-B & VLA-NW-C & CO2-1 & ULX-1\\
& \multicolumn{1}{c}{(\micron)} & \multicolumn{6}{c}{($10^{-21}$\,W\,m$^{-2}$)}\\
\hline
\hline
\multicolumn{8}{c}{\htwo\ lines} \\
\hline
\\
(0,0)S(1) & 17.0348 & 2.78 $\,\pm\,$0.60 & 19.64$\,\pm\,$0.49 &  --- & 4.07$\,\pm\,$0.99 & 2.53$\,\pm\,$0.51 &  --- \\
(0,0)S(2) & 12.2786 &  --- & 7.81$\,\pm\,$0.71 &  --- &  --- &  --- &  --- \\
(0,0)S(3) & 9.6649 & 4.00 $\,\pm\,$1.14 & 23.05$\,\pm\,$1.21 &  --- & 14.48$\,\pm\,$2.89 & 5.50$\,\pm\,$1.46 &  --- \\
(0,0)S(4) & 8.0251 &  --- & 6.13$\,\pm\,$2.02 &  --- &  --- &  --- &  --- \\
(0,0)S(5) & 6.9095 &  --- & 11.64$\,\pm\,$1.66 &  --- &  --- &  --- &  --- \\
(0,0)S(7) & 5.5112 & 35.32 $\,\pm\,$6.97 &  --- &  --- &  --- &  --- & 13.32$\,\pm\,$3.63\\
(0,0)S(8) & 5.053 & 27.21 $\,\pm\,$6.84 &  --- &  --- &  --- & 16.27$\,\pm\,$4.62 & 19.58$\,\pm\,$3.90\\
\\
\hline
\hline
\end{tabular}
\end{center}
\vspace{-\baselineskip}
\end{table*}
\begin{table*}[h!]
\caption{Integrated line fluxes from the convolved cubes in the SE-region 1\farcs3 diameter apertures with S/N$\,\geq\,3$ \label{tab:seflux} }
\begin{center}
\begin{tabular}{lrccccc}
\hline
\hline
\\
\multicolumn{1}{c}{Line} & \multicolumn{1}{c}{Rest $\lambda$} & JWST-SE-1 & JWST-SE-2 & JWST-SE-3 & SE & VLA-SE\\
& \multicolumn{1}{c}{(\micron)} & \multicolumn{5}{c}{($10^{-21}$\,W\,m$^{-2}$)}\\
\hline
\hline
\multicolumn{7}{c}{\htwo\ lines} \\
\hline
\\
%
(0,0)S(1) & 17.0348 & 25.47 $\,\pm\,$0.70 & 28.93$\,\pm\,$0.73 & 31.38$\,\pm\,$0.75 & 27.73$\,\pm\,$0.77 & 23.27$\,\pm\,$0.70 \\
(0,0)S(2) & 12.2786 & 10.35 $\,\pm\,$1.08 & 12.94$\,\pm\,$0.65 & 15.51$\,\pm\,$0.74 & 12.65$\,\pm\,$1.23 & 10.57$\,\pm\,$0.74 \\
(0,0)S(3) & 9.6649 & 31.47 $\,\pm\,$1.45 & 36.38$\,\pm\,$1.61 & 34.83$\,\pm\,$1.21 & 33.88$\,\pm\,$1.60 & 27.84$\,\pm\,$1.43 \\
(0,0)S(4) & 8.0251 & 7.94 $\,\pm\,$1.79 & 11.87$\,\pm\,$2.30 & 14.11$\,\pm\,$1.64 & 14.41$\,\pm\,$1.61 & 10.08$\,\pm\,$1.87 \\
(0,0)S(5) & 6.9095 & 19.68 $\,\pm\,$1.60 & 17.44$\,\pm\,$1.64 & 23.26$\,\pm\,$1.18 & 20.04$\,\pm\,$1.07 & 14.54$\,\pm\,$1.28 \\
(0,0)S(7) & 5.5112 & 25.12 $\,\pm\,$6.96 & 10.29$\,\pm\,$3.38 &  --- & 28.25$\,\pm\,$6.07 & 8.13$\,\pm\,$2.51 \\
(0,0)S(8) & 5.053 &  --- &  --- &  --- & 15.65$\,\pm\,$4.97 &  --- \\
\\
\hline
\hline
\end{tabular}
\end{center}
\vspace{-\baselineskip}
\end{table*}

\bigskip
\bigskip

\section{Comparison of the \htwo\ population diagram fits with \tu\,=\,2000\,K and \tu\,=\,3500\,K\label{sec:tupper}}

\begin{figure*}[t!]
\includegraphics[width=0.5\linewidth]{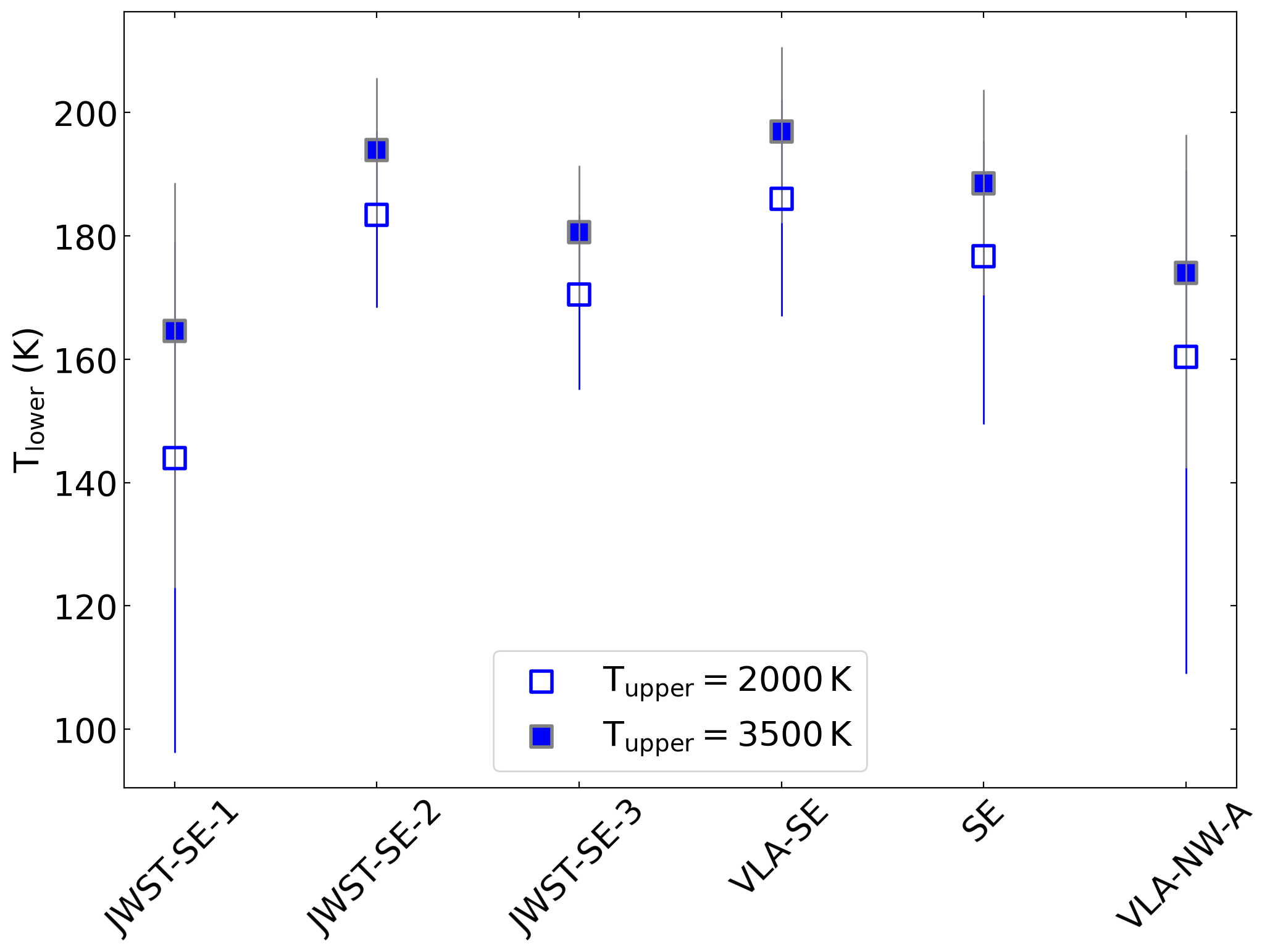}
\includegraphics[width=0.5\linewidth]{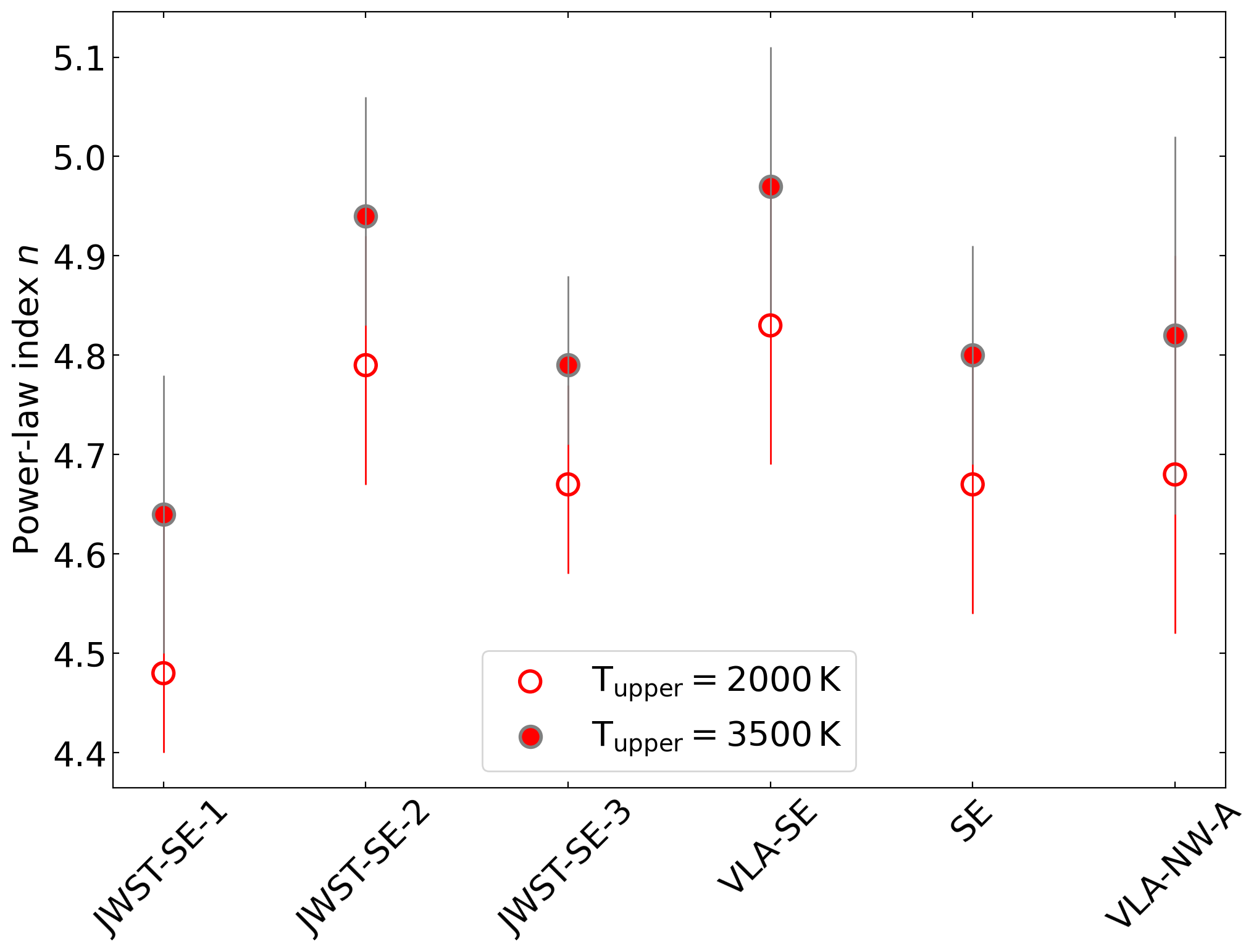} \\
\includegraphics[width=0.5\linewidth]{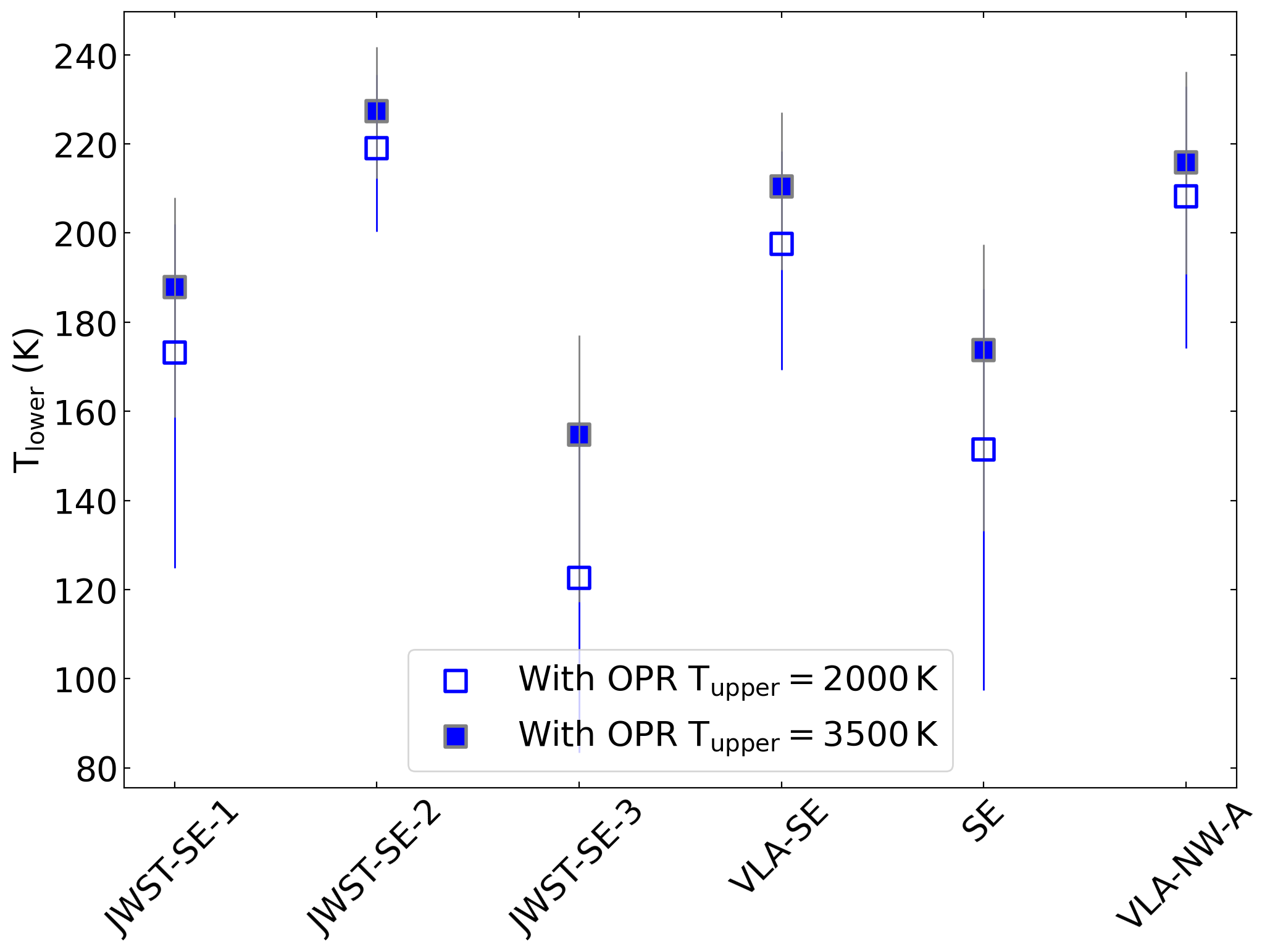} 
\includegraphics[width=0.5\linewidth]{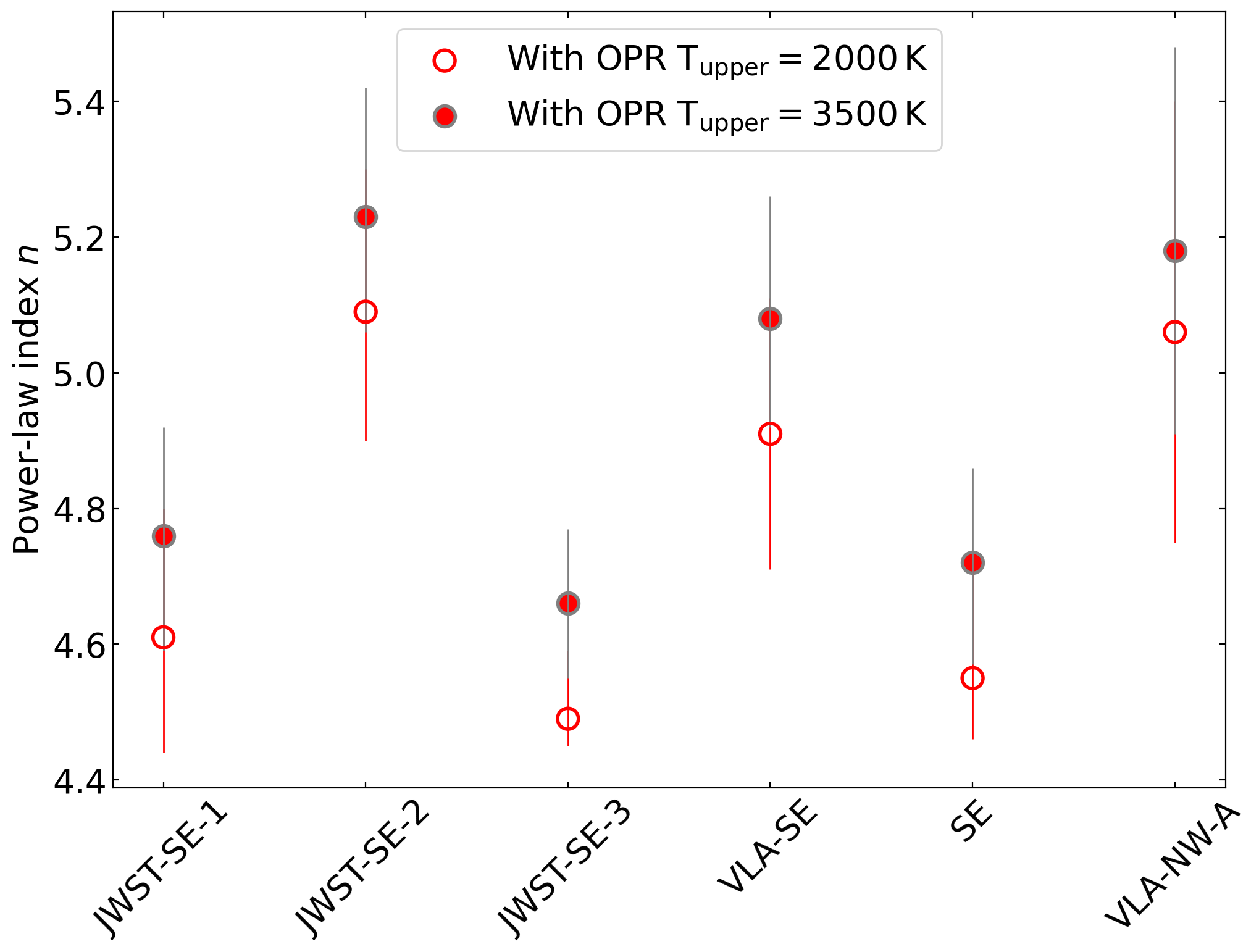} \\
\caption{Comparison of the \htwo\ population diagram best-fit parameters with \tu\,=\,2000\,K and \tu\,=\,3500\,K.
The lower panels including ``With OPR'' in the legends correspond to the 3-parameter fits,
and the upper ones to the 2-parameter fits with OPR\,=\,3.
Filled symbols correspond to \tu\,=\,3500\,K, and open ones to \tu\,=\,3500\,K.
\tl\ comparisons are shown by blue squares, and $n$ by red circles.
}
\label{fig:tupper}
\end{figure*}

We have also fit the  \htwo\ population diagrams with \tu\,=\,3500\,K, and find very similar results.
Table \ref{tab:popfits3500} gives the best-fit values, analogous to Table \ref{tab:popfits}, and
Figure \ref{fig:tupper} compares for the two \tu\ values the best-fit \tl\ and the power-law index $n$.
The similarities of the \htwo\ column densities and masses within the apertures, as well as the OPR, are 
discussed in the main text.

\begin{table*}[h!]
\caption{Best-fit results for \htwo\ population diagrams with \tu\,=\,3500\,K\label{tab:popfits3500}}
\begin{center}
\begin{tabular}{lcccccccrrr}
\hline
\hline
\multicolumn{1}{c}{Region} &
\multicolumn{1}{c}{Number} &
\multicolumn{1}{c}{Max} &
\multicolumn{1}{c}{\tl} &
\multicolumn{1}{c}{$n$} &
\multicolumn{1}{c}{OPR} &
\multicolumn{1}{c}{log$_{10}$} &
\multicolumn{1}{c}{log$_{10}$} \\
& \multicolumn{1}{c}{points} &
\multicolumn{1}{c}{$J_\mathrm{low}$} &
\multicolumn{1}{c}{(K)} &
&& \multicolumn{1}{c}{(\Ntot/cm$^{-2}$)$^{a}$} &
\multicolumn{1}{c}{(\htwo/\msun)$^{b}$} &
\multicolumn{1}{c}{AIC$^{c}$} &
\multicolumn{1}{c}{BIC$^{d}$} &
\multicolumn{1}{c}{$\chi^2$} \\
\multicolumn{1}{c}{(1)} &
\multicolumn{1}{c}{(2)} &
\multicolumn{1}{c}{(3)} &
\multicolumn{1}{c}{(4)} &
\multicolumn{1}{c}{(5)} &
\multicolumn{1}{c}{(6)} &
\multicolumn{1}{c}{(7)} &
\multicolumn{1}{c}{(8)} &
\multicolumn{1}{c}{(9)} &
\multicolumn{1}{c}{(10)} &
\multicolumn{1}{c}{(11)} \\
\hline
\hline
\multicolumn{10}{c}{OPR\,=\,3}\\
\hline
JWST-SE-1 & 6 & 7 & 165$^{+42}_{-24}$ & 4.64$^{+0.14}_{-0.14}$ & $-$ & 18.65 $\,\pm\,$ 0.35 & 2.86 $\,\pm\,$ 0.35 & 4.69 & 4.28 & 6.74 \\
JWST-SE-2 & 6 & 7 & 194$^{+12}_{-12}$ & 4.94$^{+0.11}_{-0.12}$ & $-$ & 18.54 $\,\pm\,$ 0.06 & 2.76 $\,\pm\,$ 0.06 & 7.54 & 7.12 & 10.82 \\
JWST-SE-3 & 5 & 5 & 181$^{+12}_{-11}$ & 4.79$^{+0.08}_{-0.09}$ & $-$ & 18.65 $\,\pm\,$ 0.07 & 2.86 $\,\pm\,$ 0.07 & 6.00 & 5.22 & 7.45 \\
SE & 7 & 8 & 189$^{+18}_{-15}$ & 4.80$^{+0.11}_{-0.11}$ & $-$ & 18.54 $\,\pm\,$ 0.12 & 2.76 $\,\pm\,$ 0.12 & 12.83 & 12.73 & 24.73 \\
VLA-SE & 6 & 7 & 197$^{+15}_{-14}$ & 4.97$^{+0.13}_{-0.14}$ & $-$ & 18.44 $\,\pm\,$ 0.09 & 2.65 $\,\pm\,$ 0.09 & 3.05 & 2.63 & 5.12 \\
VLA-NW-A & 5 & 5 & 174$^{+32}_{-22}$ & 4.82$^{+0.18}_{-0.20}$ & $-$ & 18.49 $\,\pm\,$ 0.26 & 2.71 $\,\pm\,$ 0.26 & 4.70 & 3.92 & 5.75 \\
\hline
\multicolumn{10}{c}{Fitting OPR}\\
\hline
JWST-SE-1 & 6 & 7 & 188$^{+29}_{-20}$ & 4.76$^{+0.17}_{-0.16}$ & 3.61$^{+0.39}_{-0.44}$ & 18.49 $\,\pm\,$ 0.22 & 2.71 $\,\pm\,$ 0.22 & 3.15 & 2.52 & 3.73 \\
JWST-SE-2 & 6 & 7 & 227$^{+15}_{-14}$ & 5.23$^{+0.17}_{-0.19}$ & 3.62$^{+0.24}_{-0.26}$ & 18.38 $\,\pm\,$ 0.06 & 2.60 $\,\pm\,$ 0.06 & 3.64 & 3.01 & 4.05 \\
JWST-SE-3 & 5 & 5 & 155$^{+38}_{-22}$ & 4.66$^{+0.11}_{-0.11}$ & 2.59$^{+0.18}_{-0.19}$ & 18.83 $\,\pm\,$ 0.27 & 3.05 $\,\pm\,$ 0.27 & 3.01 & 1.84 & 2.75 \\
SE & 7 & 8 & 174$^{+41}_{-24}$ & 4.72$^{+0.15}_{-0.14}$ & 2.68$^{+0.26}_{-0.30}$ & 18.64 $\,\pm\,$ 0.35 & 2.86 $\,\pm\,$ 0.35 & 14.48 & 14.31 & 23.49 \\
VLA-SE & 6 & 7 & 210$^{+19}_{-16.7}$ & 5.08$^{+0.16}_{-0.18}$ & 3.30$^{+0.29}_{-0.30}$ & 18.36 $\,\pm\,$ 0.10 & 2.58 $\,\pm\,$ 0.10 & 3.58 & 2.95 & 4.01 \\
VLA-NW-A & 5 & 5 & 216$^{+25}_{-20}$ & 5.18$^{+0.27}_{-0.30}$ & 3.86$^{+0.43}_{-0.44}$ & 18.26 $\,\pm\,$ 0.11 & 2.47$\,\pm\,$ 0.11 & -4.02 & -5.19 & 0.67 \\
\hline
\hline
\end{tabular}
\end{center}
\begin{flushleft}
$^{a}$\,Calculated from Eq. \eqref{eqn:ntot}. \\
$^{b}$\,Calculated from Eq. \eqref{eqn:h2mass}. \\
$^{c}$\,Calculated according to \texttt{SciPy/optimize}: AIC$\,=\,N \ln(\chi^2/N) + 2 N_\mathrm{fit}$ where $N_\mathrm{fit}$ is
the number of fitted parameters, and $N$ is the number of data points.  \\
$^{d}$\,Calculated according to \texttt{SciPy/optimize}: BIC$\,=\,N \ln(\chi^2/N) + \ln(N) N_\mathrm{fit}$.  \\
\end{flushleft}
\end{table*}

\bibliography{izw18,btdrefs_4izw18}{}
\bibliographystyle{aasjournal}



\end{document}